\documentclass[sigconf]{acmart}

\usepackage{subcaption}

\usepackage{cancel}
\usepackage{multirow}

\usepackage{stfloats}
\usepackage{acmart-taps}
\newcommand{\addl}[1]{{#1}}

\newenvironment{add}  % 环境名还是 add（和原命令一致，方便使用）
{\color{black}}        % 环境开始时：设置文字颜色为蓝
{}                    % 环境结束时：不额外操作（自动恢复默认颜色）

\newcommand{\faddl}[1]{\textcolor{black}{#1}}

\newenvironment{fadd}  % 环境名还是 add（和原命令一致，方便使用）
{\color{black}}        % 环境开始时：设置文字颜色为蓝
{}                    % 环境结束时：不额外操作（自动恢复默认颜色）

{\end{sout}}  % 环境结束时：关闭删除线

\AtBeginDocument{%
  }

\copyrightyear{2026}
\acmYear{2026}
\setcopyright{cc}
\setcctype{by}
\acmConference[CHI '26]{Proceedings of the 2026 CHI Conference on Human Factors in Computing Systems}{April 13--17, 2026}{Barcelona, Spain}
\acmBooktitle{Proceedings of the 2026 CHI Conference on Human Factors in Computing Systems (CHI '26), April 13--17, 2026, Barcelona, Spain}
\acmPrice{}
\acmDOI{10.1145/3772318.3790345}
\acmISBN{979-8-4007-2278-3/2026/04}

\newcommand{\methods}{HiSync}

\begin{document}

\title[HiSync]{HiSync: Spatio-Temporally Aligning Hand Motion from Wearable IMU and On-Robot Camera for Command Source Identification in Long-Range HRI}
\author{Chengwen Zhang}
\orcid{0009-0003-4285-7192}
\affiliation{%
  \institution{Department of Computer Science and Technology, BNRist}
  \institution{Tsinghua University}
  \city{Beijing}
  \country{China}
}
\email{zcw25@mails.tsinghua.edu.cn}

\author{Chun Yu}
\orcid{0000-0003-2591-7993}
\authornote{Also with Key Laboratory of Pervasive Computing, Ministry of Education.}
\affiliation{%
  \institution{Department of Computer Science and Technology, BNRist, College of AI}
  \institution{Tsinghua University}
    \city{Beijing}
  \country{China}
}

\author{Borong Zhuang}
\orcid{0009-0003-3369-8134}
\affiliation{%
  \institution{Department of Computer Science and Technology \\
  Tsinghua University}
  \city{Beijing}
  \country{China}
}

\author{Haopeng Jin}
\orcid{0009-0005-7247-5552}
\affiliation{%
  % \institution{Tsinghua University}
  \institution{Beijing University of Posts and Telecommunications}
  \city{Beijing}
  \country{China}
}

\author{Qingyang Wan}
\orcid{0009-0007-3362-9605}
\affiliation{%
  \institution{Academy of Arts \& Design \\
  % \institution{
  Tsinghua University
  }
  \city{Beijing}
  \country{China}
}

\author{Zhuojun Li}
\orcid{0000-0003-4374-9452}
\affiliation{%
  \institution{Department of Computer Science and Technology}
  \institution{Tsinghua University}
  \city{Beijing}
  \country{China}
}

\author{Zhe He}
% \author{Zhoutong Ye}
% \author{Yu Mei}
% \author{Chang Liu}
\orcid{0000-0001-5874-1096}
\affiliation{%
  % \institution{Tsinghua University}
  \institution{Department of Computer Science and Technology}
  \institution{Tsinghua University}
  \city{Beijing}
  \country{China}
}

\author{Zhoutong Ye}
\orcid{0009-0009-9929-4734}
\affiliation{%
  % \institution{Tsinghua University}
  \institution{Department of Computer Science and Technology}
  \institution{Tsinghua University}
  \city{Beijing}
  \country{China}
}

\author{Yu Mei}
\orcid{0009-0008-3126-2974}
\affiliation{%
  % \institution{Tsinghua University}
  \institution{Department of Computer Science and Technology}
  \institution{Tsinghua University}
  \city{Beijing}
  \country{China}
}

\author{Chang Liu}
\orcid{0000-0002-1444-0993}
\affiliation{%
  % \institution{Tsinghua University}
  \institution{Department of Computer Science and Technology}
  \institution{Tsinghua University}
  \city{Beijing}
  \country{China}
}

\author{Weinan Shi}
\authornotemark[1]
\authornote{Corresponding author.}
\orcid{0000-0002-1351-9034}
\affiliation{%
  \institution{Department of Computer Science and Technology}
  \institution{Tsinghua University}
    \city{Beijing}
  \country{China}
}
\email{swn@tsinghua.edu.cn}

\author{Yuanchun Shi}
\orcid{0000-0003-2273-6927}
\authornotemark[1]

\affiliation{%
  \institution{Department of Computer Science and Technology, BNRist}
  \institution{Tsinghua University}
    \city{Beijing}
  \country{China}
}

\affiliation{%
  \institution{Qinghai University}
  \city{Xining}
  \country{China}
}

\renewcommand{\shortauthors}{Zhang et al.}

\begin{abstract}

Long-range Human-Robot Interaction (HRI) remains underexplored. Within it, Command Source Identification (CSI) --- determining who issued a command --- is especially challenging due to multi-user and distance-induced sensor ambiguity. 
We introduce \methods, an optical-inertial fusion framework that treats hand motion as binding cues by aligning robot-mounted camera optical flow with hand-worn IMU signals. 
We first elicit a user-defined (N=12) gesture set and collect a multimodal command gesture dataset (N=38) in long-range multi-user HRI scenarios.
Next, \methods~ extracts frequency-domain hand motion features from both camera and IMU data, and a learned CSINet denoises IMU readings, temporally aligns modalities, and performs distance-aware multi-window fusion to compute cross-modal similarity of subtle, natural gestures, enabling robust CSI. 
% with mechanisms that are designed to 降低IMU信号噪声，时间对齐，根据距离综合不同窗口的信息，并最终输出 cross-modal similarity of subtle, natural gesture motions, enabling robust CSI. 
In three-person scenes up to 34\,m, \methods~ achieves \addl{92.32\%} CSI accuracy, outperforming the prior SOTA by \addl{48.44\%}. 
\addl{\methods~ is also validated on real-robot deployment.} By making CSI reliable and natural, \methods~ provides a practical primitive and design guidance for public-space HRI.

\end{abstract}





% \afterpage{ % 核心命令：让图片在当前页之后的页面顶部显示
%   \begin{figure*}[t!] % t! 强制优先放在页顶，不浮动到后面
%     \centering
%     \includegraphics[width=0.9\textwidth]{pics/Teaser_revised.jpg}
%     \caption{\textbf{Demonstration of \methods}. Multiple people perform similar gestures at a distance of 34 m. \addl{The figure illustrates the application scenario: Person 1 (blue link) controls the quadruped, while Person 2 (orange link) simultaneously controls the drone. Other bystanders act as visual distractors. Each robot receives an inertial stream from its paired device. \methods~ enables the robot to identify its bound command issuer, effectively rejecting distractors regardless of their kinematic similarity.}}
%     \label{fig:teaser}
%   \end{figure*}
% }

% \todo{teaser30m waving}


\maketitle

% motion as fiducials

% \noindent
% \begin{minipage}{\textwidth}
%   \centering
%   \includegraphics[width=0.9\textwidth]{pics/Teaser_revised.jpg}
%   \captionof{figure}{\textbf{Demonstration of \methods}. Multiple people perform similar gestures at a distance of 34 m. \addl{The figure illustrates the application scenario: Person 1 (blue link) controls the quadruped, while Person 2 (orange link) simultaneously controls the drone. Other bystanders act as visual distractors. Each robot receives an inertial stream from its paired device. \methods~ enables the robot to identify its bound command issuer, effectively rejecting distractors regardless of their kinematic similarity.}}
% \end{minipage}

\begin{figure*}[ht]
  \centering
  \includegraphics[width=\textwidth]{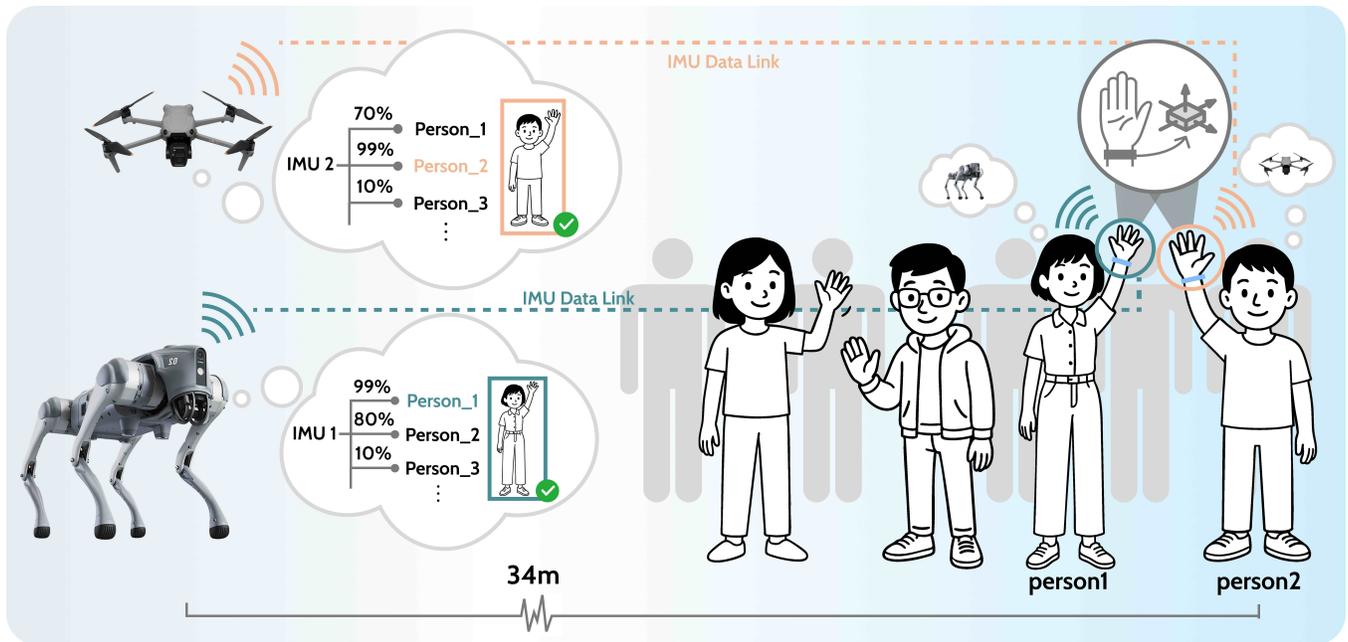}
  % \caption{The Demonstration of \methods. Multiple people perform similar gestures at a distance of 34 m. \methods~ enables the robot to identify its bound command issuer even under such long-range and multi-person conditions.}
  \caption{\textbf{Demonstration of \methods}. Multiple people perform similar gestures at a distance of 34 m. \addl{The figure illustrates the application scenario: Person 1 (blue link) controls the quadruped, while Person 2 (orange link) simultaneously controls the drone. Other bystanders act as visual distractors. Each robot receives an inertial stream from its paired device. \methods~ enables the robot to identify its bound command issuer, effectively rejecting distractors regardless of their kinematic similarity.}}
  \Description{HiSync scenario: two robots identify their bound human controllers among multiple people at 34 meters. On the left, a drone and a quadruped robot each connect via a color-coded inertial measurement unit data link (blue to Person_1, orange to Person_2) to a group of four people on the right who all raise one hand in similar gestures. Insets above each robot show likelihood tables from its inertial measurement unit over three candidate users; the quadruped’s table gives 99 percent to Person_1, while the drone’s gives 99 percent to Person_2, and the selected person in each table is outlined and marked with a check. Greyed human silhouettes behind the four foreground people represent visual distractors, and thought bubbles above Person_1 and Person_2 contain icons of the quadruped and drone, indicating which robot each user intends to control.}
  \label{fig:teaser}
\end{figure*}

\section{INTRODUCTION}

% 研究背景，强调场景越来越重要

% 主流的交互使用相机的方法

% 为了解决这个问题，人们提出了一些什么新颖方法，
% we aim to ...  and  introduce
% 
% 存在两个问题，识别不到，判断不了是谁

% 远距离场景是存在的-导致了一些近距离没有的问题-所以我们要提到这个事情

% 未来场景下，人跟机器人（场景描述）

% 机器人会越来越多，各种形态，交互的能力和空间范围越来越广。Consider summoning an autonomous car across a parking lot, directing an inspection drone outdoors, or calling a service robot in a large venue: all require intuitive and reliable interaction at distances. 交互过程中要求机器人对操作主体做认证。在现在的普遍认知下，没有多人多机的方式，现在的近距离是怎么解决。但是随着交互空间的扩大，很多方法会失效，在远距离中机器人很多，人很多，之前就不好使了 （为什么产生了远距离的场景）

% 第二段远距离的challenge

% 为了弥补这些我们提出了

% =======
% As embodied AI systems such as drones, quadrupeds, and service robots increasingly enter everyday environments, 

% HRI是一个raising 的领域，因为之后机器人会越来越多，各种形态，交互的能力和空间范围越来越广，机器人会逐渐进入我们的日常生活. Consider summoning an autonomous car across a parking lot, directing an inspection drone outdoors, or calling a service robot in a large venue These scenarios are characterized by long ranges (10m+), dynamic multi-user contexts, and complex environments  all require intuitive and reliable interaction at distances. The demand for far-range human-robot interaction is becoming pressing.

Human-robot interaction (HRI) is a vibrant and fast-growing area within HCI \cite{hosseini2023towards,villanueva2021robotar}. However, the vast majority of HRI research  \cite{hostettler2025real,kamikubo2025beyond,lee2022unboxing,zhou2023tactorbots,lyu2025signaling} focuses on near-range or single-person settings --- within $10$\,m where the operator is visually salient and easy to disambiguate.
At the same time, advances in robotics and real-world deployments are driving interactions into larger public spaces and more diverse embodiments \cite{matheus2025long,wang2024multimodal}. 
Many emerging use cases require interaction at a distance. For example, summoning an autonomous car across a parking lot, directing a drone outdoors, or calling a service robot in a large venue \cite{bamani2024ultra,zhong2025skybound,chang2024must}. These settings introduce long distances ($\geq 10$\,m), dynamic multi-user crowds, and cluttered, uncertain environments, demanding interfaces that remain reliable from afar.

A foundational perceptual capability in such long-range scenarios is \textbf{Command Source Identification (CSI)} --- determining who in the scene issued the command. In near-range HRI, CSI is often trivial or implicitly handled because the operator is close, prominent, and rarely confused with bystanders \cite{yura2025multi,bandi2025action,park2023clara}. In long-range contexts, however, distance-induced visual ambiguity and simultaneous gestures from multiple users make CSI difficult. This gap motivates our work: we extend HRI beyond near-range by focusing on robust CSI as a prerequisite for reliable long-range interaction.
\faddl{To scope this work, we consider scenarios in which a single user has already established a one-to-one connection with the robot, while other users only act as visual distractors.}

\begin{figure}[h]
  \centering
  \includegraphics[width=\linewidth]{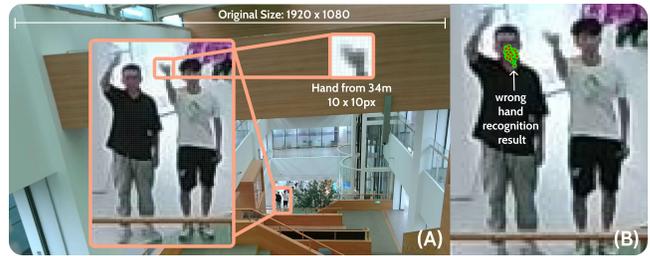}
  \caption{\textbf{Visual Ambiguity at a Distance of 34 m.} \addl{Figure (A) shows a real sample from our dataset (1920 $\times$ 1080 resolution). The inset highlights that the hand region occupies fewer than 10 $\times$ 10 pixels. Even with a manually zoomed-in view of the detection result like (B), YOLOv11x~\cite{Jocher_Ultralytics_YOLO_2023} fails to identify the hand. This demonstrates the inherent visual ambiguity in long-range interactions.} } 
\Description{This figure shows two images demonstrating visual ambiguity in recognizing hand gestures at long range. (A) A real sample from our dataset, captured at a resolution of 1920x1080, showing a hand gesture at a distance of 34 meters. (B) An example of incorrect hand recognition despite manual zoom-in, highlighting the difficulty of accurate gesture identification over long distances.}
  \label{fig:ambiguity}
\end{figure}

We identify three key challenges introduced by long-range CSI, which were not covered by existing literature.
% Through the literature review above, we identify three key challenges introduced by long-range CSI.
\textbf{(C1) Unnatural Interaction}. Existing methods \cite{wang2019person, olson2011apriltag, sun2020when} require users to perform substantial movements or set up additional devices, which can be unnatural and burdensome.
\textbf{(C2) Visual ambiguities}. At long ranges, hands and subtle wrist motions collapse to only a few pixels on typical robot-mounted cameras; appearance cues such as texture, skin tone, and fingers are degraded, and hand estimators become unreliable (Fig. \ref{fig:ambiguity}).
% in open multi-user scenes, background flow, compression, and motion blur corrupt visual motion; hand-worn IMUs pick up locomotion, grip jitter, and soft-tissue artifacts; minor cross-sensor lags and distance-dependent motion scaling further smear temporal structure, making raw time-domain matching brittle.
\textbf{(C3) Data noise}.
At long ranges, both optical flow and IMU signals are plagued by noise. Visual flow is distorted by background clutter and motion blur \cite{baker2011database}, while IMU readings are affected by inherent noise and grip jitter \cite{suvorkin2024assessment}. Moreover, sensor lags and sample-rate mismatch further complicate reliable CSI \cite{furgale2013unified}.
We introduce \methods, which tackles the challenges of command source identification in long-range human-robot interaction. 
Firstly, to address \textbf{C1}, \addl{we use common low-cost on-robot camera and IMU as CSI sensors.} Moreover, we focus on hand gestures as the interaction modality, as they are intuitive and commonly used in daily life. To this end, we conduct a formative study to elicit a user-defined gesture set from 12 participants.

% To this end, we conduct a formative study to elicit a user-defined gesture set from 12 participants.

Secondly, to solve \textbf{C2}, \methods~ is designed as an optical-inertial fusion framework that treats hand motion as a binding cue: it aligns robot-mounted camera optical flow with hand-worn IMU traces to disambiguate the command issuer. By treating RGB as optical flow and matching candidates on motion cues, \methods~ remains reliable when visual details are blurred.
% Secondly, to solve \textbf{C2}, we introduce \methods, a framework for robust long-range CSI. It uses \textbf{Optical-Inertial} perception that leverages optical flow from robot-mounted cameras and hand-worn IMU signals as CSI  . 
% Third, to handle \textbf{C3}, \methods maps both optical flow and IMU traces to the frequency domain and designs a CSINet within which, Quality-Aware Feature Modulation reweights IMU spectral features by quality cues; IMU-Anchored Cross-Modal Attention aligns flow-spectral sequence to the IMU time reference, reducing time jitter, missed detections, and phase drift; and Scale-Aware Multi-Window Fusion aggregates multiple window lengths conditioned on distance, stabilizing decisions cross ranges. Together, these modules raise yield a more stable similarity under high noise

Thirdly, to handle \textbf{C3}, \methods~ maps optical flow and IMU to the frequency domain, leveraging a novel CSINet that incorporates the following mechanisms. Quality-Aware Feature Modulation reweights IMU spectral features by quality cues to suppress frames with low signal-to-noise ratio (SNR). IMU-Anchored Cross-Modal Attention aligns flow-spectral sequences to the IMU time reference, mitigating time jitter, missed detections, and phase drift. Scale-Aware Multi-Window Fusion aggregates predictions across multiple window lengths conditioned on distance, stabilizing decisions across ranges. Together, these modules yield a more robust cross-modal similarity in noisy long-range settings.
% It also transforms time-domain data to the frequency domain, featuring a \textbf{Spectral Neural Network}, aiming to handle \textbf{C2}. 

% Furthermore, we curate, to our knowledge, the first large-scale dataset of long-range gestural interactions. Evaluations on this dataset show that \methods~ sustains reliable CSI performance from 3m to 34m with 97.82\% overall accuracy, outperforming SOTA 27.3\%, and at distances up to 34 m, remains accuracy of 94.31\% (SOTA 43.88\%), demonstrating its applicability in open, multi-user environments. This dataset also enables a reproducible benchmark in this underexplored domain.

Furthermore, to address the lack of public datasets for CSI in long-range multi-user settings, we curate, to our knowledge, the first large-scale multimodal dataset of long-range gestural interactions. 
\faddl{In this dataset, each frame has at least 2 people gesturing at the on-robot camera. Only one person was equipped with an IMU, while bystanders acted as visual-only distractors. 
We acknowledge that this setup simplifies the CSI problem compared to scenarios where multiple users simultaneously transmit IMU data to the same robot. 
However, evaluations on this setting still show that HiSync demonstrates the feasibility of optical-inertial binding for CSI and exhibits robustness to visual modality interference.
}

% Our work is based on the hypothesis that the robot and the user have already established a one-to-one connection.

% However, HiSync addresses a critical gap: Visual-Inertial Binding. While the robot receives the command signal, it does not inherently know which person in its camera field of view corresponds to that signal.

%试试这个，这个我放到Discussion里面感觉合适，这里不做过多的辩解
%那能否故事侧重一点binding，太对了（）但是现在改不了
%CSI的I天然带了 knowing where the command came from，所以不用大改
%你把CSI相关的贡献稍微往那边扯一点，然后承认说虽然实验没有考虑IMU信号冲突，但仍然验证了binding的可行性，以及对视觉模态的干扰的鲁棒性。根据该这两段局部可行，我思考一下
% However, HiSync still tackles the critical challenge of visual-inertial binding. Without binding visual input to inertial data, the robot would receive a command issued as an inertial signal without knowing where the command came from. Therefore, the visual-inertial binding capability afforded by HiSync enables the robot to correctly respond to distant inertial commands.

On this dataset, \methods~ achieved 97.82\% overall accuracy from 3--34\,m, exceeding the previous SOTA by 27.3\% points. Even at 34\,m, it maintained 94.31\% accuracy, while the previous SOTA achieved only 43.88\%. These results demonstrate \methods's applicability to open, multi-user environments. The dataset also establishes a reproducible benchmark for this underexplored problem.

In summary, our main contributions are as follows:
\begin{itemize}
% \item We provide a systematic account of natural long-range gestures, showing how they adapt to distance.  
% \item We introduce HiSync, an optical-inertial fusion architecture that leverages spectral motion cues to achieve robust CSI under visual ambiguity and sensor noise.  
% \item We release a large multimodal dataset (38 participants, 3-34\,m) and establish a benchmark for reproducible evaluation of long-range gestural CSI. 

\item We present a systematic user-defined natural gesture set for multiple robot types at long range, showing how gesture properties scale with distance.
% \item We introduce \methods, an optical-inertial fusion architecture that leverages spectral motion cues from both noisy 的视觉和IMU信息 to achieve robust CSI 在远距离下.
\item \addl{We introduce \methods, an optical-inertial fusion architecture that leverages spectral motion cues extracted from low-cost and widely deployed sensors (on-robot camera and wearable IMU) to achieve robust CSI at long range.}
% \item We carefully benchmark and discuss \methods, and also release a large multimodal dataset covering 3-34\,m ranges and multi-gesture types collected on 38 real users, empowering future long-range HRI research.
\item We evaluate \methods~and release a large, annotated multimodal dataset\footnote{https://github.com/OctopusWen/HiSync} spanning 3-34\,m and multiple gesture categories, collected from 38 participants, to enable reproducible research evaluation on long-range HRI.
\item \addl{We deploy \methods~ on a quadruped robot to validate its real-world feasibility. Field experiments confirm robust performance. User studies demonstrate superior usability and low burden compared to standard interfaces.}

% We validate the system's viability through an online deployment on a quadruped robot. Results show high robustness to network jitter and active interference in open spaces, outperforming traditional controllers in user preference at long ranges.

\end{itemize}
% 技术贡献点：motion（频域） as fiducials 提出这个问题来解决跨模态同步问题 或者光惯融合

Together, these contributions establish gesture-based command source identification as a distinct research challenge in long-range HRI, offering both empirical insights and technical foundations that open new avenues for designing robust, scalable human-robot interaction systems.

% RQ1. How do people naturally summon embodied systems across 3-30 m, and how do gesture properties vary with distance and device type?
% RQ2. Can a calibration-free fusion of inertial and visual sensing achieve robust, low-power recognition of far-range summoning gestures, even under multi-person interference?

% To address these questions, we make three contributions:

% Behavioral study. We present the first systematic account of natural summoning gestures toward drones, quadrupeds, and service robots across distances up to 30 m, revealing how amplitude and temporal patterns scale with range and differ by device.

% Robust recognition framework. We introduce a real-time system that fuses IMU and optical flow without precise calibration, achieving robustness to noise and multi-person interference while remaining efficient enough for edge deployment.

% Benchmark dataset. We release the first far-range summoning dataset, covering indoor/outdoor environments, 3-34 m distances, and six core gestures, enabling reproducible evaluation in this underexplored interaction space.

% Together, these contributions establish far-range summoning gestures as a distinct research domain in human-robot interaction, bridging the gap from natural user behavior to deployable recognition technology.

\section{RELATED WORK}

\subsection{Long-range Human-Robot Interaction}

Long-range human-robot interaction (HRI) has explored multiple input channels, including microphones for speech or prosodic cues \cite{velner2020intonation,cauchard2015drone,yoshimura2019extending,zhang2025can,cambre2019one}, screen-based or on-robot displays for directing and confirming actions \cite{chen2023comparing}, hand-held or wearable controllers for precise teleoperation at a distance \cite{wang2025systematic,trinitatova2023study,suzuki2022augmented,cichon2018digital,li2023restype}, and free-hand gestures detected by onboard or remote cameras \cite{dang2025user,bhatnagar2023long,male2021recognition,bamani2024ultra}. Among these, hand gesture interaction is often favored for its low barrier and social legibility across users and settings \cite{leusmann2025approach,hosseini2023towards,BaileyJohnson2020_HumanCenteredGestures,cauchard2015drone}.

Yet most gesture-based HRI systems are designed for close-range use, typically within arm's reach to a few meters,
% most gesture-based HRI systems target close-range use—within arm's reach to a few meters—
where tracking and disambiguation are simpler and environments are more controlled \cite{delpreto2020plug,leusmann2025approach,cichon2018digital}. Only a small subset explicitly addresses far-field perception, and these works largely prioritize recognition robustness over interaction breadth \cite{bhatnagar2023long,male2021recognition,bamani2024ultra}. Despite these technical explorations, studies situated in public spaces or crowd settings remain sparse, and few examine how people would naturally gesture to summon and direct robots from tens of meters away without prior training \cite{cauchard2015drone,leusmann2025approach,hosseini2023towards}. This gap motivates our focus on free-hand commands for long-range, multi-user scenarios: we first conduct a formative study to elicit a user-defined gesture set and characterize distance-mediated communication, which then informs our system design and evaluation.

% prior work commonly assumes a single active user and a pre-defined gesture vocabulary, focusing on improving detection under distance, view, and illumination changes rather than resolving multi-user ambiguity \cite{delpreto2020plug}. 

% \subsection{Gesture-based Command Source Identification}

% HiSync的核心目标是实现远距离多人手势HRI的Command Source Identification。To the best of our knowledge，远距离手势CSI这一任务为首次作为明确任务被系统提出，与之最接近的是Person Localiztion，Re-ID等目标识别与追踪任务，虽然CSI与这些任务的侧重点不同，但是在技术路线上有高度相似性，围绕这些任务展开的工作具备高度参考价值。To summarize previous research，we have compiled related work on ring-based input in Table 1.

%  现有可用于 CSI 的技术路线主要有三类：纯视觉深度学习、显式标记/信标、以及视觉-惯性（VI）融合。纯视觉方法，如\cite{wei2018person,zhou2019omni,li2020learning,li2019global,chen2019abd}基于深度学习的方法利用特征匹配从而实现从多个候选人中找出目标，然而这些方法往往依赖建立目标人员信息数据库，无法脱离先验，所用的数据集多在中近距离且内容上遮挡较少，对于在30m以上的超远距离,这类生物特征极易退化或混淆\cite{gupta2025mimicgait,knoche2021susceptibility,huang2024occluded}。显式标记的方法，如视觉标记\cite{olson2011apriltag,kalaitzakis2021fiducial,berral2024deeparuco++}和UWB标签\cite{xianjia2021applications，kwon2024uwb}等，可比较有效的解决纯视觉方法的先验依赖和环境干扰问题，然而需要提前部署基站或者让用户佩戴特殊标签，这限制了方法的应用范围，并损害交互自然性。为了更好的解决这个问题，学界也出现了利用视觉-惯性融合的方法， Henschel 等 \cite{henschel2019simultaneous} 以视觉预测朝向与背部 IMU 方位进行匹配完成身份绑定；Sun 等 \cite{sun2020when} 则通过对比学习将候选人的局部光流-朝向-尺度与手机 IMU信号投射到共享嵌入空间，在无外观先验下识别佩戴者。上述工作主要验证了惯性-视觉融合在长时间遮挡和人物离开视野后可进行识别具有鲁棒性，体现出VI融合在远距离手势CSI的潜力，然而When We First Met也表明在人物像素稀疏时，鲁棒的匹配仍是一大挑战。本工作为解决这一问题作出了探索

 \subsection{Gesture-based Command Source Identification}

The core goal of \methods~ is \emph{command source identification (CSI)} for long-range, multi-user, gesture-based HRI --- given several people and a detected command, determine who issued it \cite{cauchard2015drone,leusmann2025approach,hosseini2023towards,wang2025computing}. To the best of our knowledge, far-field hand-gesture CSI has not been explicitly defined and evaluated as a standalone task in prior HRI literature, which motivates our formulation of CSI as distinct from person localization, person search, and ReID \cite{wei2018person,zhou2019omni,li2020learning,li2019global,chen2019abd,salinas2024multimodal}. Closely related pipelines in vision emphasize detection-association-metric learning for persistent identity, whereas CSI binds an ephemeral command to its current actor without long-term identity enrollment \cite{wei2018person,li2019global,chen2019abd}. For completeness, we summarize input modalities that inform multi-user gesture settings in Table~\ref{tab:ring-input}.

Existing routes for CSI fall into three families --- vision-only deep learning~\cite{chen2019abd}, \addl{device-assisted localization~\cite{kwon2024uwb,xianjia2021applications}}, and visual-inertial (VI) fusion \cite{henschel2019simultaneous,sun2020when}. \textbf{Vision-only} ReID/person-search match learned appearance features but typically require a gallery and mid-range conditions; at $\geq 30$\,m, sparse resolution, occlusion, and confounds make identity brittle without priors \cite{wei2018person,zhou2019omni,li2020learning,li2019global,chen2019abd,gupta2025mimicgait,knoche2021susceptibility,huang2024occluded}.
\addl{\textbf{Device-assisted localization} leverages "privileged data" from user to bind identity, yet struggles in long-range ad-hoc scenarios. High-precision solutions require expensive hardware (e.g., LiDAR~\cite{furst2021hperl}) and pre-deploying infrastructure in the scene (e.g., AprilTag, ArUco and UWB ~\cite{olson2011apriltag,kalaitzakis2021fiducial,berral2024deeparuco++,xianjia2021applications,kwon2024uwb,xiao2022exploring}). Ubiquitous signals like WiFi~\cite{wang2019person}, GPS~\cite{merry2019smartphone} and mmWave~\cite{lee2023hupr} lack the fine-grained spatial resolution required to disambiguate multiple users in common public HRI scenarios, where individuals often stand closely at a distance.
% : markers (e.g., AprilTag and ArUco) and UWB tags~\cite{olson2011apriltag,kalaitzakis2021fiducial,berral2024deeparuco++,xianjia2021applications,kwon2024uwb,xiao2022exploring} provide robust binding yet demand infrastructure or wearables; GPS~\cite{merry2019smartphone} lacks fine-grained precision; WiFi~\cite{wang2019person} and mmWave~\cite{lee2023hupr} are limited to room-scale sensing; and LiDAR~\cite{furst2021hperl} is prohibitively expensive and sparse at distance.
}
\textbf{VI} association links on-body IMU with visual motion (e.g., heading matching and contrastive phone-IMU alignment), but still degrades when people occupy few pixels or gestures are brief \cite{henschel2019simultaneous,sun2020when}. 
Building on this line, \methods~ binds the command via hand-motion spectra, aligning robot-mounted cameras with \addl{low-cost and widely deployed} IMU within the command window 
to avoid identity enrollment while preserving uninstrumented, free-hand interaction \cite{cauchard2015drone,leusmann2025approach,hosseini2023towards,park2021learning}.

\begin{table*}[htbp]
\centering
\caption{Representative works related to long-range gesture CSI.}
\label{tab:CSI_overview}
\begin{tabular}{p{2.5cm} p{3cm} p{4cm} p{4.2cm}}
\toprule
\textbf{Work} & \textbf{Sensors} & \textbf{Data Modalities} & \textbf{Supported Distance} \\
\midrule
PTGAN\cite{wei2018person} & Camera & RGB images & 5--20\,m\\
OSNet\cite{zhou2019omni} & Camera & RGB images & 5--20\,m\\
CASE-Net\cite{li2019global} & Camera & Video & 5--20\,m\\
ABD-Net\cite{chen2019abd} & Camera & RGB images & 5--20\,m \\
\midrule
AprilTag\cite{olson2011apriltag} & Camera + Fiducial Tag & Visual Markers & 3--10\,m\\
DeepArUco++\cite{berral2024deeparuco++} & Camera + Fiducial Tag & Learned Fiducial Detection & 3--10\,m \\
UWB\cite{xianjia2021applications} & UWB Tags & RF Ranging &  30\,m~\addl{*Need Base Stations}\\
\addl{GPS\cite{merry2019smartphone}} & \addl{GPS Receiver} & \addl{Global Location} & \addl{*Error Range:7--13\,m}\\
\addl{Person-in-WiFi\cite{wang2019person}} & \addl{MIMO WiFi Tx\&Rx} & \addl{RF} & \addl{3--6\,m}\\
\addl{HuPR\cite{lee2023hupr}} & \addl{mmWave Radar} & \addl{mmWave} & \addl{3--6\,m}\\
\addl{HPERL\cite{furst2021hperl}} & \addl{Expensive LiDAR} & \addl{Laser} & \addl{5--20\,m}\\
\midrule
Simultaneous\cite{henschel2019simultaneous} & Camera + IMU & Video + Inertial Orientation & 5-10\,m \\
VIPL\cite{sun2020when} & Camera + IMU & Optical Flow + IMU & <10\,m \addl{*Need Walking} \\
\midrule
\addl{\methods} & \addl{Camera + IMU} & \addl{Optical Flow + IMU on Spectral} & \addl{3--34\,m}\\

\bottomrule

\end{tabular}
\label{tab:ring-input}
\Description{A table comparing related sensing works against the proposed HiSync method across sensors, data modalities, and distance. The columns list the Work name, Sensors used, Data Modalities, and Supported Distance. The rows are grouped into three categories: vision-only methods, hardware-specific approaches (such as Fiducial Tags, Ultra-Wideband, Global Positioning System, and Radar), and sensor fusion. While vision methods typically range from 5 to 20 meters and hardware solutions have specific constraints like needing base stations, the proposed HiSync method is highlighted in the final row. Utilizing Camera plus Inertial Measurement Unit (IMU) sensors with Optical Flow, HiSync demonstrates the widest supported distance of 3 to 34 meters, exceeding prior Camera-IMU works which are limited to under 10 meters.}
\end{table*}

\subsection{Vision-Inertial Fusion Approaches}

Recent vision-inertial (VI) work, \addl{like VIPL~\cite{sun2020when}, learns shared embeddings that align optical flow with IMU signals via contrastive training. Nevertheless, its reliance on walking scenarios with 5-second full-body motion, assumption of tight video-IMU synchronization, and short-range limitation (<10\,m) collectively restrict its deployment in real-world HRI settings.} And inertial signals also suffer long-term noise and bias \cite{wall2006characterization,jaskot2010inertial}; IMU-only estimates are especially sensitive \cite{henschel2019simultaneous}. Without quality-aware gating, such errors propagate through fusion and reduce robustness.

Beyond synchronization, alignment is intrinsically hard due to distance: optical flow and IMU differ in units, sampling rates, and noise profiles, so simple shared embeddings underperform without additional constraints or priors \cite{cheshmi2025improving,moon2023imu2clip,ray2025improving,liu2025core4d}. Temporal window length strongly shapes VI performance \cite{niu2018vipl,sun2020when}, yet most systems fix a single window and under-discuss its impact across users and motions. These gaps motivate fusion that is synchronization-tolerant, quality-aware, and adaptive in the temporal context.

\section{FORMATIVE STUDY}

    % \begin{table}[htbp]
    % % 不同距离下用户preference的手势，径向和横向的比例
    %   \caption{Comparison of Gesture Recognition Methods}
    %   \label{tab:commands}
    %   \begin{tabular}{ccl}
    %     \toprule
    %     Paper & Device & Range\\
    %     \midrule
    %     \texttt{{\char'134}author} & 100& Author \\
    %     \texttt{{\char'134}table}& 300 & For tables\\
    %     \texttt{{\char'134}table*}& 400& For wider tables\\
    %     \midrule
    %     Ours & & \\
    %     \bottomrule
    %     \label{tab:related}
    %   \end{tabular}
    % \end{table}

Existing user-defined gesture sets for HRI rarely account for long-range, open, multi-user settings, limiting their usefulness for Command Source Identification (CSI) \cite{hosseini2023towards}. To inform algorithm and system design under these conditions, we conducted a formative study to \textbf{(1)} elicit CSI-capable, user-preferred gestures for long-range HRI and \textbf{(2)} examine how these gestures change with distance to guide the design and optimization of long-range CSI systems.

% To support Gesture-based Command Source Interaction (CSI) in long-range, multi-user scenarios, we first identify the gestures users naturally employ when interacting with robots, then analyze their characteristics to drive method design. However, prior user-defined gesture sets have largely overlooked the influence of distance and open multi-user contexts \cite{hosseini2023towards}, thereby limiting their practical applicability. To address this gap, we conducted the formative study to investigate \textbf{(1)} the types of user-defined gestures people prefer for HRI while enabling CSI, and \textbf{(2)} how these gestures evolve across distances in ways that could guide the design and optimization of long-range CSI systems.

% 有两个motivation，一个是研究用户希望对不同的设备使用什么指令，另一个是用户在不同距离下会采用什么样的手势（主要）

% To examine how humans perform gesture-based interactions with embodied AI systems under diverse conditions, we conducted a formative study with 

To elicit natural gestures, we employed three real robotic platforms as interaction targets: a quadruped robot, a human-height wheeled robot, and a drone. Participants were instructed to issue gesture-based commands to these robots across varying distances and in multi-person settings.

\subsection{Participants}

We recruited 12 participants (8 male, 3 female, 1 preferred not to disclose), aged 20-31 years ($M_{\text{age}}=24.4$, $SD=3.9$). They represented diverse national, cultural, and educational backgrounds. Five participants had non-STEM backgrounds and reported little to no prior knowledge of robotics or robotic perception, while the other 7 participants' STEM backgrounds ranged from unrelated fields to expert in robotics. Two participants were left-handed. All participants received compensation of \$14 per hour.

\subsection{Apparatus}
\label{sec:formative:apparatus}

\begin{figure}[h]
  \centering
  \includegraphics[width=\linewidth]{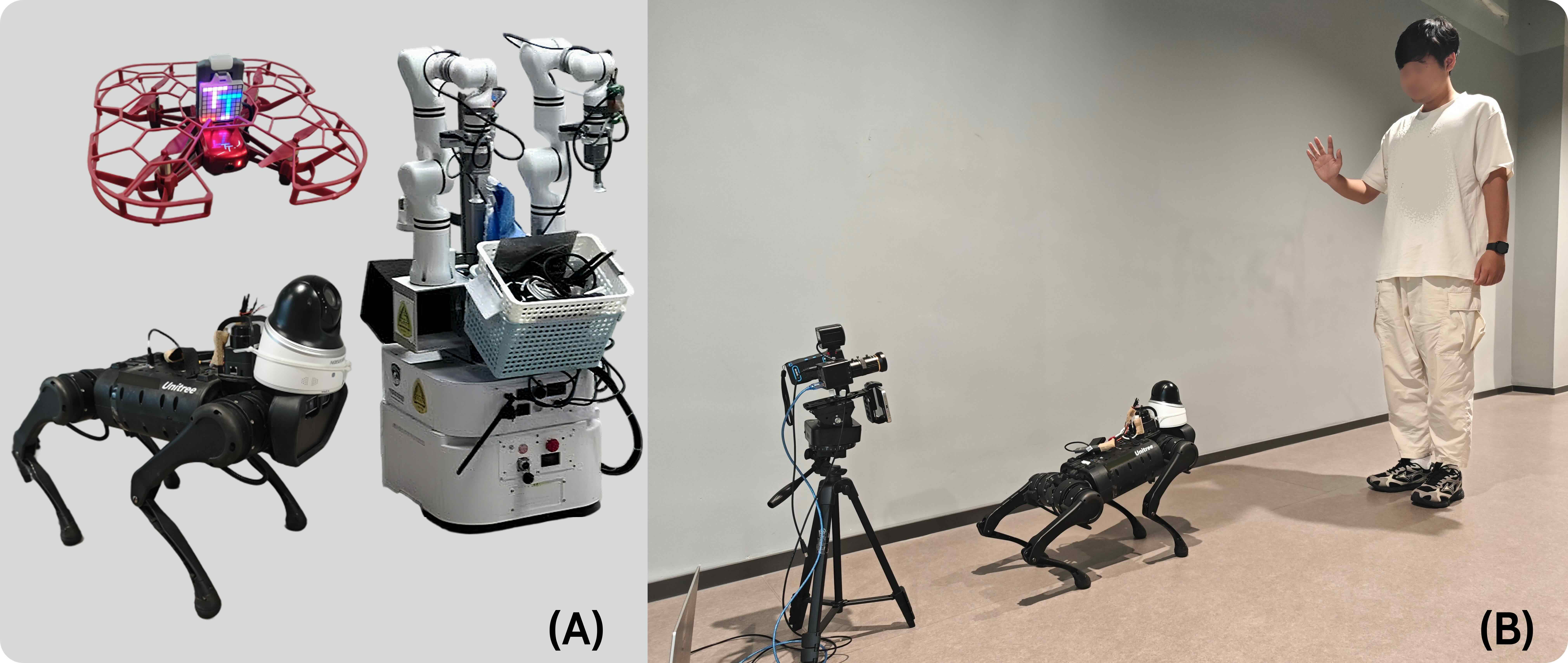}
  \caption{\textbf{Apparatus of Formative Study.} (A) Three robot forms used in the study. (B) Example of a participant performing a gesture toward a quadruped robot.}
  \label{fig:apparatus}
    \Description{Apparatus of Formative Study. (A) Three types of robots used in the study: a drone, a quadruped robot, and a dual-arm robotic system. (B) An example of a participant performing a gesture toward the quadruped robot, demonstrating the setup used in the study for gesture-based interaction.}
\end{figure}

A robot's morphology and height can shape both its perceived social role and its perceptual capabilities \cite{hiroi2016influence, rae2013influence}, and following the design rationale in \cite{lyu2025signaling}, we employed three robot embodiments that map to distinct perceptual perspectives: downward-looking (drone), eye-level (human-height wheeled platform), and upward-looking (quadruped).

% Because a robot's morphology and height can shape both its perceived social role and its perceptual capabilities \cite{hiroi2016influence, rae2013influence}, and following the design rationale of \cite{lyu2025signaling}, we employed three distinct robot embodiments --- drone, quadruped robot, and ground-based wheeled robot --- corresponding to three perceptual perspectives: downward-looking, eye-level, and upward-looking.

Specifically, as shown in Fig.~\ref{fig:apparatus}(a), we used a Unitree A1 quadruped robot, an AgileX Cobot wheeled platform (approximately human height), and a DJI Tello Talent drone (equipped with propeller guards) as representative embodied AI systems. The quadruped robot carried a front-mounted white camera, the wheeled robot and the drone were outfitted with LED lights to facilitate visual localization by participants --- thereby improving users' ability to identify the intended interaction target \cite{ginosar2023first}.

% 具体来说，As shown in Fig. \todo{X}, we used a Unitree A1 quadruped robot, a AGILE·X cobot 四轮机器人 and a DJI Tello Talent drone（带螺旋桨保护罩） as representative embodied AI systems. The quadruped robot was equipped with a white camera and a mounting bracket, while the drone featured LED lighting to facilitate visual localization by participants—thereby enhancing their ability to identify the intended interaction target\cite{ginosar2023first}.

\subsection{Settings}

% 及他们分别代表的三种视角

We manipulated two independent variables: robot embodiment (three forms) and interaction distance (six levels ranging from 3\,m to 30\,m), resulting in $3 \times 6 = 18$ experimental conditions. To minimize fatigue-related drift in participants' gestures, we randomized the trial order for each participant. Drone trials were conducted outdoors, while the quadruped and wheeled robots were tested in a corridor. Each condition involved two participants simultaneously, and in both environments, occasional passersby provided natural background activity and potential sources of distraction. Additional cameras were positioned near the participants to record their movements for subsequent analysis. All robots were remotely operated by a research team member, who remained out of the participants' immediate view to avoid influencing their gestures.

% 场景描述+实验组 + WoZ 用什么相机录制

\subsection{Procedure}

The experiment comprised two stages, designed to capture both the commands participants wished to issue to different robots and the gestures they employed to convey these commands across distances. At the outset, we measured participants' upper-arm, forearm, palm, and index-finger lengths to provide quantitative references for subsequent analysis. Participants were then introduced to a background scenario (see supplementary materials) crafted to elicit natural and spontaneous gesturing.

In Stage 1 (Command Elicitation), each robot was positioned nearby and teleoperated to showcase its range of motion and interaction capabilities. Participants were asked to propose the commands they would want the robot to execute. They continued until they judged the command set to be comprehensive. All proposed commands were recorded and later categorized by experts.

In Stage 2 (Distance-Conditioned Gesturing), the same robots were placed at distances of 3, 5, 10, 15, 20, and 30 meters (the drone was evaluated only at $\geq$ 10\,m for safety). Participants were instructed to perform gestural realizations of the previously elicited commands while verbally articulating their intended meaning. To preserve ecological validity, we occasionally introduced deliberate pauses before the robot responded, simulating recognition failures and allowing observation of participants' recovery behaviors.

% 整个实验过程分为两个阶段，分别探究用户研究用户希望对不同的设备使用什么指令，另一个是用户在不
% 同距离下会采用什么样的手势

% 我们首先测量了用户的大臂、小臂、手掌、食指长度，作为定量分析的重要参考。then, participants were provided with a background scenario (\todo{see supplementary materials}) designed to encourage them to produce gestures as naturally as possible. 

% 为了探究用户希望对不同的设备使用什么指令，我们首先将设备置于用户附近，并操纵各种运动，充分展示设备的运动交互能力，
% They were then asked what types of commands they would like to issue to the devices. These commands were recorded in sequence and later categorized by domain experts.

% 在用户声称已经提出的命令已经足够全面后，我们接下来探究用户在不同距离下会采用什么样的手势
% Next, the target device was positioned at varying distances (\todo{3 m, 5 m, 10 m, 15 m, 20 m, and 30 m}). Participants were instructed to perform the pre-conceived command while verbally articulating the intended meaning.

% Occasionally, deliberate pauses were introduced before executing the response to simulate recognition failures and to observe participants’ subsequent reactions.

% \subsection{Results and Analysis}

\subsection{Gestures Defined By Users}
\label{sec:formative:gesture}

\begin{figure}[h]
  \centering
  \includegraphics[width=0.8\linewidth]{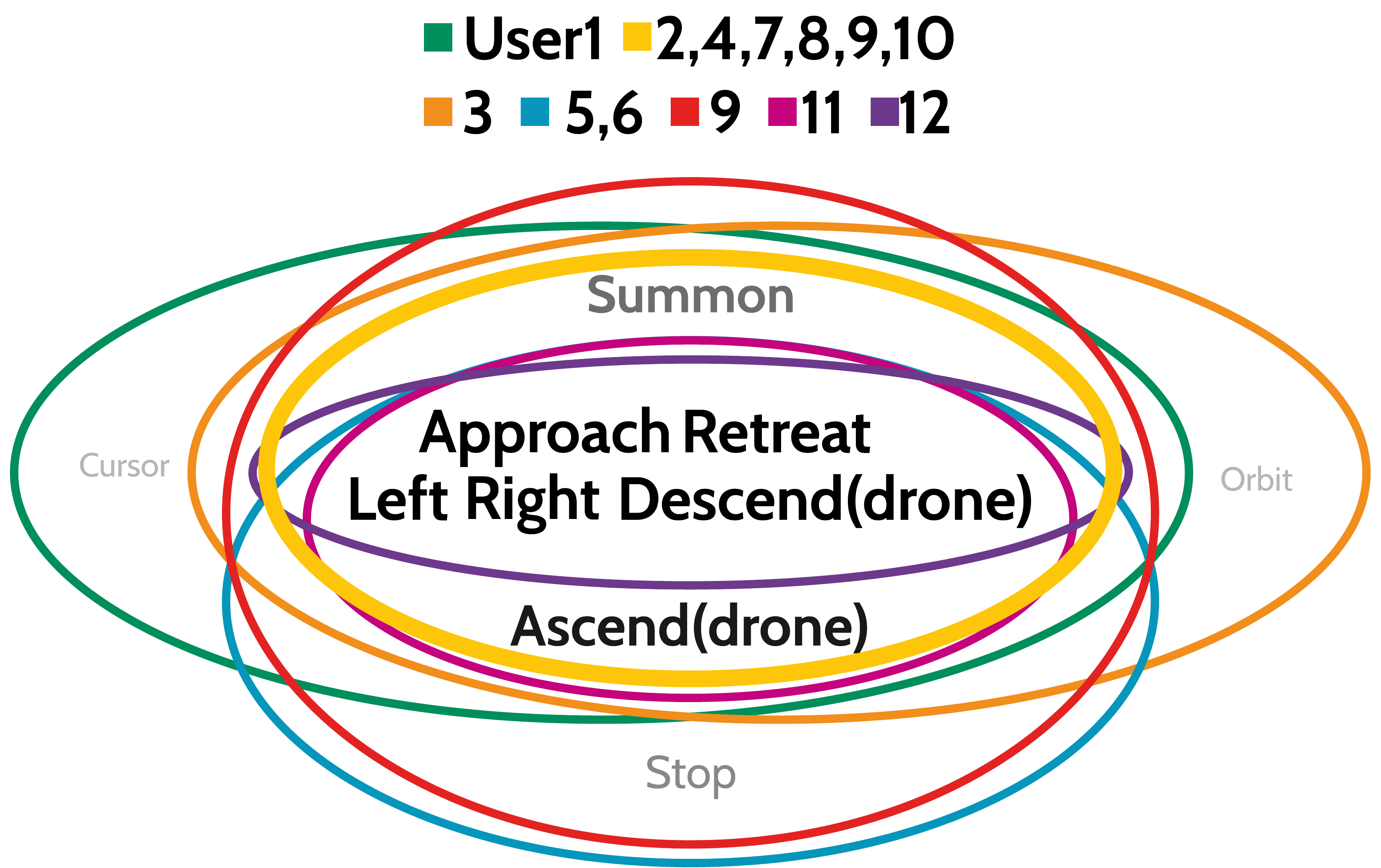}
  \caption{\textbf{Command Vocabulary Proposed by Participants.}}
  \label{fig:command}
    \Description{This figure shows a Venn diagram representing the overlap of gesture commands proposed by participants. Each circle corresponds to a specific command (e.g., "Approach," "Retreat," "Left," "Right," "Ascend," "Descend," etc.), with areas of overlap indicating shared vocabulary across different users. The diagram visually depicts the commands related to drone control and general navigation, highlighting how different participants proposed similar or distinct gestures for specific actions.}
\end{figure}

\begin{figure}[h]
  \centering
  \includegraphics[width=\linewidth]{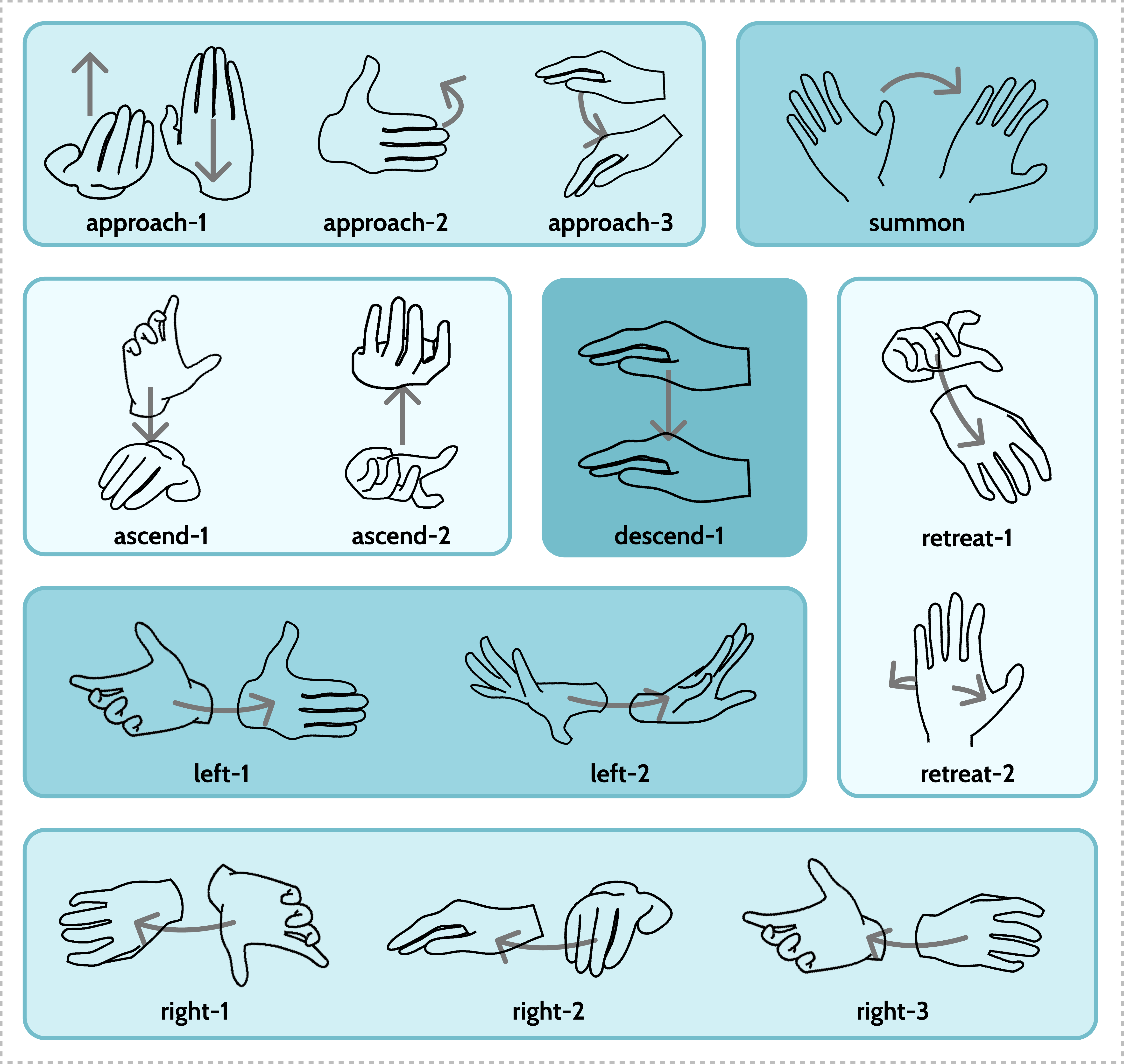}
  \caption{\textbf{Illustration of the User-defined Gesture Set.}}
  \label{fig:gesture}
  \Description{This figure shows the set of gestures proposed by participants for interacting with robots. The gestures are organized into rows by their corresponding actions: "approach," "summon," "ascend," "descend," "retreat," "left," and "right." Each action includes multiple variants (e.g., "approach-1," "ascend-1," etc.), demonstrating the diversity of user-defined hand movements for controlling robot behavior.}
\end{figure}

Fig.~\ref{fig:command} shows the command vocabulary proposed by participants in Stage 1. From these proposals, we distilled a core set comprising seven commands --- ``Approach'', ``Retreat'', ``Left'', ``Right'', ``Ascend'' (drone), ``Descend'' (drone) and ``Summon''. The first six motion commands unambiguously specify movement direction, whereas ``Summon'' exhibited two interpretations: (i) attracting the robot's attention and expecting an acknowledgment, and (ii) instructing the robot to continuously approach and stop near the operator (3m on average). We thus expose it as a user-configurable behavior to align system response with user intent. Several low-frequency suggestions were excluded, including ``Orbit'' for drones (N=1) and ``Cursor''-like steering (N=2). 
% ``Stop'' (N=3) was also excluded: participants framed it as compensating for limited obstacle avoidance, and most (N=7) expected such halting to be handled autonomously as a system-level safety behavior.
We excluded ``Stop'': a few participants (N=3) used it to compensate for limited obstacle avoidance, but most (N=7) expected autonomous handling, and it is ill-suited to CSI as a static pose.

% Fig. \todo{} 展示了Stage 1中用户提出的各种指令需求。我们从中总结出了最为常见的几种指令，包括 Approach	Retreat	Left	Right	Summon	，以及对drone 还有 extra 的 Ascend	Descend。具体来说 Approach	Retreat	Left	Right  Ascend	Descend 都是清晰的表达了对运动方向的控制。 summon对不同的人略有不同，有些用户认为是需要机器人pay attention to them and then expected some feedback. While others 认为是直接将机器人召唤到面前( about 3m on average). 另外，还有各有一名用户提出希望drone环绕飞行，以及握拳像拖拽光标一样操控方向，由于少所以不采纳。还有少量用户(N=3)提出 stop 命令，并解释为对机器人避障能力的不自信，由于这个是静态的手势且大多数用户(N=7)同意此动作应该由机器人自主完成，所以没采纳

In Stage 2, the user-defined gesture set is illustrated in Fig.~\ref{fig:gesture}. Participants showed high within-class consistency for gestures labeled ``Left'' and ``Retreat'', whereas ``Right'' often involved an additional wrist rotation. Several participants preferred to use both hands or to switch hands during operation. We further observed cross-class visual similarity among ``Left'', ``Approach'' and ``Descend'', indicating potential ambiguity for our system. \addl{Results demonstrate high consistency in gesture execution across different robot embodiments when following the same control protocol. This finding is further corroborated by user feedback (P1, P2, P5, P8, P11, P12).}

% On Stage 2, we found that Gestures for ``move left'' and ``move backward'' were generally consistent across participants, whereas ``move right'' often involved an additional wrist rotation. Several participants preferred to use both hands or to switch hands during operation. Notably, gestures for ``move left,'' ``come here,'' and ``move down'' were often visually similar, suggesting potential ambiguity in interpretation.

% 然后是手势的特点

% For the quadruped robot, participants frequently expressed the need for commands such as moving forward, backward, left, right, moving directly to the user, or following the user. 

% For the drone, participants used gestures for moving in six directions—forward, backward, left, right, up, and down—as well as for moving directly to the user. One participant additionally proposed a gesture for the drone to circle around them.

\subsection{Quantitative Results}

\begin{figure}[htbp]
    \centering
    \includegraphics[width=\linewidth]{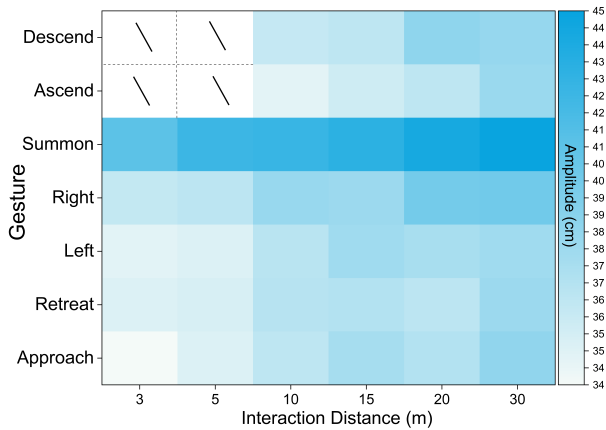}
    \caption{\addl{\textbf{Gesture Amplitude vs. Interaction Distance.} The color intensity represents the average amplitude. A general trend of motion amplification is observed as distance increases. Note: Drone-specific gestures (Ascend, Descend) were not evaluated at close ranges (3-5 m) due to safety constraints.}}
    \label{fig:amp_dis}
    \Description{Heatmap of gesture amplitude across interaction distances. A blue heatmap plots average gesture amplitude (color bar 34–45 centimeters) versus interaction distance (3–30 meters) for seven gestures on the vertical axis. Darker cells, especially for the Summon gesture, show that users increase motion amplitude at longer distances. Cells for Ascend and Descend at 3–5 meters are crossed out, indicating that these drone-specific gestures were not evaluated at close range.}
\end{figure}

% \begin{figure}[htbp]
%   \centering 
  
%     \begin{minipage}[b]{0.49\textwidth}
%     \centering
%     \includegraphics[width=\linewidth]{pics/amp_bjq.pdf}
%     \subcaption{Example of Invariance; User 1;
%     Amplitude vs. Distance;\\
%     Error bars denote the standard deviation of different devices.}
%     \label{subfig:stable}
%   \end{minipage}
%   \hfill 
%   \begin{minipage}[b]{0.49\textwidth}
%     \centering
%     \includegraphics[width=\linewidth]{pics/amp_cxx.pdf}
%     \subcaption{Example of Gradual Increase; User 2; Amplitude vs. Distance;\\
%     Error bars denote standard deviation of different devices.}
%     \label{subfig:increase}
%   \end{minipage}
%   \hfill

%   \begin{minipage}[b]{0.90\textwidth}
%     \centering
%     \includegraphics[width=\linewidth]{pics/amp_all.png}
%     \subcaption{Amplitude vs. Distance on Average of All Users (Min - Max).}
%     \label{subfig:all}
%   \end{minipage}
  
%   \caption{\textbf{Quantitative Results of Formative Study.}}
%   \label{fig:amp_dis}
% \end{figure}

% Pearson Correlation Coefficients across Gesture Categories. Average correlations between 3D radial and 2D tangential motion exceed 0.925 for all classes. This consistently high alignment confirms 2D flow as a reliable proxy for 3D dynamics, independent of specific kinematic trajectories. Error bars denote standard deviation.

\begin{figure}[htbp] 
  \centering 
  \begin{minipage}[b]{0.49\textwidth}
    \centering
    \includegraphics[width=\linewidth]{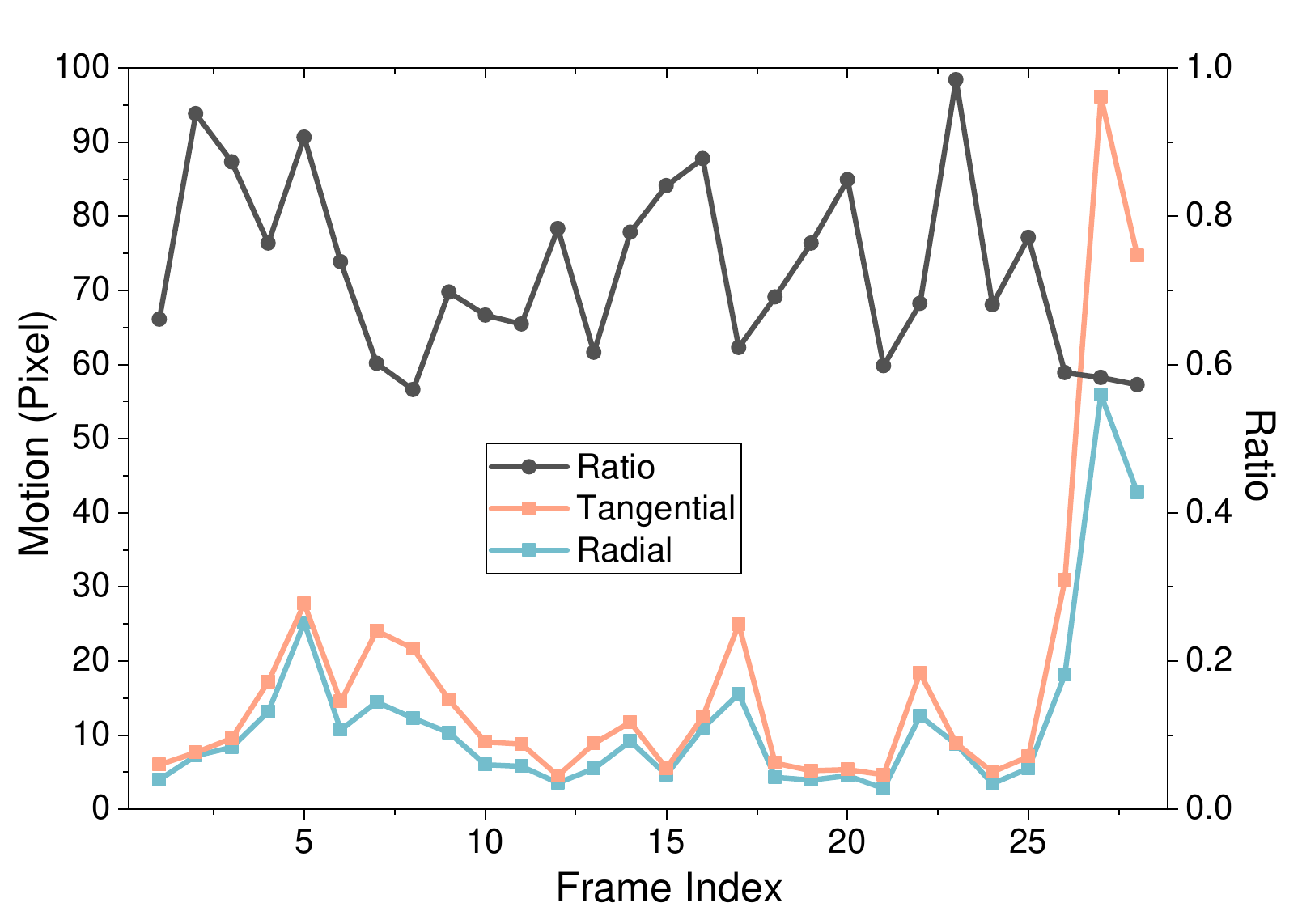}
    \subcaption{\textbf{Illusition of 3D Radial and 2D Tangential Motion.} \addl{A sampled sequence showing 2D tangential motion (orange) tightly aligns with 3D radial magnitude (blue). Although with fluctuating ratio, this high correlation validates monocular 2D flow as a reliable proxy for 3D dynamics.}}
    \label{subfig:increase}
  \end{minipage}
  \hfill 
  \begin{minipage}[b]{0.49\textwidth}
    \centering
    \includegraphics[width=\linewidth]{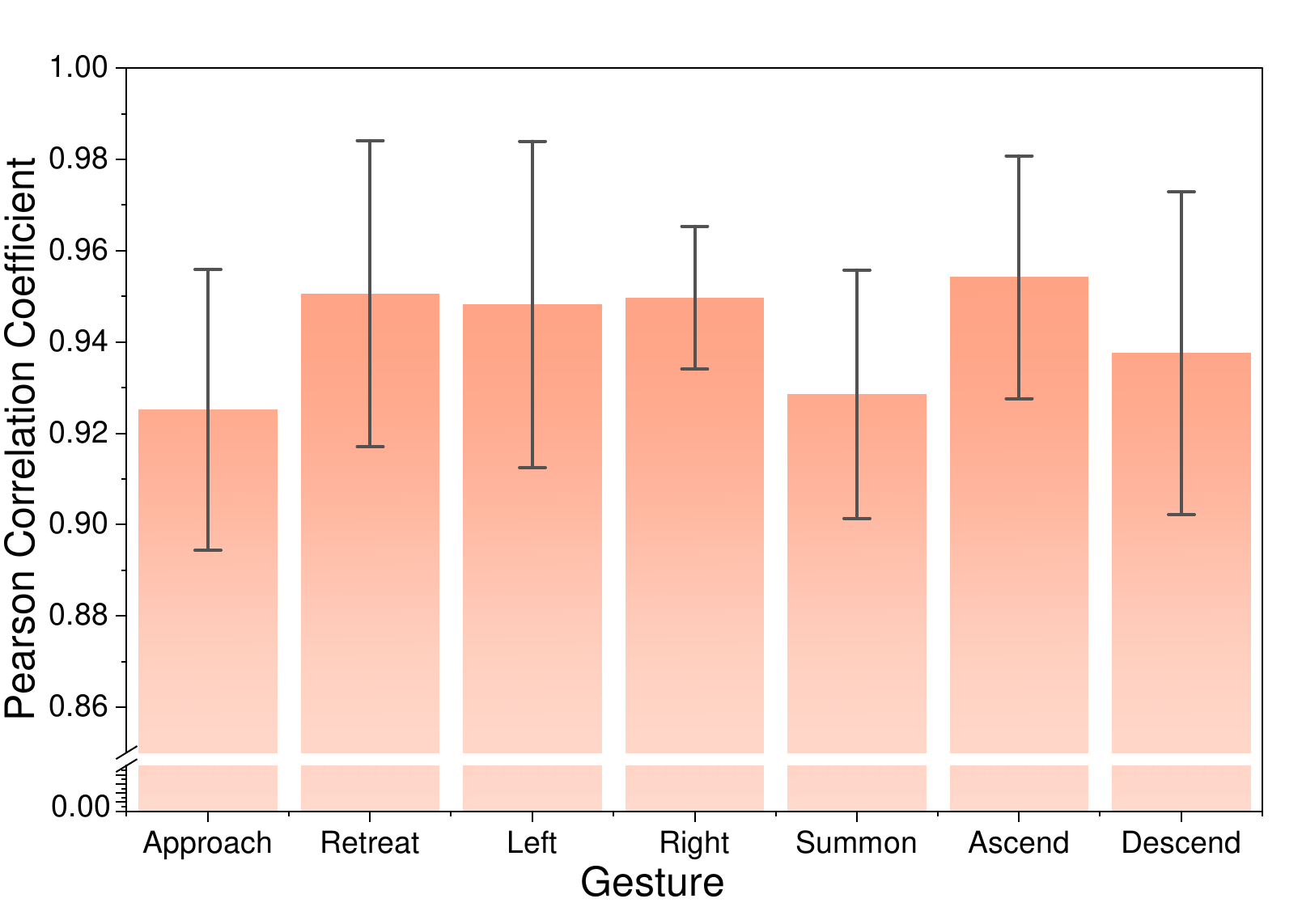}
    \subcaption{\textbf{Pearson Correlation Coefficient.} \addl{Average correlations between 3D radial and 2D tangential motion exceed 0.925 for all classes. This consistently high alignment confirms 2D flow as a reliable proxy for 3D dynamics, independent of specific gesture trajectories. Error bars denote standard deviation.}}
    \label{subfig:stable}
  \end{minipage}
  
  \caption{\textbf{Correlation between 3D Radial and 2D Tangential Motion.}}
  \label{fig:correlation}
  \Description{This figure includes two parts: Panel (a) is a line graph titled "Illustration of 3D Radial and 2D Tangential Motion" plotting motion magnitude over frame indices. An orange line representing tangential motion and a blue line representing radial motion follow nearly identical trajectories and peak simultaneously, illustrating their synchronization. A gray line plots the fluctuating ratio between them on a secondary y-axis. Panel (b) is a bar chart titled "Pearson Correlation Coefficient" displaying average correlation values for seven gesture types: Approach, Retreat, Left, Right, Summon, Ascend, and Descend. All categories exhibit high coefficients exceeding 0.92, with error bars indicating standard deviation, confirming the strong relationship between the two motion types across different gestures.}
\end{figure}

Using each participant's limb measurements (Sec.~\ref{fig:dataset}) and an estimated skeleton model \cite{loper2023smpl}, we compute gesture amplitude as the displacement of the middle finger-tip. We then analyze amplitude over different robot embodiments and participants (Sec~\ref{sec:formative:quant:device}), interaction distance (Sec~\ref{sec:formative:quant:distance}), and time under a non-response condition (Sec~\ref{sec:formative:quant:time}). Finally, we empirically validate that 2D pixel motion serves as a reliable proxy for 3D gesture (Sec~\ref{sec:formative:quant:correlation}).

% We use 测量得到的用户的骨骼长度作为参照，和估计 得到用户的2D骨骼模型\todo{cite}，估算Participants 的动作的幅度，以中指指尖的最大位移计。分析了动作随着不同device, distance, 以及在设备无响应情况下随时间的变化。以及验证了用2D的像素运动表征用户3D geasture motion的合理性

\subsubsection{Amplitude over Embodiments and Participants} 
\label{sec:formative:quant:device}
Our gesture set exhibited high within-participant consistency across robot embodiments, suggesting cross-platform generalizability (detailed in App.~\ref{appendix:amp}). By contrast, we observed huge between-participant differences in amplitude profiles (minimum: User 12 at 16.53\,cm; maximum: User 5 at 70.64\,cm), indicating strong idiosyncratic kinematics that can support CSI. 
% Complete per-participant curves are provided in Appendix\todo{X}.

\subsubsection{Amplitude over Distance} 
\label{sec:formative:quant:distance}
As shown in Fig.~\ref{fig:amp_dis}, gesture amplitude generally increased with interaction distance. However, this trend was not universal: for example, User 1 (detailed in App.~\ref{appendix:amp}) exhibited nearly distance-invariant amplitude. We also observed that participants without STEM backgrounds rarely consider the robot's sensing limits; they tended to use more conservative motions even with a long interaction distance, thereby potentially degrading recognition accuracy. These behaviors impose stricter requirements on \methods's ability to capture subtle motion cues.
% As shown in Fig.\ref{fig:amp_dis}\subref{subfig:all}, 总体上来说，gesture Amplitude tended to increase with distance，但是并不是每位用户都是如此，例如 User 1 (Fig.\ref{fig:amp_dis}\subref{subfig:stable})，他的Amplitude和distance无关。我们发现，非STEM背景的被试几乎不会从传感器感知能力的角度进行考虑，他们通常会用较小的动作进行手势,leading to reduced discriminability between commands and potentially lowering recognition reliability。对我们系统捕捉suble motion Feature的能力提出了更高的要求。

% （越熟悉的人越会从感知的角度进行）但是这让我们系统

\subsubsection{Correlation between 3D Radial and 2D Tangential Motion}
\label{sec:formative:quant:correlation}

To investigate whether 2D camera data can effectively represent gesture motion information in 3D space, we computed the radial motion magnitude from optical flow data. We then compared it with the tangential optical flow. As shown in Fig.~\ref{fig:correlation}, the radial motion exhibited a very strong correlation with the magnitude ($\rho > 0.925$). This finding indicates that, under natural conditions, participants rarely performed movements strictly perpendicular to the camera's focal plane. Consequently, a 2D camera is sufficient to capture the essential motion characteristics of users' gestures.

% radial motion magnitude from optical flow data (\todo{Appendix})

% \subsubsection{Findings} 
% Across both devices, gesture amplitude tended to increase with distance, although the effect was not pronounced. However, when the device did not respond, participants consistently amplified their gestures, presumably to increase recognizability. At longer distances, gesture distinctiveness decreased, leading to reduced discriminability between commands and potentially lowering recognition reliability.

\begin{add}

\section{TASK DEFINITION}

We address the challenge of Command Source Identification (CSI) in long-range, multi-user environments. In our setup, we identify a specific \textbf{target operator} ($O_{target}$) who holds interaction authority and is equipped with a paired IMU.

As illustrated in Fig.~\ref{fig:teaser}, the system input consists of a single continuous inertial stream $\mathcal{I}_{target}$ transmitted from the target’s device, and a visual stream $\mathcal{V}$ capturing $N$ individuals $\mathcal{P} = \{P_1, ..., P_N\}$. Crucially, this scene may include bystanders ($P_{j}, j \neq k$) who are actively performing gestures \textbf{(e.g., controlling their own separate robots or interacting with human peers)}.

In this formulation, these bystanders \textbf{constitute visual-only negative samples}. Since their inertial data is not transmitted to \emph{this} robot, the system must distinguish the true commander solely by cross-referencing the received $\mathcal{I}_{target}$ with the visual motion fields of all candidates. The goal is to identify the target $O_{target}$ by validating which candidate's visual motion spectrally aligns with the reference inertial signal, thereby rejecting visual distractors regardless of their gestural similarity to $O_{target}$.
    
\end{add}

\section{DATASET}
\label{sec:dataset}

To address the absence of public datasets for CSI under long-range, multi-user conditions, we construct a multimodal dataset grounded in user-defined natural hand gestures, with synchronized monocular RGB flow and finger-worn IMU streams to enable and evaluate long-range CSI approaches.

\subsection{Dataset Characteristics}

\begin{figure}[htbp]
    \centering
        \includegraphics[width=0.6\linewidth]{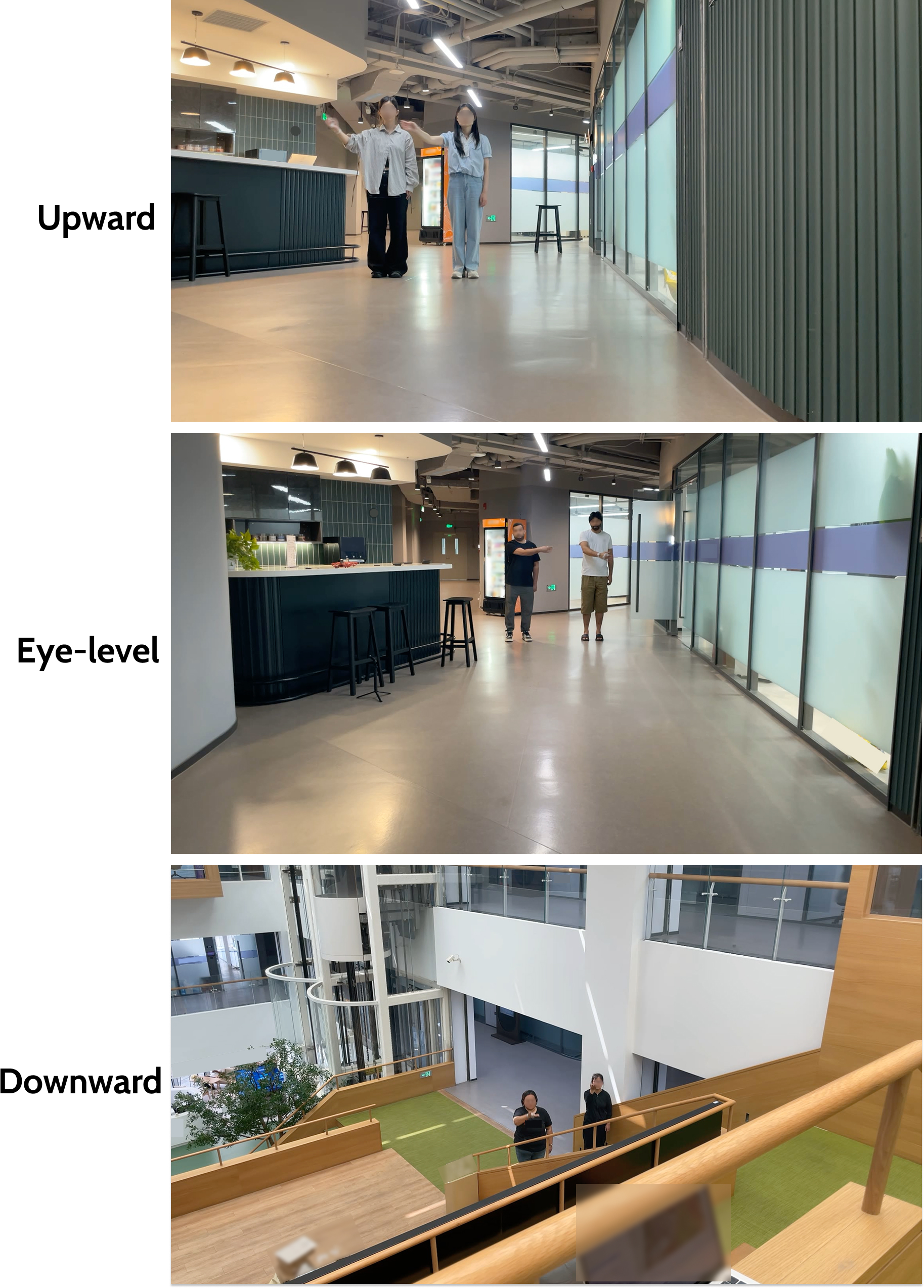} 
        \caption{\textbf{Visualization of data perspectives.} \addl{The three camera angles—Upward, Eye-level, and Downward—are designed to emulate the perceptual perspectives of a quadruped, service robot, and drone, respectively.}}
        \label{fig:prespective}
        \Description{This figure shows three perspectives used in the study: (1) "Upward" view, showing participants walking in a hallway from below; (2) "Eye-level" view, capturing participants from a standard standing perspective in a corridor; (3) "Downward" view, showing participants from a high vantage point, looking down at a lower floor.}
\end{figure}

\begin{figure}[htbp]
        \centering
        \includegraphics[width=\linewidth]{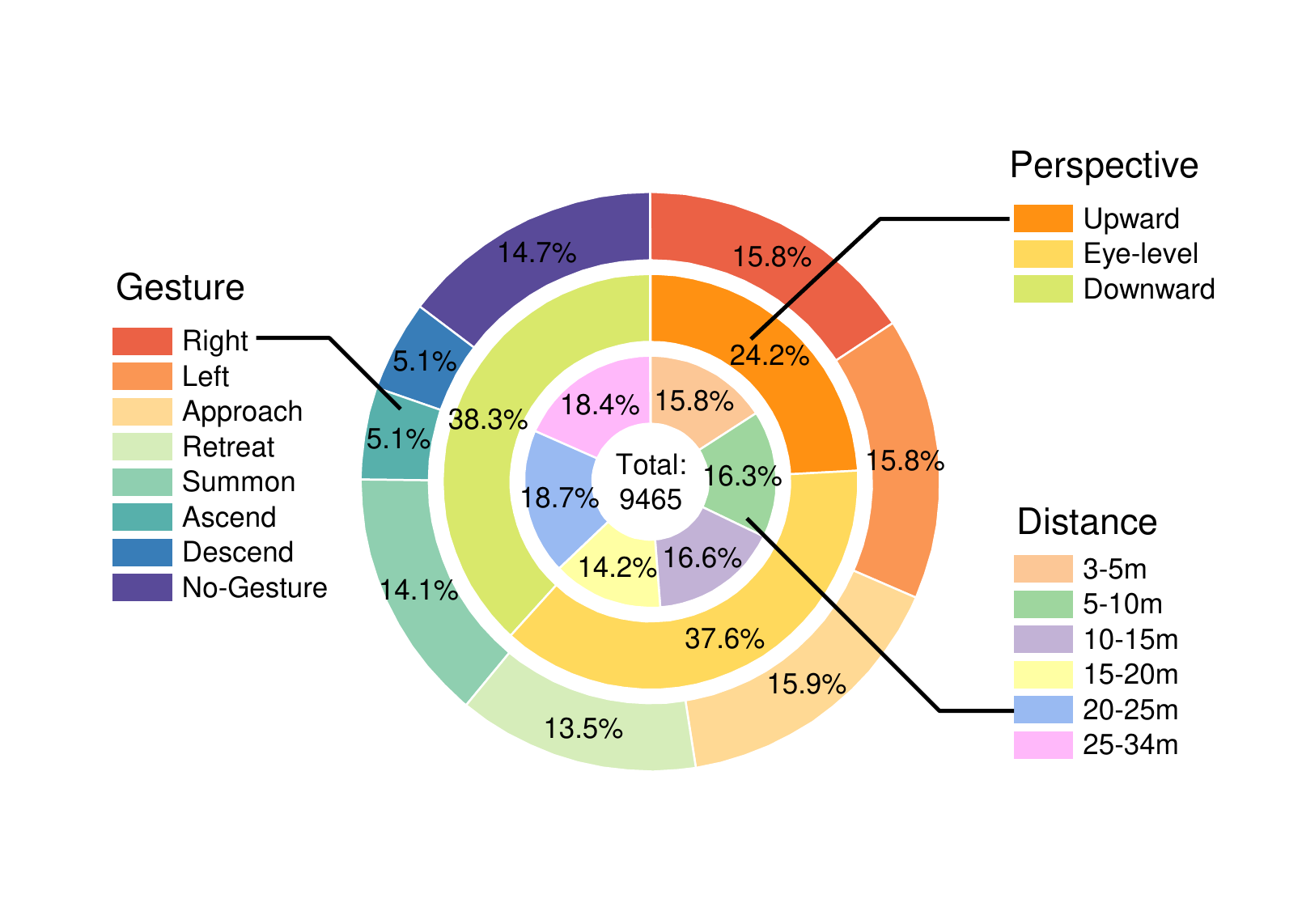}
        \caption{\textbf{Distribution of Dataset.} \addl{The chart visualizes the composition of 9,465 sequences, demonstrating a relatively uniform coverage across all dimensions to minimize bias. The rings, from outer to inner, represent gesture categories, camera perspectives, and interaction distances.}}
        \label{fig:dataset_statistic}
        \Description{This figure presents a circular diagram showing the distribution of the dataset across three factors: gesture, perspective, and distance. The outer ring represents gesture distribution, with categories like "Right," "Left," "Approach," "Retreat," and "No-Gesture." The middle ring shows perspective distribution, including "Upward," "Eye-level," and "Downward." The inner ring illustrates the distance distribution, with ranges from 3-5 meters to 25-34 meters. The total number of data points is 9,465, and each section’s proportion is indicated.}
\end{figure}

\begin{figure*}[h]
  \centering
  \includegraphics[width=0.7\linewidth]{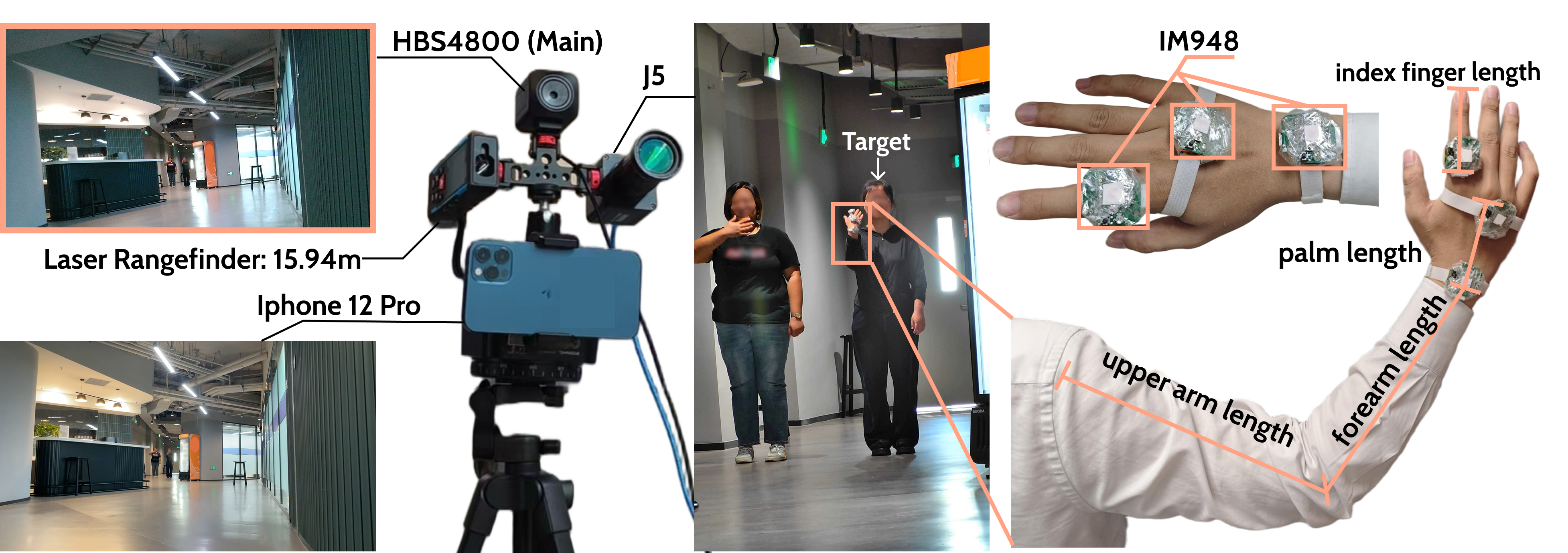}
  \caption{\textbf{Dataset Characteristics.} \addl{(Left) Recording apparatus with cameras and laser ranging, paired with visualizations of their data. The orange-framed inset displays the main industrial camera view used for experiments. (Right) Detailed sensor placement and biometrics. IMUs are positioned to simulate smart ring, handheld phone, and smartwatch form factors, alongside anthropometric measurements collected.}}
  \label{fig:dataset}
  \Description{Recording setup and sensor placement for the HiSync gesture dataset. The left panel shows a multi-sensor rig on a tripod with an industrial camera, laser rangefinder, and smartphone, plus an inset of the captured hallway view and a distant participant at the target location. The right panel zooms in on a participant’s extended arm and hand, with multiple inertial measurement units attached to the index finger, wrist, and hand to emulate different wearable form factors, alongside annotated body segment lengths for palm, index finger, forearm, and upper arm.}
\end{figure*}

\textbf{Gesture Taxonomy.}
We distilled seven gesture classes (Sec.~\ref{sec:formative:gesture}) plus a ``no-gesture'' class composed of unconstrained free-motion sequences. The class distribution is shown in Fig.~\ref{fig:dataset_statistic}.

\textbf{Participants.}
We recruited 38 participants (23 male, 14 female, 1 preferred not to disclose; mean age = 24.66 years, variance = 11.1), including two left-handed individuals. Participants self-reported diverse professional backgrounds and cultural origins. Additionally, we collected measurements of their upper arm, forearm, palm, and index finger lengths.

\textbf{Visual.}
Fig.~\ref{fig:prespective} shows that data were recorded from three viewpoints: eye-level, overhead, and low-angle. Video data were captured using two industrial cameras that are commonly used by robots (Model HBS4800 and Model J5), at a resolution of 1920×1080 and 30 fps. The field of view (FOV) of the main Camera HBS4800 was 75°, while another J5 was a telephoto camera with a focal length range of 10-50 mm, enabling the capture of high-definition close-up views of participants for extracting ground-truth motion trajectories. In addition, we used a smartphone (iPhone 12 Pro) camera with a larger FOV and higher image quality, also recording at 1920×1080 and 30 fps. 

\textbf{Inertial.}
For inertial sensing, we employed three IM948 9-axis IMUs that logged tri-axial accelerometer, gyroscope, and magnetometer signals; orientation quaternions were then derived. The sensors were affixed at three hand locations --- wrist, palm, and the base of the index finger (Fig.~\ref{fig:dataset}) --- to emulate three common consumer form factors: a smartwatch, a phone held in hand, and a smart ring \cite{he2025writingring}, respectively.

\textbf{Synchronization and Labels.}
Each recording consists of three time-synchronized video streams and three IMU signals. Most sequences intentionally include more than two participants to introduce multi-person interference, while all IMUs are worn by a single (target) participant. Ground-truth annotations specify the target participant's distance to the camera, camera viewpoint, gesture class, and the 2D bounding box of the target participant. Labels were produced by an automated detection pipeline and subsequently refined via manual verification to ensure high annotation fidelity.

% Each recording consists of three time-synchronized videos and three IMU streams. Most sequences intentionally include more than two participants to introduce multi-person interference, 所有的IMU都只在一个Participant身上. Labels 包括 用户到相机的距离、视角、手势的种类、佩戴IMU的Participant的bbox，labels were first generated via automated detection and then refined through manual verification to ensure annotation accuracy.

\textbf{Dataset Size.}
The dataset comprises 9465 gesture sequences, totaling 452055 frames of RGB data and corresponding data across all viewpoints and sensors. Data from 8 participants were reserved as the test set, while the remaining 30 participants formed the training set. In total, the training set contains 7530 sequences and 1935 sequences for test set.

\subsection{Dataset Collection}
Participants were instructed to treat the distant camera as the target device and to perform their naturally preferred gesture for each command. 
\addl{The first 12 dataset participants are the same as those in the formative study. The additional 26 participants did not introduce any new gesture types with non-negligible frequency (i.e., appearing in $\ge 1\%~\text{sequences}$) beyond the gesture set in the formative study (Fig.~\ref{fig:gesture}), thus ensuring the consistency between the formative elicitation and the dataset gesture.}
Participants repeated each gesture five times to ensure sufficient coverage and consistency. 

Data were collected in outdoor environments. To simulate aerial viewpoints for drone interactions, we leveraged building height differences to capture gestures from low-, medium-, and high-angle perspectives. In each recording session, at least two participants were present: one designated participant wore an IMU sensor for gesture tracking, \addl{while the other(s) moved freely or actively issued the same commands (accounting for 70.4\% of the data).}
This setup naturally introduced negative samples and simulated multi-user interference. Passersby sometimes entered the field of view, adding uncontrolled motion and occlusion that increased the ecological validity of the dataset.  

Recordings captured both natural-scale gestures and adaptive variations. Participants initially performed gestures at a comfortable amplitude and speed, but if the target device did not respond, they may naturally amplify the movement amplitude or frequency. Commands could also be alternated within a recording, allowing us to capture a broad spectrum of gesture expressions, including subtle, exaggerated, and transitional forms like their occurrence in real world.

\subsection{Data Processing}
% \todo{（track可能还要写一下）}

% \textbf{Hand ROI Annotation}. We first used YOLOv11x to detect human bodies in each RGB frame, lowering the confidence threshold to XX to ensure higher recall. Within each detected human bounding box, we applied YOLOv11x-Pose to estimate the human skeleton. The approximate hand location was then derived from the corresponding keypoints, and the surrounding 3×3 region was set as hand ROI.

\textbf{Target Human Annotation}. For ground-truth identification of the target participant, we used the skeleton keypoints 
``Right Shoulder'', ``Left Shoulder'', ``Right Hip'', ``Left Hip'' to form a body region. This region was then compared with predefined clothing color profiles of participants, and the person with the smallest color difference was selected as the ground-truth subject. \addl{Two researchers checked the results independently and resolved all disagreements collaboratively through discussion afterwards.}

\textbf{Temporal Synchronization}. We utilized the on-chip CPU perf of the IMU for time alignment. After connecting the IMU to the computer used for video recording, we performed a linear regression to map IMU timestamps to the corresponding video frame times. \addl{The final time error is less than one frame of the camera (about 33\,ms).}

\subsection{Privacy Protection}
All participants signed an informed consent form. They all consented to the release of their recorded gesture data and associated metadata, and were informed of their right to withdraw their data from the dataset at any time. To ensure participant anonymity, all faces in RGB videos were blurred~\cite{haarcascades}.

\section{\methods~ALGORITHM}

CSI using IMUs and cameras faces three main challenges. 
\textbf{Firstly}, both modalities are highly noisy: IMU signals suffer from sensor drift and motion artifacts, while cameras at long range produce ambiguities as people occupy only a few pixels with blurred outlines. 
\textbf{Secondly}, IMUs capture acceleration whereas cameras capture positional trajectories. In theory, trajectories can be integrated from IMU data or accelerations derived by differentiating positions, but in practice such transformations are highly inaccurate, leading to inconsistent feature spaces \cite{scaramuzza2019visual,woodman2007introduction}. 
\textbf{Thirdly}, precise temporal synchronization between IMU and camera streams is difficult, as wireless transmission (e.g., Bluetooth) introduces delays and clock drift that accumulate over distances. Unsynced data makes trivial feature matching infeasible for CSI.

% Pairing information between IMUs and cameras has three main difficulties.
% Firstly, 两个模态的信息噪声都非常大， IMU signals are inherently noisy due to sensor drift and motion artifacts, while cameras at long range produce ambiguities because people appear with few pixels and blurred outlines。
% Secondly，IMU数据体现的是加速度信息，而摄像头数据体现的是位置轨迹信息。尽管理论上可以通过IMU数据积分得到轨迹，或者通过位置差分得到加速度，但是在实践中\cite{} 处理得到数据非常的不准，二者的数据分布空间并不一致。
% Thirdly, in practice it is difficult to synchronize IMU and camera streams precisely, as wireless transmission (e.g., Bluetooth) introduces unstable delays and clock drift that accumulate over long distances.

% Firstly, IMU数据体现的是加速度信息，而摄像头数据体现的是位置轨迹信息。尽管理论上可以通过IMU数据积分得到轨迹，或者通过位置差分得到加速度，但是由于IMU不可避免的噪声问题以及远距离下的相机 ambiguities，整个手部的位置信息较为粗略，同样含有较大的噪声，所以二者的数据分布空间并不完全一致。Secondly, IMU表征的是三维空间的运动，而视觉只能捕捉到二维的运动信息，depth camera或者深度估计算法都无法在远距离大范围场景中work。同样的，基于前述的噪声问题，也无法进行降维匹配。Thirdly, in practice it is difficult to synchronize IMU and camera streams precisely, as wireless transmission (e.g., Bluetooth) introduces unstable delays and clock drift that accumulate over long distances.

% 系统需要解决两个问题，谁在操纵、操纵指令是什么。覆盖宽的距离范围、机器人形态以及环境。这就需要能匹配上。我们创造性的使用motion的频域信息作为fiducials，实现在RGB中找到和IMU对应的运动信息，同时实现定位操作者和识别操作指令。

% intro：freq-based 方法能有什么能力

% 时间不同步会对常规感知作为干扰。频域上处理规避这个问题。
% 能做到规避干扰 （误触），同时实现感知能力（recall）

% 所以我们使用了光惯混合的技术来同时实现定位操作者和识别操作指令。

\subsection{Overview}

% Fig.~\todo{} summarizes \methods, a system for gesture-based command-source identification (CSI) in multi-user, long-range human-robot interaction. \methods fuses RGB video with hand-worn IMU streams and operates in the frequency domain to obtain stable, modality-comparable motion signatures. The pipeline comprises (i) a Motion Feature Extractor \todo{ref} that estimates per-person image-plane velocities from optical flow and reads 3-axis linear accelerations from the IMU, transforming both into spectral representations (e.g., STFT/PSD); and (ii) CSINet, a spectral network designed for cross-modal matching under loose synchronization and low SNR. Within CSINet, a quality-aware FiLM layer reweights unreliable frequency channels \cite{perez2018film}; an IMU-anchored cross-modal attention module performs soft temporal alignment between IMU and visual spectra; a similarity head yields dimensionless match scores by combining normalized L2 and cosine distances; and a distance-prior-guided multi-scale window fusion aggregates evidence across windows whose sizes vary with the camera-estimated range. The model outputs a score for each visible person; the highest-scoring candidate is selected as the command source and passed to the robot controller for downstream behaviors.

\begin{figure*}[h]
  \centering
  \includegraphics[width=0.8\linewidth]{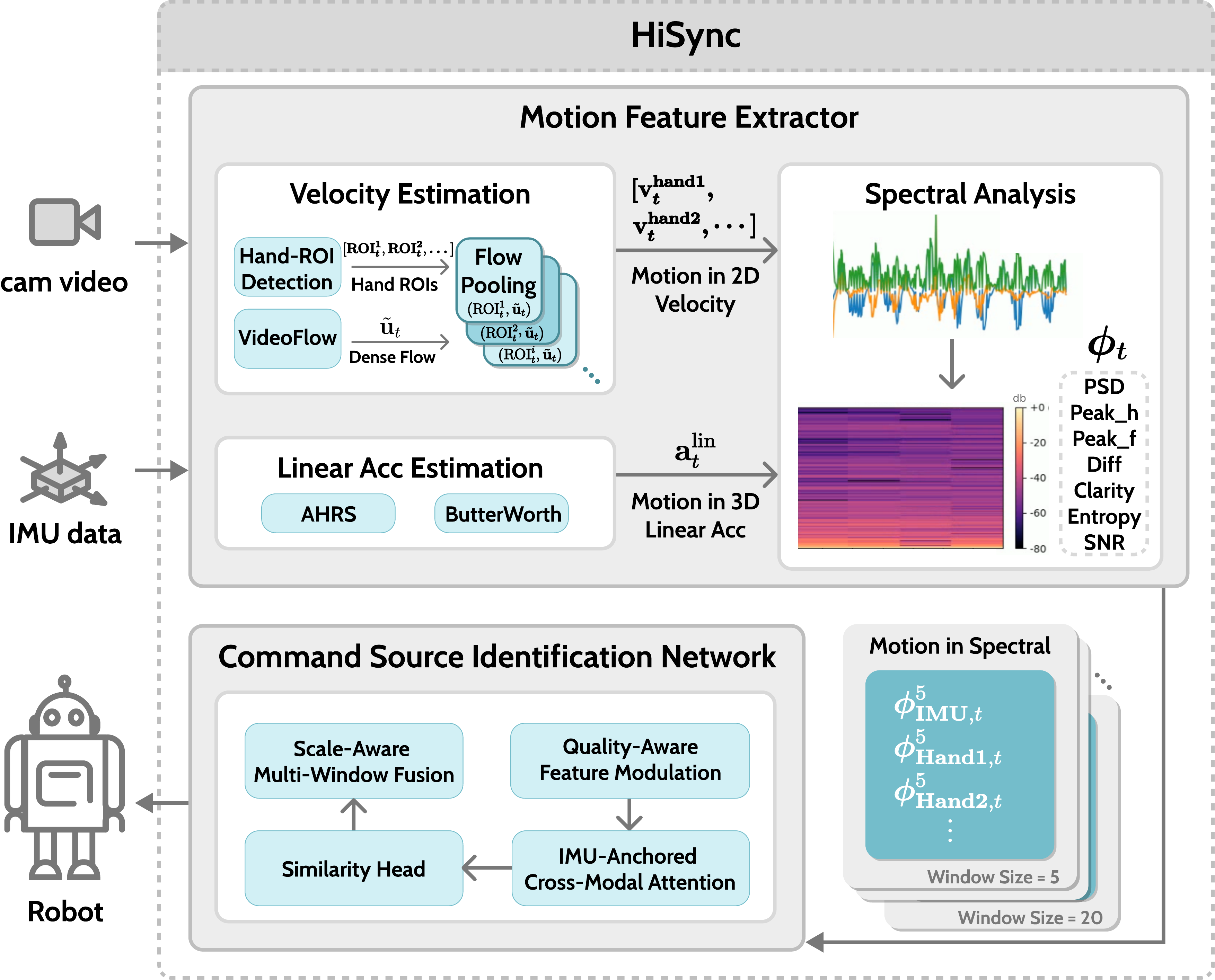}
  \caption{\textbf{Overview of \methods.} \addl{The framework fuses robot-mounted RGB video and wearable IMU signals in the frequency domain for robust Command Source Identification (CSI). It consists of two main stages: (1) a Motion Feature Extractor that estimates and transforms visual velocity and inertial acceleration into spectral representations; and (2) CSINet, which aligns and matches these cross-modal features using Quality-Aware Feature Modulation, IMU-Anchored Attention, and Scale-Aware Multi-Window Fusion to identify the target operator.}}
  % \Description{\todo{画一条33\%的参考线}}
  \label{fig:pipeline}
  \Description{Block diagram of HiSync pipeline from camera and IMU signals to command source identification. Camera video and wearable inertial data enter a Motion Feature Extractor, where hand regions are tracked with dense optical flow and inertial accelerations are estimated, then both streams are transformed into spectral motion features. These spectral features, organized over multiple temporal windows for each hand and inertial unit, are processed by the Command Source Identification Network—combining scale-aware multi-window fusion, quality-aware feature modulation, IMU-anchored cross-modal attention, and a similarity head—to predict the robot’s target operator.}
\end{figure*}

Fig.~\ref{fig:pipeline} provides an overview of \methods, whose goal is gesture-based command-source identification in multi-user, long-range human-robot interaction. \methods~fuses robot-mounted RGB video and hand-worn IMU streams and reasons in the frequency domain to obtain robust CSI results. The pipeline has two components: (i) a Motion Feature Extractor (Sec.~\ref{methods:motion_feature_extractor}) that estimates per-person 2D image-plane velocity from RGB video and 3D linear accelerations from the IMU, then transforms both into spectral representations, and (ii) CSINet (Sec.~\ref{sec:methods:CSINet}), a spectral neural network tailored to cross-modal matching under loose synchronization and low SNR. Within CSINet, a Quality-Aware Feature Modulation recalibrates unreliable spectral channels; an IMU-Anchored Cross-Modal Attention module performs soft temporal alignment between the IMU spectra and each person's visual spectra; a Similarity Head produces dimensionless match scores by combining normalized $\ell2$ and cosine distances; and a Scale-Aware Multi-Window Fusion aggregates evidence over window sizes that differ with camera-estimated range. The network outputs a score for every visible person; the arg-max is taken as the command source and forwarded to the robot controller for downstream tasks.

% Fig \todo{} presents the pipeline of our algorithm. The overall goal of \methods is to 解决 geasture-based Command source identification in multi-user, long-range human-robot interaction。我们创造性的使用了光惯混合方法以及将运动数据在频域上进行分析来处理这个task， 
% 我们的系统包含两个个部分Motion Feature Extractor\cite{}, Command Source Identification Network(CSINet), 。系统的原始输入是摄像头的RGB图片和IMU的运动数据，从中分别提取出2D的velocity motion data和3D的linear accelerate motion data，并分别转换为频域上的motion特征数据。CSINet是一个针对频域数据设计的spectral Neural Network，包含以质量感知通道重标定（FiLM）、以 IMU 为锚的跨模态软对齐（cross-modal attention）、无量纲相似度评分（L2 + cosine）、以及由距离先验驱动的多尺度窗口融合；以极小开销显著提升多人、远距离、低质量场景下的 CSI 稳健性。最终输出了画面中每个人的运动和IMU信息的相似度，得到CSI结果，并被传递给机器人以支持进一步的交互响应。

\subsection{Motion Feature Extractor}
\label{methods:motion_feature_extractor}

% \todo{总结这个的子结构和输入输出，可能还有motivation}

\subsubsection{2D Velocity Estimation}
\label{sec:methods:velocity}

Our objective is to estimate \emph{image-plane hand velocities} for every visible person from the RGB stream. We decompose the task into two stages: (a) \textbf{hand-ROI localization} and (b) \textbf{hand velocity estimation}.

\paragraph{Hand-ROI Detection.} 
Long-range imagery suffers from a small person scale and frequent keypoint dropouts. We therefore adopt a two-stage pose pipeline.
\addl{First, a high-recall person detector ( YOLOv11x \cite{Jocher_Ultralytics_YOLO_2023}, conf $> 0.2$ ) first crops regions and then passed to YOLOv11x-pose to obtain a rough skeleton. However, monocular 3D Human Pose Estimation (HPE) remains unreliable at such distances, even state-of-the-art methods \cite{zhang2025pose,xu2025optimizing} fail to capture usable motion cues for CSI. We thus define a \emph{hand region of interest (ROI)} with side length
$
s_t^{(i)} \;=\; \alpha \, h^{(i)}_{t,\text{bbox}},
$
where $h^{(i)}_{t,\text{bbox}}$ is the height of person $i$'s bounding box at time $t$ and $\alpha=0.1$ is selected via experiment. When hand keypoints are missing or low-confidence, we propagate the last valid ROI with short-term velocity prediction and, if needed, infer a proxy ROI from the elbow-to-wrist direction.}

% Since even SOTA monocular 3D HPE~\cite{zhang2025pose,xu2025optimizing} proves unreliable for CSI at such distances, we strictly utilize a 2D hand ROI. Its size adapts to distance via $s_t^{(i)} = \alpha \, h^{(i)}_{t,\text{bbox}}$, where $h^{(i)}_{t,\text{bbox}}$ is the bounding box height. To handle missing keypoints, we propagate the ROI using short-term velocity prediction or, if needed, infer it from the elbow-to-wrist direction.

\paragraph{VideoFlow.}
To capture fine 2D motion $\tilde{\mathbf{u}}_t$ under low pixel counts, we compute dense optical flow with \emph{VideoFlow-MOP}~\cite{shi2023videoflow}, which pairs a \emph{Tri-frame Optical Flow (TROF)} module --- jointly predicting flows from the middle frame to its preceding and succeeding frames --- with a \emph{Motion Propagation (MOP)} module that passes motion cues across adjacent triads to enlarge the temporal receptive field and improve accuracy.
For each time $t$, let $\mathbf{u}^{+}_t$ denote the flow from frame $t$ to $t{+}1$, and $\mathbf{u}^{-}_t$ the flow from $t$ to $t{-}1$. We form a per-pixel velocity field
$
\tilde{\mathbf{u}}_t \;=\; \tfrac{1}{2}\,\big( \mathbf{u}^{+}_t \;-\; \mathbf{u}^{-}_t \big).
$

\addl{\paragraph{Flow Pooling.} We estimate the hand velocity $\mathbf{v}^{\text{hand},(i)}_{t}$ for person $i$ by mean pooling $\tilde{\mathbf{u}}_t$ within the hand ROI.}
Moreover, if the camera exhibits residual motion (e.g., slight pan/tilt), we estimate a global background flow $\mathbf{g}_t$ via robust homography fitting on rigid regions and compensate for the $\mathbf{v}^{\text{hand},(i)}_{t}$.

% The resulting per-person 2D velocity magnitudes are then forwarded to the spectral transform in the \emph{Spectral Analysis} for downstream cross-modal matching.

% 这部分的目的是从RGB视频流中提取出每个人的IMU佩戴部位（手部）在image-plane上的速度估计值。
% 我们将手部的velocity estimation分成两个部分，估计手部ROI区域以及图像光流速度估计。
% 我们使用了YOLOv11x-pose来进行human pose estimation, however, although it adopts the transformer mechanism, 30m的人还是不在其识别范围内。所以我们先试用YOLOv11x识别到人体可能存在区域，并将置信度阈值降低到0.2以获得尽可能高的recall。然后将人体区域chop出来输入YOLOv11x-pose，得到比较rough的skeleton，将用户手部的关节点附近3x3的区域作为hand region of interest（hand ROI）。
% 然而受制于分辨率的影响，即使是目前SOTA的Monocular 3D Human Pose Estimation（HPE）方法\cite{zhang2025pose,xu2025optimizing}也无法做到精准的识别，即使能识别的误差也较大，无法捕捉到有效的运动信息来完成CSI任务\todo{引用实验结果}。因此我们提出了使用光流来捕获2D的运动细节，我们使用VideoFlow-MOP处理RGB图像得到图像上每个像素点的运动速度，VideoFlow是一种多帧双向光流估计模型，对于每三帧相邻图片，提出了TRi-frame Optical Flow (TROF)模块，同时预测从中间帧到前后两帧的光流，当输入帧数大于三帧时，复用三帧模块TROF，额外引入一个运动传递（Motion Propagation）模块（MOP），通过在相邻的三帧模块之间传递运动信息，扩大了时序维度的感受野，提高光流估计的准确性。in the \textbf{Flow Aggregation} 我们将TROF中的双向光流信息的average作为当前帧的速度估计，并且根据hand ROI内光流的均值计算对应手部区域的速度估计值作为最终得到的motion data in 2D velocity of each human。

\subsubsection{3D Linear Acceleration Estimation}

\begin{add}
% \subsubsection{3D Linear Acceleration Estimation}
We estimate linear acceleration $\mathbf{a}^{\text{lin}}_t$ from a $150\,\text{Hz}$ 9-DoF IMU. We compute attitude $q_t$ via \textit{Madgwick AHRS}~\cite{madgwick2011estimation}, fusing gyroscope ($\boldsymbol{\omega}_t$), magnetometer ($\mathbf{m}_t$), and accelerometer ($\mathbf{a}^{\mathrm{meas}}_t$). We reject dynamic disturbances by scaling the accelerometer gradient with Madgwick parameters: base gain $\beta=0.035$ and weight $w_a = \mathrm{clip}(1 - |\lVert \mathbf{a}^{\mathrm{meas}}_t\rVert - g|/3, 0, 1)$. World-frame linear acceleration is then obtained by removing pre-calibrated sensor bias $\mathbf{b}_a$ and gravity $\mathbf{g}_w$: $\mathbf{a}^{\text{lin}}_t = \mathbf{R}(q_t)(\mathbf{a}^{\mathrm{meas}}_t - \mathbf{b}_a) - \mathbf{g}_w$, followed by a $15\,\text{Hz}$ Butterworth low-pass filter.
\end{add}

% and magnetometer updates are gated whenever the field magnitude deviates from the calibrated norm by more than $\pm 15\%$ (otherwise yaw is propagated open-loop by gyro integration). 
% For completeness, the equivalent body-frame form is
% \begin{equation}
% \mathbf{a}^{\text{lin},b}_t \;=\; \bigl(\mathbf{a}^{\mathrm{meas}}_t - \mathbf{b}_a\bigr) - \mathbf{R}_t^\top \mathbf{g}_w .
% \end{equation}

% using zero-phase filtering (offline) to preserve temporal alignment with vision. IMU timestamps are aligned to video frames by linear interpolation so that $\mathbf{a}^{\text{lin},w}_t$ can be fused with the 2D velocity estimates.

% First对RGB和IMU信息进行数据预处理。RGB用VideoFLow 预处理RGB得到光流信息，同时用YOLO得到rough的人体skeleton信息，获得手部速度的估计。九轴IMU数据经过降噪后得到手部ACC的估计，然后通过CIR和scale-aware attention algorithm to enable command source identification

% \textbf{IMU} 对于九轴IMU，使用Madgwick 滤波得到

% \textbf{RGB} 对于输入的RGB信号

\subsubsection{Spectral Analysis}
\label{sec:methods:spectral}
% 上面估计得到的motions in 2D velocity 和 motion in 3D linear accelration，如果直接用他们进行精准相似度匹配，会遇到前述的数据噪声, 数据分布空间不匹配和时序不匹配的问题\todo{ref}。所以本办法将数据转换到spectral域上再进行下一步处理，在频域上可以让加速度和速度的具有可对比性，并且可以有效解决时间不对齐和噪声问题。

Directly matching raw 2D image-plane velocities from RGB with 3D linear accelerations from the IMU is brittle in long-range, multi-person settings due to sensor and estimation noise and loose cross-modal synchronization. We therefore transform both streams into the frequency domain. Working in frequency space (i) provides a common, modality-comparable basis for both velocity and acceleration, (ii) reduces sensitivity to temporal offsets, and (iii) suppresses uncorrelated noise through windowing and band-wise aggregation. This spectral projection serves as the frontend for CSINet.

\begin{add}
\textbf{Validation of Comparing Acceleration and Velocity on Spectral.} 
Let $v(t)$ be a velocity component and $a(t)=\dot v(t)$ the corresponding acceleration. By the \textit{differentiation property} of the Fourier transform $\mathcal{F}$, their spectra are related by: $A(\omega) = \mathcal{F}\{\dot v(t)\} = j\omega\,V(\omega),$
where $A(\omega)$ and $V(\omega)$ are the Fourier transforms of $a(t)$ and $v(t)$. Consequently, for any angular frequency $\omega\neq 0$:
\begin{equation}
    |A(\omega)| = |\omega|\,|V(\omega)|, \quad \angle A(\omega) = \angle V(\omega) + \frac{\pi}{2}\mathrm{sgn}(\omega).
\end{equation}
Here, $|\cdot|$ is the complex magnitude, $\angle(\cdot)$ is the principal phase and $\mathrm{sgn}(\cdot)$ denotes the sign function. Therefore, defining a normalized acceleration spectrum $\tilde A(\omega) \triangleq A(\omega) / (j\omega)$ yields $\tilde A(\omega) \equiv V(\omega)$. This confirms that acceleration features inherently encode velocity patterns, differing only by a known frequency-dependent scaling and phase shift.
\end{add}

% \begin{add}

% where $x(t)$ denotes a real- or complex-valued time-domain signal, $\omega\in\mathbb{R}$ is the angular frequency (rad/s) and $j=\sqrt{-1}$, the differentiation property gives $A(\omega)=\mathcal{F}\{\dot v(t)\}(\omega)=j\omega\,V(\omega)$,
% where $V(\omega)=\mathcal{F}\{v(t)\}$ and $A(\omega)=\mathcal{F}\{a(t)\}$. Hence, for $\omega\neq 0$,
% \begin{equation}
% |A(\omega)|=|\omega|\,|V(\omega)|,\qquad
% \angle A(\omega)=\angle V(\omega)+\tfrac{\pi}{2}\,\mathrm{sgn}(\omega),
% \end{equation}
% with $|\cdot|$ the complex magnitude, $\angle(\cdot)$ the principal phase, and
% $\mathrm{sgn}(\omega) = \{ 1, \omega > 0; \; 0, \omega = 0; \; -1, \omega < 0 \}.$
% Therefore, defining a phase/gain-normalized acceleration spectrum $\tilde A(\omega)\triangleq \frac{A(\omega)}{j\omega}$
% yields $\tilde A(\omega)\equiv V(\omega)$ \emph{(noise-free, perfectly aligned, same spatial component)}, i.e., no residual phase difference and identical magnitude.
% \end{add}

% \vspace{0.1cm}

\textbf{Validation of Utilizing Spectral to Avoid Time Shift}.
\label{sec:validation_of_time_shift}
\emph{Spectral features are shift-tolerant within a window}.
If a global time shift $\tau$ exists between $v$ and $a$ (e.g., from sensor/video misalignment), then $A(\omega)=e^{-j\omega\tau}\,j\omega V(\omega)$; subtracting the linear phase $e^{-j\omega\tau}$ restores the result.

% \vspace{0.1cm}

% \textbf{Validation of Spectral-domain Denoising.}
% Direct time-domain matching is vulnerable to transient spikes, low-frequency drift, and modality-specific biases. By operating in the frequency domain, we attenuate these effects through band-wise aggregation and normalization. Transient or irregular disturbances distribute their energy across many frequencies, so their influence on any single band --- and thus on the final similarity --- is limited. And many sensor noises appear as isolated spectral lines \cite{an2022narrowband}, which can be down-weighted via masking bands. Whereas quasi-constant bias, gravity leakage, and slow drift concentrate near Direct Current (DC) \cite{gulmammadov2009analysis}.
% To avoid this, we exclude a small neighborhood around $\omega=0$.
% With these designs, the spectral features are markedly less sensitive to noise, bias, and drift while preserving motion-discriminative information for cross-modal matching.

\begin{add}
\textbf{Validation of Spectral-domain Denoising.}
Unlike time-domain matching, which is vulnerable to spikes and drift, our spectral approach naturally attenuates these artifacts. Specifically, transient disturbances disperse energy across wide bands, minimizing local impact; narrowband sensor noise~\cite{an2022narrowband} is implicitly masked via band selection by the downstream neural network; and low-frequency drift or gravity leakage~\cite{gulmammadov2009analysis} is eliminated by excluding the near Direct Current neighborhood. This ensures robustness to modality-specific noise while preserving discriminative motion cues.
\end{add}

\begin{add}
    \textbf{Spectral Feature $\boldsymbol{\phi}$.} 
\label{sec:methods:PSD_feature}
    Given a length-$N$ sequence $x$, we apply a window size $w$ and compute the discrete Fourier transform (DFT) $X[k]$, where $k$ denotes the frequency bin index. We then form the power spectral density (PSD) $S[k]=|X[k]|^2$ which serves as our primary feature denoted as $\mathbf{PSD}$. From the PSD, we derive a compact set of interpretable descriptors: peak height ($p$), peak frequency ($f$), spectral clarity ($\kappa$), spectral entropy ($H$), frequency spacing ($\Delta f$), in-band SNR ($\boldsymbol{\mathrm{SNR}}$), and average power ($P_{\mathrm{avg}}$). Detailed definitions and ablations are provided in the App.~\ref{appendix:PSD_feature}. These descriptors, concatenated with the full PSD, form the final feature vector:
\end{add}

\begin{equation}
\boldsymbol{\phi}(x)=[\mathbf{PSD}; \text{Feat]}=\left[\mathbf{PSD},\, p,\, f,\, \kappa,\, H,\, \Delta f,\, \mathrm{SNR},\, P_{\mathrm{avg}}\right].
\end{equation}

Different DFT window lengths capture complementary spectral cues, much like different receptive fields in a CNN. 
We compute spectral features for window lengths \(w\in\{5,\dots,20\}\) (frames) and denote the feature vector from window \(w\) as \(\phi^{w}\). 
These multi-scale spectral descriptors offer complementary time-frequency resolutions.

\subsection{Command Source Identification Network}
\label{sec:methods:CSINet}

\begin{figure*}[h]
  \centering
  \includegraphics[width=0.8\linewidth]{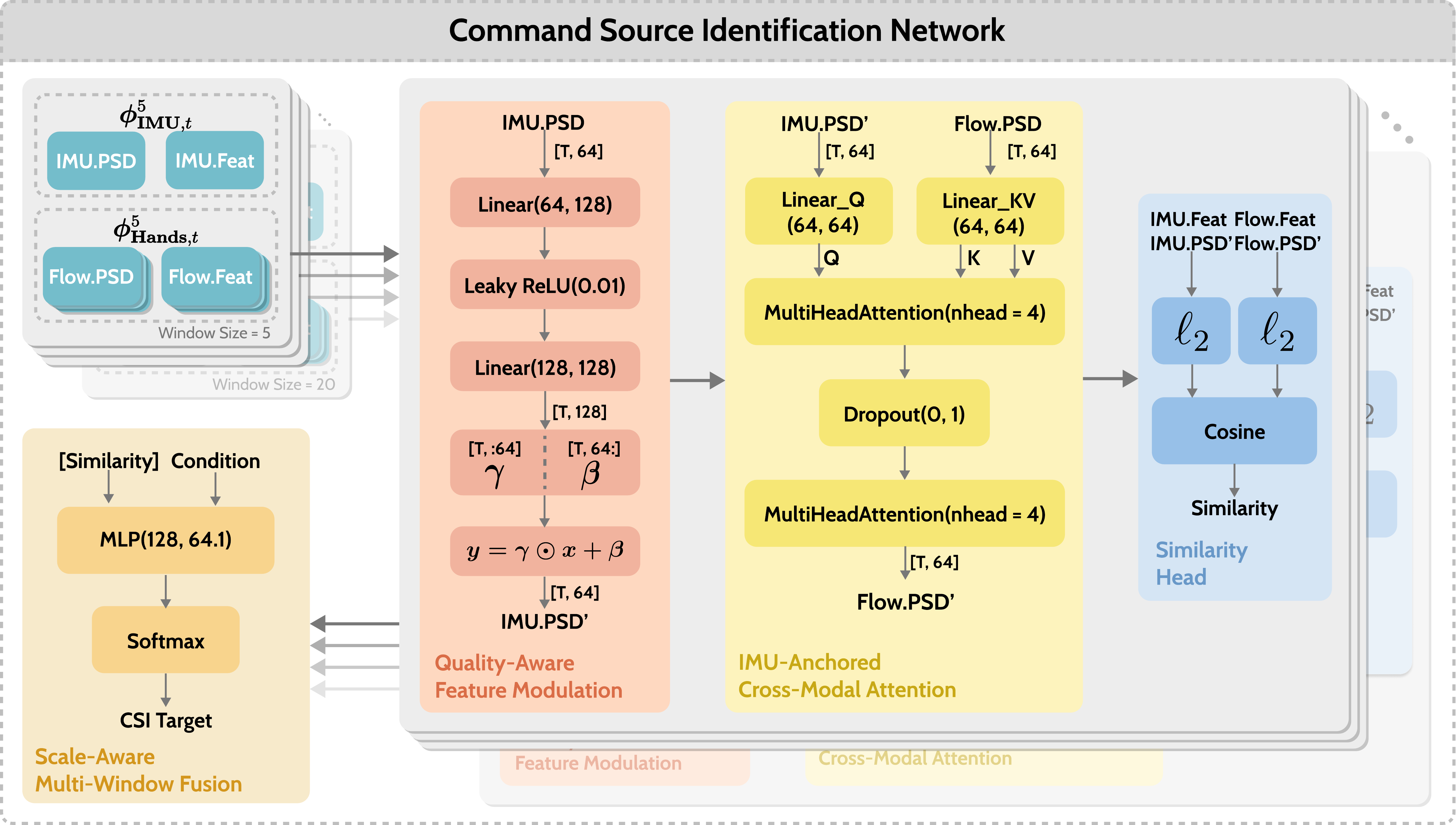}
  \caption{\textbf{Architecture of Command Source Identification Network.} \addl{Taking spectral features as input, for each window size, the network first applies Quality-Aware Feature Modulation to suppress noise in IMU signals. It then utilizes IMU-Anchored Cross-Modal Attention to align visual flow sequences with the inertial reference. Finally, the Similarity Head computes cosine scores, which are dynamically aggregated by Scale-Aware Multi-Window Fusion across multiple window sizes to output the final similarity.}}
  % \Description{\todo{画一条33\%的参考线}}
  \label{fig:CSINet}
  \Description{Architecture diagram of the Command Source Identification Network detailing the multi-stage feature processing pipeline. On the left, stacked input cards represent parallel processing for different window sizes (e.g., 5 and 20), containing Inertial Measurement Unit (IMU) and Flow Power Spectral Densities (PSD) and features. The central processing flow consists of three modules: first, "Quality-Aware Feature Modulation" passes IMU PSD through linear layers and a Leaky Rectified Linear Unit to generate affine parameters ($\gamma$ and $\beta$) for signal denoising. Second, "IMU-Anchored Cross-Modal Attention" uses the modulated IMU data as Queries and Flow PSD as Keys and Values within a Multi-Head Attention mechanism to align the modalities. Third, a "Similarity Head" applies $l_2$ normalization to the features and computes a cosine similarity score. Finally, a "Scale-Aware Multi-Window Fusion" module on the bottom left aggregates these scores using a Multi-Layer Perceptron and softmax weighting to produce the final target.}
\end{figure*}

\subsubsection{Model Architecture}

% \todo{这里每块的输入输出需要讲清楚含义和起个名字，然后在文章和图里面统一使用}
% As illustrated in Fig.~\todo{}, the Command Source Identification Network (CSINet) comprises four modules. First, a \emph{Quality-Aware Feature Modulation} layer recalibrates IMU features using per-frame quality cues. Second, an \emph{IMU-Anchored Cross-Modal Attention} aligns each candidate's visual sequence to the IMU time reference, soft-handling jitter, missed detections, and phase drift. Finally, a \emph{Similarity Head} produces a cross-modal matching score between the IMU segment and each candidate, enabling robust command source identification in crowded scenes.
As illustrated in Fig.~\ref{fig:CSINet}, the Command Source Identification Network (CSINet) operates on precomputed spectral feature vector $\phi$ and comprises four modules: (1) \emph{Quality-Aware Feature Modulation}, which recalibrates IMU spectral features using per-frame quality cues; (2) \emph{IMU-Anchored Cross-Modal Attention}, which aligns each candidate's flow-spectral sequence to the IMU time reference, mitigating jitter, missed detections, and phase drift; (3) \emph{Similarity Head} that $\ell_2$-normalizes temporally aggregated embeddings and outputs a cosine-similarity score for each IMU-candidate pair; and (4) \emph{Scale-Aware Multi-Window Fusion}, which evaluates results from multiple window lengths and fuses their scores via learned weights conditioned on a distance proxy. This design yields robust command-source identification in crowded, long-range settings.
% \todo{把他们的输入输出再写清楚一点}

% 如Fig \todo{}所示，整个Command Source Identification Network（CSINet）主要由  五个部分组成。输入是one sequence of IMU data和多个画面中出现的人的RGB data经过Motion Feature Extractor extract 分别得到的Motion in Spectral\todo{加一个公式说明}. 可以实现鲁棒的得出IMU motion data和每个RGB motion data分别的相似度，从而实现 Command Source Identification任务

% \textbf{Spectral Encoder}

% 对于输入的IMU和光流 PSD信息，首先使用同一个spectral编码器进行encoding。

% \textbf{Quality‑Aware Feature Modulation}
% 实验中观测到有时候用户的运动并不能很好体现特征，或者用户并没有在做有意义的动作，这样会形成比较差的相似度结果，为了从数据序列中提取到有效的信息
% We introduce a quality‑aware FiLM layer that conditions the spectral features on per‑frame quality descriptors. Given an psd feature $x_t ∈ R^D$ and its quality vector $q_t ∈ R^C$, a lightweight MLP predicts channel‑wise affine parameters$ (γ_t, β_t) = MLP(q_t) $. The modulated representation $y_t = γ_t ⊙ x_t + β_t$ recalibrates unreliable frequency channels (e.g., caused by occlusion or low detection confidence) before cross‑modal alignment. Compared with simple concatenation, FiLM provides multiplicative interactions with negligible parameter overhead and consistently improves robustness under low‑quality or long‑distance scenarios. FiLM被作用于IMU数据。

\textbf{Quality-Aware Feature Modulation.}
\label{methods:CSINet:FiLM}
We observed that some segments contain weak or non-purposeful motion, degrading similarity estimates. To emphasize informative frames while suppressing unreliable channels, we adopt a FiLM operator~\cite{perez2018film} on the IMU spectral features. Given a per-frame PSD embedding $x_t\in\mathbb{R}^D$ \addl{(composed of \(\text{IMU.PSD}\) and \(\text{Flow.PSD}\))}, a lightweight MLP predicts channel-wise affine parameters $(\gamma_t,\beta_t)=\mathrm{MLP}(x_t)$. The modulated representation
$
y_t \;=\; \gamma_t \odot x_t \;+\; \beta_t
$
recalibrates frequency channels (e.g., down-weights those affected by occlusion or low detection confidence) prior to cross-modal alignment. Compared with simple concatenation, Quality-Aware Feature Modulation improves robustness with low-quality. In \methods, Quality-Aware Feature Modulation is applied to IMU features, producing \(\text{IMU.PSD}'\).

% 参考\todo{cite  Feature-wise Linear Modulation}

% \textbf{Cross Attention\todo{这个名字没对}}

% 将高置信 IMU 序列作为时间参考，对每个候选人的光流频谱嵌入进行软对齐与信息筛选，缓解：帧级抖动、检测缺失、运动不同步（相位漂移）造成的对应帧不匹配问题。Cross Attention 用IMU数据经过一个线性层作为Query， Flow数据经过线性层同时作为key 和value，进行multihead attention 操作。number of head is set to 4, 采用了两个Attention层，中间以0.1的概率进行dropout。为每个候选人生成的、跨模态注意力增强的时间序列特征，用来代表“该候选人从 IMU 视角看到的运动证据”。

\textbf{IMU-Anchored Cross-Modal Attention.}
\label{methods:CSINet:attention}
Treating the IMU sequence as the temporal anchor, we perform soft alignment and evidence selection over each candidate's flow-spectrum embeddings. We compute

\begin{equation}
\begin{aligned}
& Q=W_q \cdot \text{IMU.PSD}',\quad K=W_k \cdot \text{Flow.PSD}, \\
& V=W_k \cdot \text{Flow.PSD}, \quad \text{Flow.PSD}'=\mathrm{MHA}\!\left(Q,K,V\right),  
\end{aligned}
\end{equation}

% Let $Z^{\mathrm{IMU}}\!\in\!\text{IMU.PSD}'$ and $Z^{\mathrm{Flow}}\!\in\!\text{Flow.PSD}$ denote the two sequences in a window. We compute
where $\mathrm{MHA}$ is multi-head attention with 4 heads, stacked for two layers with dropout $0.1$. The resulting $\text{Flow.PSD}'$ represents the candidate's motion ``as seen from the IMU's perspective,'' mitigating frame-level mismatch due to jitter, intermittent detection, or phase shifts.

\textbf{Similarity Head.}
\label{methods:CSINet:COSine}
Because IMU and optical flow live in different units ($\mathrm{m/s^2}$ vs.\ pixels/frame) yet share discriminative spectral amplitude, we remove scale effects by $\ell_2$-normalizing temporally aggregated embeddings and use cosine similarity as the cross-modal score:

\begin{add}
    \begin{equation}
    \mathrm{sim}\!\left(\phi_{\mathrm{IMU}},\phi_{\mathrm{Flow}}\right)
=\frac{\phi_{\mathrm{IMU}}^{\top}\phi_{\mathrm{Flow}}}{\|\phi_{\mathrm{IMU}}\|_2\,\|\phi_{\mathrm{Flow}}\|_2},
\end{equation}
$\phi$ is the spectral feature defined in \hyperref[sec:methods:PSD_feature]{Spectral Feature Selection}
\end{add}% temporal pooling from $Z^{\mathrm{IMU}}$ and $\hat Z^{\mathrm{Flow}}$. 

This score quantifies the motion-pattern consistency between the IMU segment and each candidate and is used for ranking and final identification.

% \textbf{Similarity Head}

% 如前面\todo{ref}所证明的一样 IMU data和光流数据只有幅值上的差距。而他们的量纲不同（IMU是 m， 光流是像素），为消除能量尺度影响，我们对经过 cross attention 后的时序嵌入进行 L2 归一化，并采用余弦相似度作为跨模态匹配分数：

% \[
% \mathrm{sim}(\mathbf{v}_\mathrm{IMU}, \mathbf{v}_\mathrm{Flow}) = \frac{\mathbf{v}_\mathrm{IMU}^\top \mathbf{v}_\mathrm{Flow}}{\|\mathbf{v}_\mathrm{IMU}\| \|\mathbf{v}_\mathrm{Flow}\|}
% \]
% 其中 $\mathbf{v}_\mathrm{IMU}$ 和 $\mathbf{v}_\mathrm{Flow}$ 分别为 IMU 和候选人光流的聚合特征。Similarity Head 的输出用于衡量每个候选人与当前 IMU 片段的运动模式一致性，并作为后续排序和判别依据。

% \todo{公式}，评估二者的相关联程度。具体来说，输入为IMU数据和每个候选人的数据经过了cross attention 的结果，输出表示该 IMU 片段与某候选人光流片段的跨模态匹配相似度

% \textbf{Scale-aware Fusion}
% \todo{这个窗口大小和时序的滑动窗口大小要区分开来，scale这个词怎么样？}
% 上述的过程都是针对一个窗口大小，由\todo{ref}可得，用户在不同的条件下有不同的动作行为表现，因此也适用于不同的窗口大小。考虑到实际使用中可以估算获得信息，我们引入了scale-aware的窗口混合方案，采用人的bbox的宽和画面宽度的比例作为距离的估算，形成condition。通过将相似度数据和 MLP对不同的窗口数据计算权重，然后将相似度

\textbf{Scale-Aware Multi-Window Fusion.}
\label{methods:CSINet:window}
User motion varies with gesture and distance; a single temporal extent may be sub-optimal. We therefore evaluate similarities over a set of DFT window lengths $\{w\}_{5}^{20}$. These DFT windows are different from the sliding window used to segment the streaming input sequence. And fuse them conditioned on an estimated scale $s$ (approximated by the ratio between the person's bounding-box width and the image width). An MLP takes $s$ (optionally concatenated with quality statistics) and outputs fusion weights $\alpha_m=\mathrm{softmax}(g(s))$. The final similarity is the weighted average of $\mathrm{sim}$ under various window sizes $m$. 
\begin{equation}
\mathrm{Sim}_\mathrm{final} \;=\; \sum_{m=1}^{M} \alpha_m \,\mathrm{sim}_{m},
\end{equation}
adapting the effective temporal context to distance while preserving robustness across heterogeneous motion regimes.

\subsubsection{Model Training.\\}
\label{sec:sys:training}

% 我们在 PyTorch 实现中采用 AdamW 优化器与余弦退火（CosineAnnealingLR）学习率调度进行端到端训练。训练轮数为 $50$，3轮无valid acc 下降后早停，批量大小为 $4$，初始学习率为 $1{\times}10^{-4}$，权重衰减为 $1{\times}10^{-4}$，调度器的 $\text{T}_{\max}$ 设为训练总轮次的一半且最小学习率为初始学习率的 $1\%$。为保证可复现性，我们在每轮结束后记录损失与准确率，并基于验证集的\emph{严格准确率}（预测候选人与标注相同）选择性能最佳的检查点用于后续推理。

\hspace{-0.5em}\textbf{Training.}
We train CSINet end-to-end in PyTorch using AdamW with cosine annealing. We run for $50$ epochs with early stopping after $3$ epochs without validation improvement, batch size $4$, initial learning rate $1\text{e}-4$, and weight decay $1{\times}10^{-4}$. The scheduler uses $T_{\max}=25$ (half of total epochs) and $\eta_{\min}=0.01\times$ the initial learning rate. 
% For reproducibility, we log loss and accuracy after each epoch and select the checkpoint with the highest \emph{strict accuracy} on the validation set for subsequent inference.

% \paragraph{批处理与掩码。}
% 为适配多人、变长序列的设置，我们按批对片段进行\emph{对齐—填充}：在批内对时间步数、尺度数与候选人数量对齐到最大值，并构造帧掩码与人员掩码以屏蔽填充项。时间维特征采用\emph{掩码均值池化}得到片段级嵌入；随后对 IMU 与跨模态对齐后的 Flow 表示进行 $\ell_2$ 归一化以提高数值稳定性与对比学习效果。

\textbf{Batching and masking.}
To accommodate multi-person scenes and variable-length sequences, we perform batchwise \emph{align-and-pad}: within each mini-batch, we pad time steps, the number of temporal scales, and the number of candidates to their within-batch maxima and construct frame and person masks to ignore padded entries. Temporal features are aggregated via \emph{masked mean pooling} to obtain segment-level embeddings. We then apply $\ell_2$ normalization to the IMU representation and to the attention-aligned flow representation to improve numerical stability and the effectiveness of contrastive learning.

\begin{add}
\textbf{Loss.}
The principal component is the InfoNCE contrastive loss, computed over per-segment, per-candidate scores. To prevent overfitting and ensure generalization, we incorporate an $L_2$ regularization term on the network parameters $\theta$. For the $i$-th sample with similarity logits $\mathbf{s}_i=[s_{i,1},\dots,s_{i,P_i}]$ and positive index $y_i$, the contrastive term is defined as:
\begin{equation}
\mathcal{L}_{\text{InfoNCE}}
= - \frac{1}{B} \sum_{i=1}^{B}
\log \frac{\exp\!\big(s_{i,y_i}/\tau\big)}
{\sum_{j=1}^{P_i} \exp\!\big(s_{i,j}/\tau\big)} + \lambda \|\theta\|_2^2,
\end{equation}

In practice, we clip logits to a safe range and sanitize NaN/Inf values to prevent gradient pathologies.
\end{add}

% \paragraph{训练目标。}
% 主损失为基于片段—候选人打分的 InfoNCE 对比损失。对第 $i$ 个样本，模型产生其与所有候选人的相似度向量 $\mathbf{s}_i = [s_{i,1},\dots,s_{i,P_i}]$，其中正样本索引为 $y_i$。损失为
% \[
% \mathcal{L}_{\text{InfoNCE}}
% = - \frac{1}{B} \sum_{i=1}^{B}
% \log \frac{\exp\big(s_{i,y_i}/\tau\big)}
% {\sum_{j=1}^{P_i} \exp\big(s_{i,j}/\tau\big)} ,
% \]
% 其中温度参数 $\tau=0.3$。当样本中仅有单一候选人时，采用边界损失 $\max(0,\,1-s_{i,y_i})$。实现上我们对 logits 做范围裁剪并在出现 NaN/Inf 时进行安全替换，以防止梯度异常。

% \paragraph{尺度聚合与先验。}
% 模型在每个频谱尺度上产生一组相似度，再做尺度聚合得到最终打分。我们实现了基于质量向量的\emph{尺度注意力}与\emph{质量门控}，并支持按人条件（如距离占比）进行调制；在最终配置中，为提升训练稳定性，我们对各尺度打分采用\emph{均匀平均}聚合，并保留可学习门控/注意力用于消融分析。

% \paragraph{难例构造与负样本增强。}
% 为提高对干扰者的判别性，我们在训练阶段引入\emph{跨序列负样本增强}：从其他序列中随机采样若干人的时间序列，按候选人维度附加到当前样本，使每个批次包含更丰富的“外部负例”（默认追加 $5$ 个）。该过程中保持原始正样本索引不变。

\textbf{Augmentation.}
To strengthen discrimination against distractors, we adopt \emph{cross-sequence negative augmentation}: for each training instance we append time series from randomly sampled people in other sequences along the candidate dimension, so each batch contains 5 external negatives. The original positive index $y_i$ is preserved throughout.

\section{EXPERIMENTS on DATASETS}

% \todo{参考VIPL的写法}

% \subsection{Settings}
% we use our dataset \todo{ref} to evaluate 我们系统的效果。specially， we train the methods on 30人的数据，在8个人的数据上做验证。共进行了5次随机数种子不同的实验.

\begin{add}

\subsection{Evaluation Protocol}
\label{sec:exp:protocol}

We evaluate \methods~ on our long-range gestural interaction dataset (Sec.~\ref{sec:dataset}). Unless otherwise noted, all results are reported as the mean over 5 runs with different random seeds.

\subsubsection{Implementation and Training Strategy}
Since the official implementations of the baselines~\cite{bhatnagar2023long,bamani2024ultra,sun2020when} are not publicly available and were originally designed for shorter ranges, we re-implemented their architectures with necessary adjustments to adapt to far-range scenarios. For image-based baselines, input sequences are sampled at 5\,fps.
To ensure a rigorous subject-independent evaluation, we use data from 30 participants for training, while the remaining 8 participants are reserved as the \textbf{held-out} test set. Within the training set, we split at 80/20 for training and validation.

\subsubsection{Robustness Test Set Construction}
\label{sec:time_noise}
All experiments are evaluated on the 8 held-out participants. To rigorously assess the system's robustness against real-world synchronization failures~\cite{furgale2013unified}, we introduce artificial temporal noise into the synchronized test set. Specifically, we apply three types of temporal perturbations between the input streams:

\begin{itemize}
    \item \textbf{Global Offset.} To simulate transmission latency, we apply a global time shift $\Delta t$ for each sequence. The offset is sampled from a uniform distribution: $\Delta t \sim \mathcal{U}(-500, 500)$\,ms.
    \item \textbf{Clock Drift:} To mimic the sampling rate mismatch between independent device clocks (e.g., camera vs. IMU crystal oscillators), we apply a linear scaling factor $\alpha \sim \mathcal{U}(0.98, 1.02)$ to the timestamps.
    \item \textbf{Jitter:} To simulate packet arrival jitter and processing fluctuations, we add independent Gaussian noise to the timestamp: $\epsilon \sim \mathcal{N}(0, \sigma^2)$, where $\sigma=60$\,ms (approx. 2 frame duration).
\end{itemize}

%     \subsection{Evaluation Protocol}

%     We evaluate \methods~ on our dataset (Sec.~\ref{sec:dataset}). Unless otherwise noted, results are reported as the mean over 5 runs with different random seeds.

%     \subsubsection{Re-implement and Training.} 由于baseline 都没有进行开源，且测试的结果不能覆盖我们的距离， we tried our best to re-implement the architecture and
% introduced practical adjustments for far-range scenes。for picture input baseline \ref{sec:experiment:DNN} and \ref{sec:experiment:URGR},All inputs are sampled at 5 fps from our dataset recordings. baselines and \methods~ are train on data from 30 participants and test on 8 held-out participants. 训练数据被按照8:2 的比例划分为训练集和测试集

%     \subsubsection{Test Set Construction.} All the experiments are evaluated on 8 \textbf{held-out} participants from our dataset. 为了模拟真实的情况，我们在test set中加入了IMU和vision数据不同步的时间噪声，specifically，参考 \cite{}, 设计三种噪声- Global offset (500ms with gaussian noise 怎么表示这个noise的等级，不是一个定值，是范围的随机）
% clock drift（描述 两个设备的时钟周期不一致）
% Jilter (Gaussian noise
\end{add}

\subsection{Baselines}

% 为了展示vision-based model在远距离上的感知能力，我们采纳了两份远距离手势识别的工作作为baseline，同时和CSI任务的VIPL\todo{cite}进行了对比。以及将我们的模型修改为使用Skeleton来表征视觉运动信息以此来验证我们使用spectral domain的数据在远距离CSI任务上的优势

To assess the perceptual limits of vision-based models at long range, we compare against two representative long-distance gesture recognition research and a cross-modal CSI method (VIPL), and we additionally construct a skeleton-based variant of our pipeline to probe the benefit of spectral representations for CSI.

\subsubsection{DNN~\cite{bhatnagar2023long}}
\label{sec:experiment:DNN}
% 以 RGB 视频为输入，通过初步特征提取器提取低维时空特征，借助 Patch 选择子网络筛选含手势信息的关键区域并早期丢弃背景特征，再由 Patch 手势分类器对选中区域进行手势分类。我们将其作为纯视觉方案的baseline，它说的Long-Distance 指的是4m的距离，also As there is not open-source code release，为了将测试扩展到30m的区间，we try our best to re-implement the module proposed 同时针对原文没考虑到的远距离场景做了一些优化，对于10m以上的场景，适当减小patch size，并增大patch选择的loss 权重。数据由sequential数据集按照sample rate 5 抽帧组成。
This vision-only baseline takes an RGB picture as input. A shallow spatiotemporal stem extracts low-dimensional features; a patch-selection subnetwork highlights gesture-relevant regions and suppresses background; and a patch-level classifier predicts the gesture class. The original work reports ``long-distance'' results at roughly 4\,m. We introduced smaller patch sizes beyond 10\,m and a higher loss weight on patch selection to encourage tighter focus.

\subsubsection{URGR~\cite{bamani2024ultra}}
\label{sec:experiment:URGR}
% 针对人机交互场景下的超远距离（最远 25 米）手势识别，仅采用普通 RGB 摄像头，提出含 HQ-Net（增强低分辨率图像）与 GVIT（融合 GCN 和改进 ViT）的 URGR 深度学习框架。其最远距离到25m，对于更远的场景，我们使用了SOTA 的Image Restoration模型\todo{SeeSR}作为其HQ-Net的替代，并自行训练了GVIT。video sample rate is also 5
URGR targets ultra-long-range (up to 25\,m) gesture recognition with a single RGB camera by combining an image enhancement module (HQ-Net) with a GVIT backbone that fuses GCN and a modified ViT. To test distances beyond 25\,m, we replace HQ-Net with a state-of-the-art image restoration model~\cite{wu2024seesr}.

\subsubsection{VIPL~\cite{sun2020when}} 
% 该方法针对无目标预先外观信息的机器人 / 自动驾驶会合场景，先通过 YOLOv3 检测与 DeepSORT 跟踪视频中人体，提取时间超像素光流、边界框尺寸、肩关键点作为视觉特征，同时从智能手机 IMU 提取去重力 3D 线性加速度与 3D 角速度作为惯性特征，再利用含 LSTM 和卷积层的双分支网络将两类特征映射到联合特征空间，以三元组损失优化使同一人的视觉与惯性特征在该空间距离接近，测试时通过匹配与目标 IMU 特征 L2 距离最小的视觉特征实现人体定位。
\addl{VIPL tackles command source identification without prior appearance enrollment. While sharing similar input modalities with \methods, it fundamentally differs by modeling temporal motion dynamics in the time domain, heavily relying on large-scale full-body walking cues and assuming tight synchronization. Specifically, it pairs visual features (optical flow and keypoints) with raw inertial signals (acceleration and angular velocity). A dual-branch LSTM-based network maps both streams into a shared embedding space optimized by a triplet loss, selecting the person with the minimum feature distance at inference.}

% \subsubsection{简单时序模型}

\subsection{Metrics}
Unless otherwise noted, a segment is counted as correct under \emph{strict accuracy} if the predicted candidate exactly matches the annotated command source. But for the vision-only baseline DNN (Sec.~\ref{sec:experiment:DNN}) and URGR (Sec.~\ref{sec:experiment:URGR}), which is a single-frame gesture recognizer rather than a multi-person CSI model, we adapt its outputs as follows: a frame is deemed correct if its predicted gesture matches the ground-truth command issued by the true source; segment-level decisions are then obtained by majority vote over frames within the segment. Because this baseline has no capability to resolve which person performed the gesture, the above protocol effectively grants it oracle source attribution; therefore, the resulting CSI accuracy should be interpreted as an \emph{upper bound} on what a purely vision-based, single-frame recognizer could achieve in our multi-person setting.

% 我们统计了多组实验下在不同距离上的正确率的均值和方差。特别的，对于vision-based baseline \todo{ref}，他们的方法是基于单帧的，且没有CSI任务，而是专注于做手势识别，所以我们将手势识别正确就视为CSI任务正确，对于整段结果，我们统计段落的每帧的识别结果，将众数作为整段结果。由于其没有辨识多人的能力，因此这样得到的正确率将是其真实正确率的上界。

\subsection{Results and Analysis}

% DNN 全测试集（1-4m）的准确率为 89.94%，而针对 4m 子集（标注为 l3 和 r3 的录制点，距离≈4m），准确率为 85.65%

% 超分正确率降低到96% 当超过 25m的时候

% \begin{figure}[h]
%     \centering % 子图内部居中
%     \includegraphics[width=0.6\linewidth]{pics/ACC.pdf} % 图片宽度适配子图容器
%     \caption{\textbf{Results of Experiments.} CSI Accuracy vs. Distance.} % 子图标题
%     \label{fig:acc_all} % 子图标签（用于交叉引用）
% \end{figure}

\begin{figure}[h]
    \centering % 子图内部居中
    \vfill
    \vspace{-1em} 
    \includegraphics[width=0.8\linewidth]{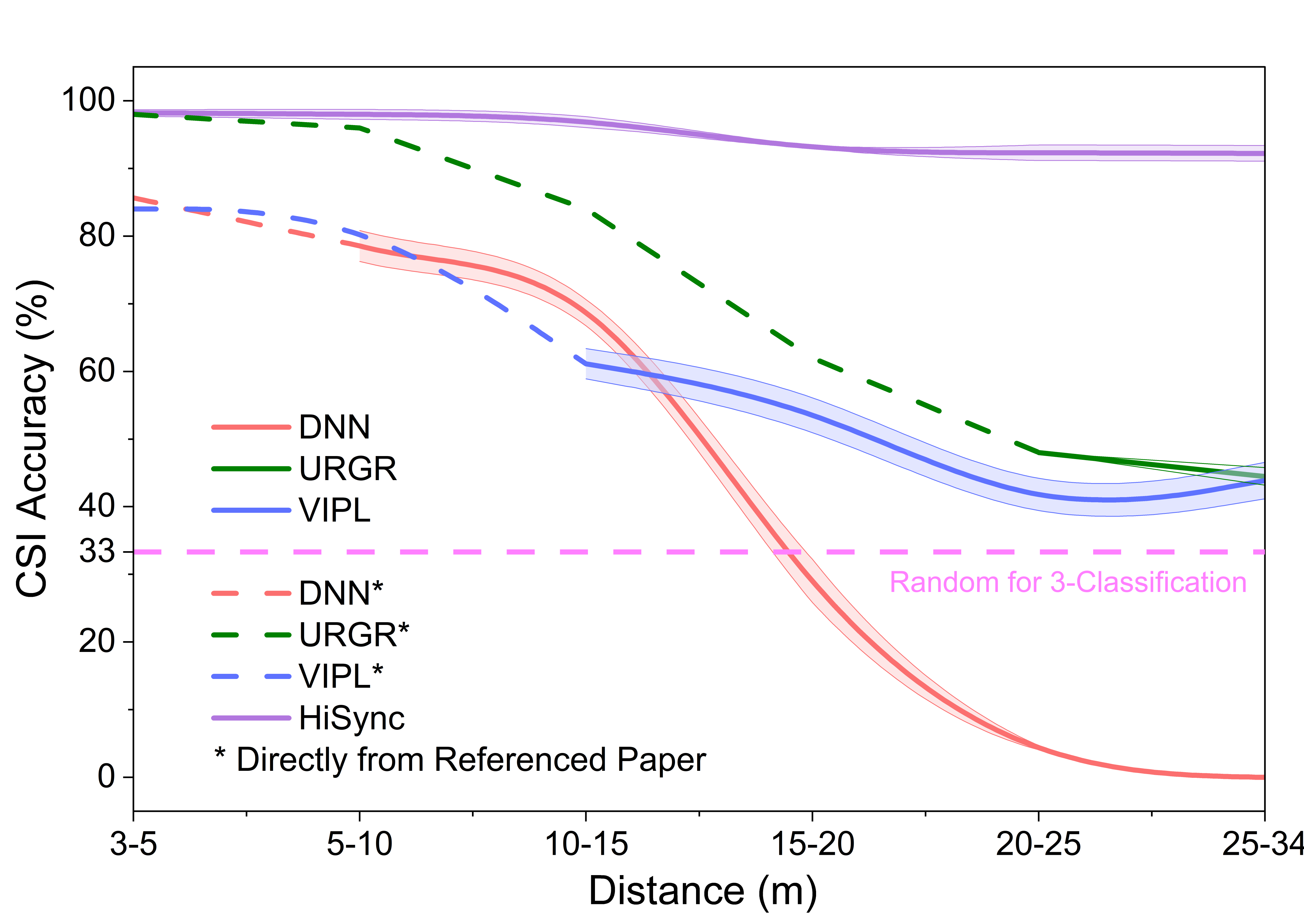} % 图片宽度适配子图容器
    \vfill
    \vspace{-1em} 
    \caption{\textbf{Results of Experiments.} \addl{This figure shows the CSI Accuracy vs. Distance. The dashed line represents the data obtained directly from the referenced paper. The results show that \methods~ outperform baselines, especially at long distances. Error bar denotes standard deviation.}} % 子图标题
    \label{fig:acc_all} % 子图标签（用于交叉引用）
    \Description{This figure presents the accuracy of Command Source Identification (CSI) across different distances from 3 to 34 meters, comparing several models. The curves show the performance of different methods: DNN (red), URGR (green), VIPL (blue), and the proposed method (purple). The dashed lines indicate results from the referenced papers, with “DNN*,” “URGR*,” and “VIPL*” marked as directly taken from previous works. The "Random" baseline is also shown as a horizontal reference line. The CSI accuracy decreases as distance increases, with the proposed method (OURS) performing better than others at longer ranges.}
\end{figure}

\begin{figure}[h]
      \includegraphics[width=0.8\linewidth]{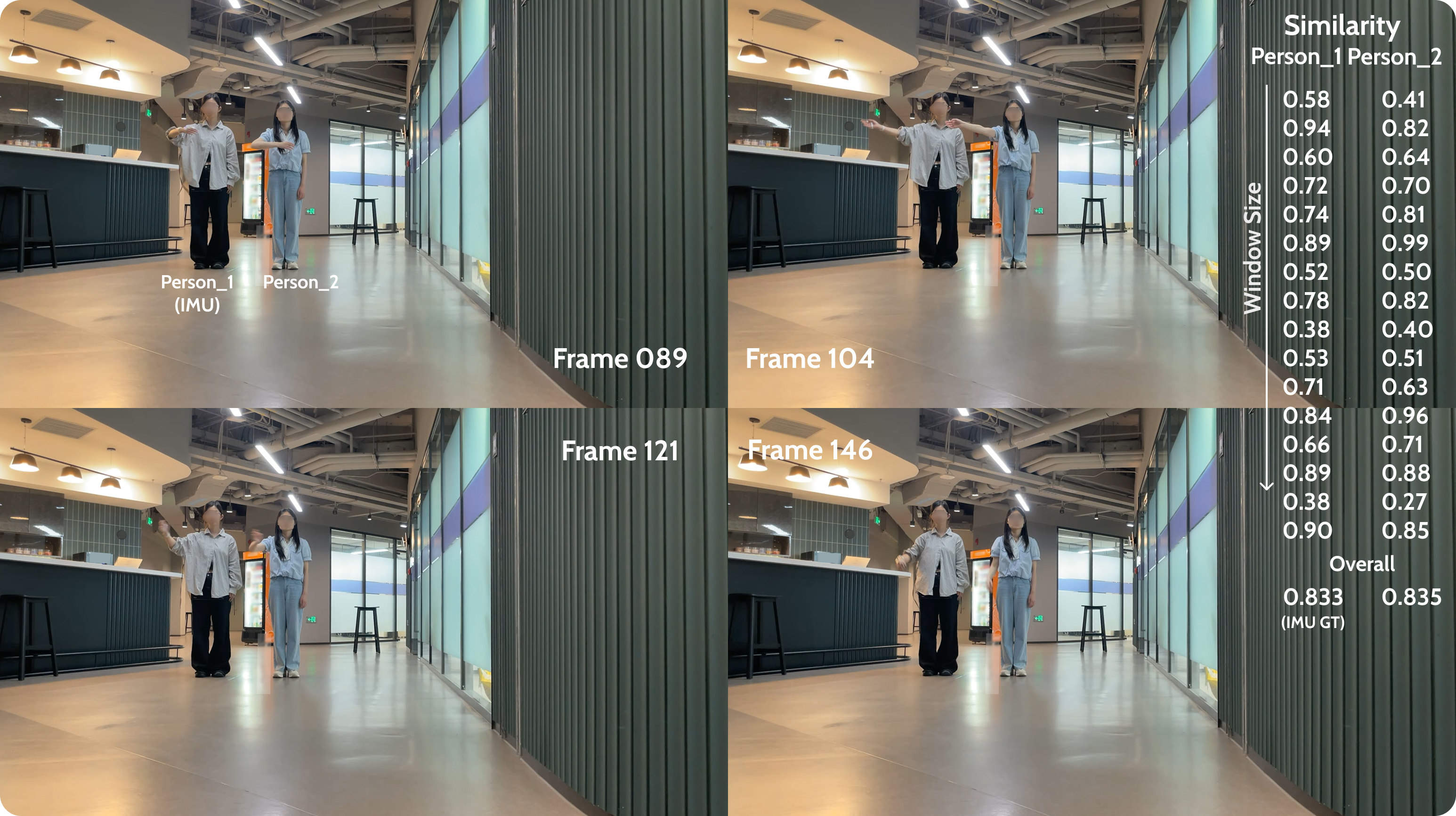}
      \vfill
      \vspace{0.5em} 
      \caption{\textbf{Failure Case.} \addl{This is a failure sequence where the two persons move in near-perfect sync. However, because the \methods~ can capture subtle differences in movements, the failure rate is very low even in such a rare case.}}
      \label{fig:fail}
      \Description{This figure shows a failure scenario where two participants (Person 1 and Person 2) move in near-perfect synchronization. The frames capture different moments (Frames 089, 104, 121, 146), with corresponding similarity scores between the two individuals. Despite the near-identical movement, the system struggles to distinguish between them, as indicated by the similarity values (ranging from 0.41 to 0.94). The overall similarity score is shown at the top right. However, because the HiSync can capture subtle differences in movements, the failure rate is very low even in such a rare case.}
\end{figure}

\subsubsection{Overall Performance}
% \methods maintains reliable command-source identification across ranges, outperforming both purely visual baselines and SOTA Optical-Inertial CSI methods. 在平均画面中用户为N=3的情况下我们的方法在10m内保持了99.77\%的正确率，超过 VIPL 论文内汇报的84\%, comparable to 纯视觉方法的上界 3-5m 100\%, 5-10m 98\%, 在10-30m内正确率 96.64\% ,而baseline方法随着距离增加效果急剧下降，我们超过了VIPL 70.34\% 。而在up to 34m的极限距离下，我们仍有 94.31\%的正确率，而baseline已经接近随机 (VIPL三分类 43.88\%  DNN手势识别八分类 xxx\%) .

\methods~sustains reliable command-source identification (CSI) across long ranges, outperforming both purely visual baselines and state-of-the-art optical-inertial methods. With about 3 people visible per frame on average and a sequence length of 3 seconds, it achieves 99.77\% CSI accuracy at $\leq 10 m$, 15.77\% over VIPL's reported 84\% and comparable with a vision-only upper bound (100\% at 3-5 m; 98\% at 5-10 m). Between 10-30 m, \methods~maintains 96.64\% accuracy, beating the VIPL baseline (70.34\%) by 26.30\% as that method degrades sharply with distance. Even at the extreme range (up to 34 m), \methods~still attains 94.31\% accuracy, whereas baselines approach random (VIPL, three-category classification: 43.88\%; vision-only gesture recognizer, eight-category: 44.48\%).

% \subsubsection{Robustness to Desync \& Noise}

% 录制的真实数据已经有一些time desync 和 data noise，为了进一步检验系统的鲁棒性,人工在

\subsubsection{Fail Case Studies}

% As Fig. \todo{Fig}, In rare cases, two participants produced \emph{kinematically near-identical} gestures (dominant frequency, harmonic ratio, and amplitude envelope), yielding almost the same cross-modal similarity to the query IMU. This creates a ``spectral collision''. 这种情况下可能需要在更长的时间跨度上考虑，detailed discuss at Sec \todo{Ref}

As shown in Fig.~\ref{fig:fail}, in rare instances two participants exhibit gestures in near-perfect sync --- sharing dominant frequencies, harmonic structure, and amplitude envelopes, thus yielding nearly equal cross-modal similarity to the query IMU, a condition we term a ``spectral collision''. In such cases, leveraging a longer temporal context is warranted; see Sec.~\ref{sec:ablation:scale} for details.

\subsection{Ablation}
\begin{figure}[h]
      \centering
      \includegraphics[width=0.8\linewidth]{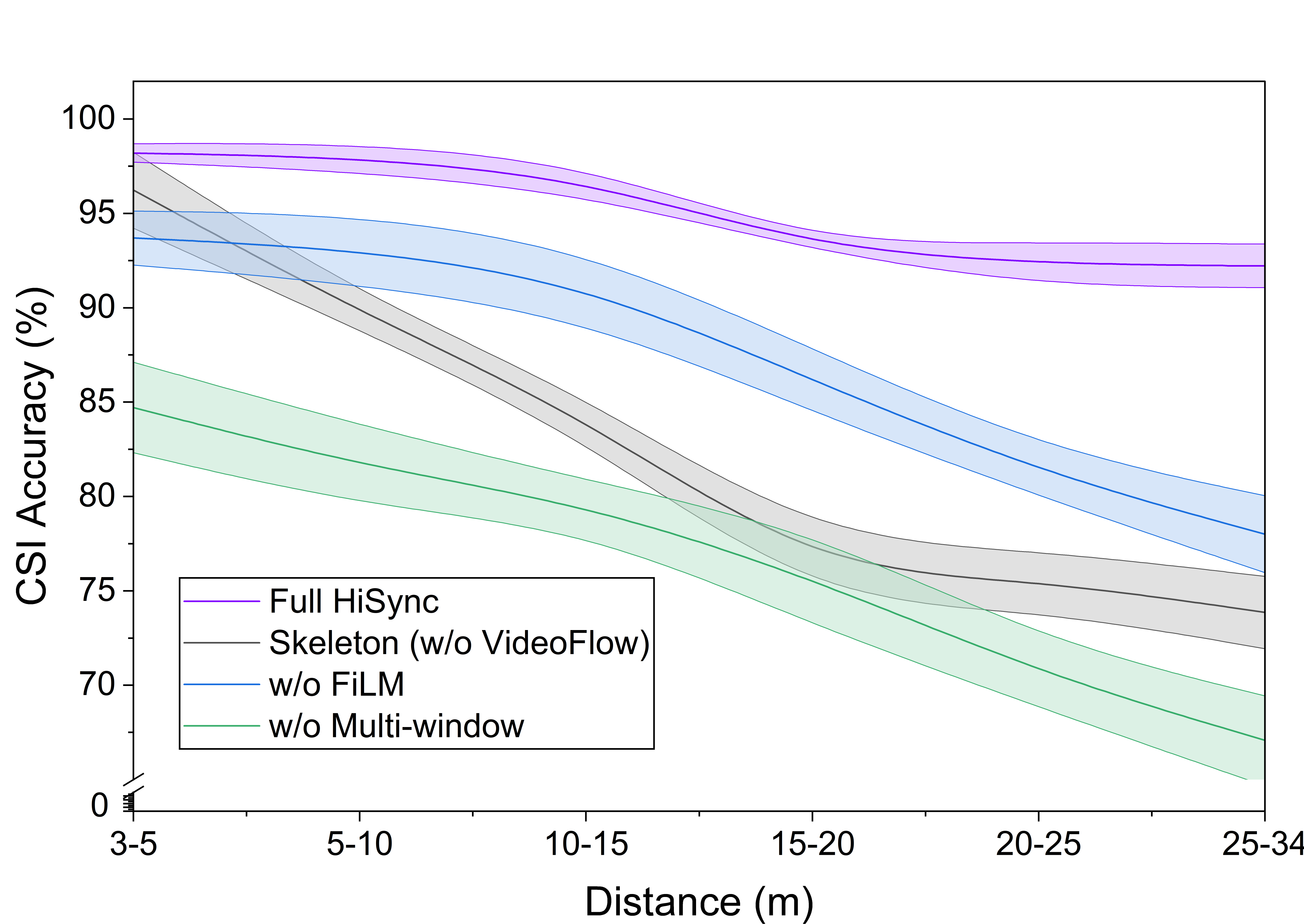}
      \caption{\textbf{Ablation Results.} \addl{We compare the full system against ablated variants across distances. The significant performance drop in incomplete models highlights the necessity of each component for maintaining robust CSI. Error bar denotes standard deviation.}}
      \label{fig:acc_ablation}
      \Description{This figure shows the performance of different models for Command Source Identification (CSI) across various distances (3-34 meters). The curves represent four models: "Full" (purple), "Skeleton" (gray), "-FiLM" (blue), and "-Multi-window" (green). CSI accuracy decreases as distance increases, with the "Full" model maintaining the highest accuracy. The figure also illustrates how removing certain components (e.g., FiLM or Multi-window) reduces performance, particularly at longer distances.}
\end{figure}

We did ablation studies on \textbf{Optical Flow} (Skeleton), \textbf{Quality-Aware Feature Modulation} (-FiLM), \textbf{Scale-Aware Multi-Window Fusion} (-Multi-window) for \methods. See Fig.~\ref{fig:acc_ablation} for quantitative comparisons.

\begin{figure}[h]
      \centering
      \includegraphics[width=0.8\linewidth]{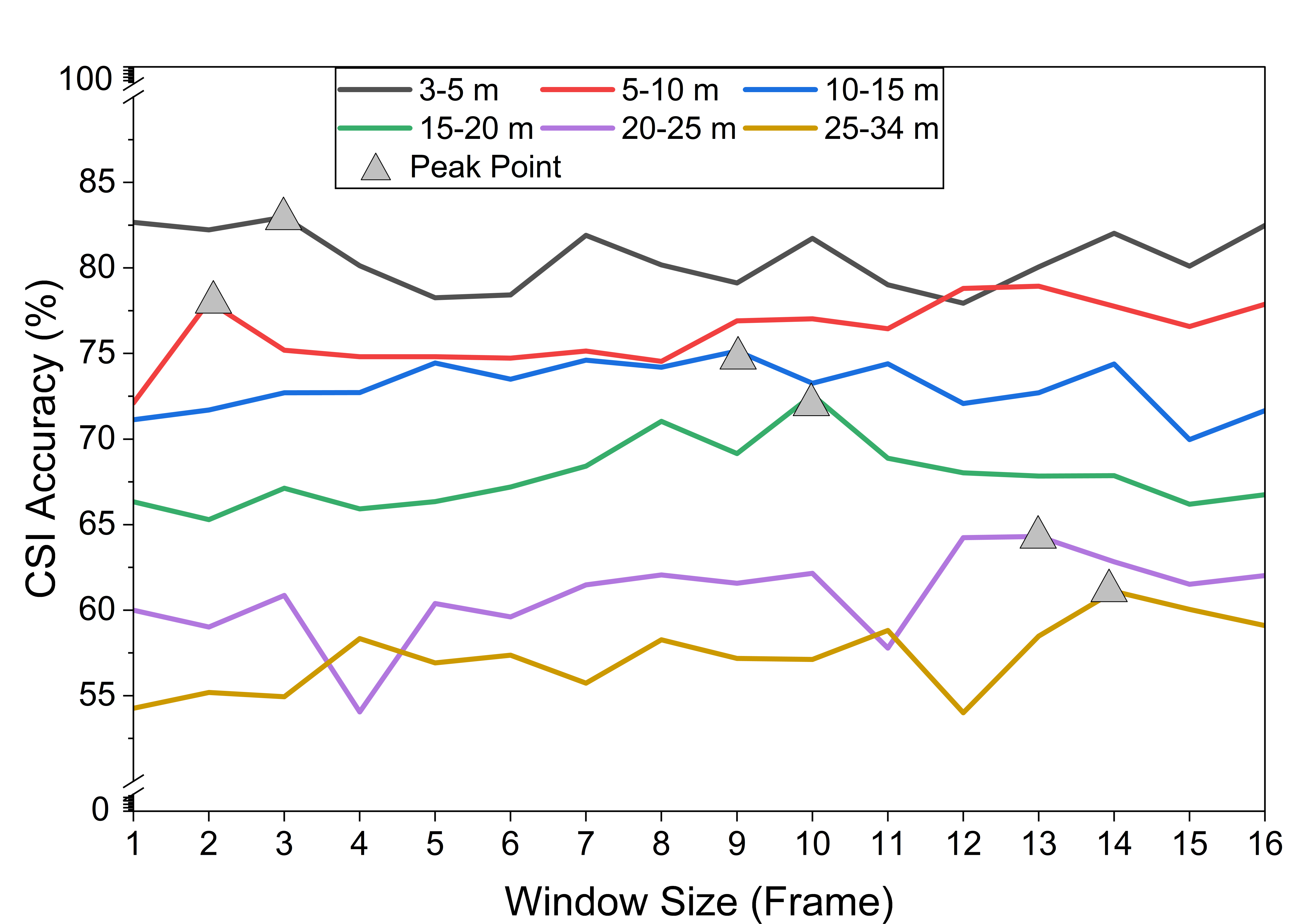}
      \caption{\textbf{Impact of Window Size on Accuracy.} \addl{This figure shows a clear trend where optimal window size increases with distance. This validates our design choice to fuse the different temporal window sizes.}}
      \label{fig:acc_window}
      \Description{This figure displays the Command Source Identification (CSI) accuracy across various window sizes (1-16 frames) and distances (3-34 meters). The curves show how CSI accuracy changes with different window sizes for each distance range, with the peak points marked by triangles. The results indicate that CSI accuracy increases with larger window sizes, with optimal performance achieved at specific window sizes for each distance range. The accuracy decreases as the distance increases, with the 3-5m range showing the highest accuracy.}
\end{figure}

% \begin{figure}[h]
%   \centering
%   \includegraphics[width=\linewidth]{pics/windowsize.pdf}
%   \caption{Gesture Amplitude vs. Time}
%   \Description{\todo{画一条33\%的参考线}}
%   \label{fig:amp_time}
% \end{figure}

\subsubsection{Ablation on Optical Flow}
% 为了验证我们使用spectral domain的数据在远距离CSI任务上的优势，此baseline采用了非常straight forward的方法利用human Skeleton作为motion信息，使用MLP来替换我们方法中的时域特征提取\todo{ref}，具体的，对于输入\todo{} 降维到64，后续不变。
To isolate the value of Optical Flow feature for CSI at long range, we replace the \textbf{Velocity Estimation} (Sec.~\ref{sec:methods:velocity}) in our method with per-person 2D skeleton trajectories and encode them using a lightweight MLP in lieu of Spectral Analysis (Sec.~\ref{sec:methods:spectral} for 2D velocity). Specifically, for each candidate, we form a concatenated keypoint trajectory vector over the window, reduce it to 64 dimensions with a linear layer, and keep all downstream components unchanged. This variant performs markedly worse at longer distances, where pose detection becomes unreliable as keypoints are often missing, jittery, or mislocalized, confirming that Optical Flow motion features provide greater robustness under low resolution and high noise.

\subsubsection{Ablation on Quality-Aware Feature Modulation}

We disable the \textbf{Quality-Aware Feature Modulation} (Sec.~\ref{methods:CSINet:FiLM}) and keep the rest of the architecture and training schedule fixed. Without Quality-Aware Feature Modulation, the model cannot down-weight low-quality or noise-dominated frequency channels, leading to consistently lower accuracy across distances and harder training convergence. Errors concentrate in segments with micro-motions that reduce SNR of IMU, indicating that Quality-Aware Feature Modulation's channel-wise modulation is critical for stabilizing learning and improving robustness under persistent IMU noise.
% 结果从一开始就不好，可能是由于IMU数据持续存在的噪声

\subsubsection{Ablation on Scale-Aware Multi-Window Fusion}
\label{sec:ablation:scale}

We replace \textbf{Scale-Aware Multi-Window Fusion} (Sec.~\ref{methods:CSINet:window}) with a single-scale baseline and sweep several window lengths for fairness. We report per-distance accuracies for each window size in Fig.~\ref{fig:acc_window}. Removing scale-aware fusion yields a substantial overall drop, and the window that maximizes accuracy varies with distance. Generally, larger windows perform better at longer ranges while shorter windows suffice at close range. Intuitively, long windows capture global motion characteristics (for example, overall swing frequency), whereas short windows better capture fine-grained cues (for example, arm length proxies and personalized way of exerting force). These results support our design choice to fuse temporal scales rather than rely on a single window.

% 测试了不同的窗口大小上的正确率

% 汇报不同窗口大小下不同动作，不同距离的正确率

\begin{add}

% Beyond proving  at conceptual level (Sec.~\ref{sec:validation_of_time_shift}),

\subsection{Evaluation on Temporal Misalignment}
\label{sec:exp:temporal_robustness}
Beyond the theoretical justification (Sec.~\ref{sec:validation_of_time_shift}), we empirically evaluate the method’s robustness against temporal misalignment under progressively severe desynchronization scenarios.

\subsubsection{Ablation on Spectral Transformation.} 
\label{sec:desync:ablation}
% To validate the necessity of the spectral transformation, we remove the Motion Feature Extractor (Sec~\ref{methods:motion_feature_extractor}. The raw IMU acceleration and visual velocity sequences are fed directly into a network 和CSINet shape一样. This serves as an ablation to verify the impact of spectral features on alignment robustness. To further compare \methods~ with traditional temporal alignment techniques, we employ \textit{Cross-Correlation (XCorr)} and \textit{Dynamic Time Warping (DTW)} to pre-align the visual and inertial streams in the time domain before feeding them into the ``w/o Spectral'' network.
To validate the necessity of the spectral transformation, we bypass the Spectral Analysis (Sec.~\ref{sec:methods:spectral}) step. Instead, the raw IMU linear acceleration and visual velocity sequences are fed directly into a network with an architecture identical to CSINet. This ``w/o Spectral'' setting serves to verify the impact of spectral features on alignment robustness. Furthermore, to benchmark against traditional temporal alignment techniques, we employ Linear Regression \cite{furgale2013unified} (Linear), Cross-Correlation \cite{knapp2003generalized} (XCorr) and Dynamic Time Warping \cite{berndt1994using} (DTW) to pre-align the visual and inertial streams in the time domain before feeding them into the ``w/o Spectral'' network.

% 我们去掉了整个 Motion Feature Extractor. 直接将时域输入到跟 CSINet 相同的backbone中，而不是转换到频域中。这个实验为了验证

% \subsubsection{Ablation on Spectral Transformation.}
% \label{sec:desync:methods}
% To validate the necessity of the spectral transformation, we construct three time-domain baselines that input raw motion sequences directly into the backbone network， replacing the Spectral Analysis module in the Motion Feature Extractor). The raw IMU acceleration and visual velocity sequences are fed directly into a network 和CSINet shape一样. This serves as a ablation to verify the impact of spectral features on alignment robustness.

% \begin{itemize}
%     \item \textbf{w/o Spectral (Time-Domain Only):} The raw IMU acceleration and visual velocity sequences are fed directly into the network without any frequency domain transformation. This serves as a baseline to verify the impact of spectral features on alignment robustness.
%     \item \textbf{Pre-Alignment (XCorr / DTW):} To assess whether traditional temporal alignment techniques can substitute for spectral analysis, we employ \textit{Cross-Correlation (XCorr)} and \textit{Dynamic Time Warping (DTW)} to pre-align the visual and inertial streams in the time domain before feeding them into the ``w/o Spectral'' network.
% \end{itemize}

% \subsubsection{Temporal Sync Baselines.}
% To througoutly evaluate, 我们用时序对齐中常见的方法Linear, XCorr, DTW 预先处理后放入-Spectral的网络中处理

% \subsubsection{测试集实验组.}
% 除了原本已经同步的测试集 $T_{clean}$ 和 Sec.~\ref{sec:time_noise} (记为 $T_3$)，还增加了两个噪声稍弱的测试集。$T_1$ 仅有 global offset  $\Delta t \sim \mathcal{U}(-200, 200)$\,ms。 $T_2$ 在 $T_3$去掉了jilter。

\subsubsection{Progressive Noise.}
\label{sec:desync:datasets}
% To simulate real-world conditions ranging from ideal to severe, we construct four test sets with varying degrees of temporal noise. In addition to the synchronized set $T_{clean}$ and the severe noise set $T_{3}$ (defined in Sec.~\ref{sec:exp:protocol}), 还增加了两个噪声稍弱的测试集。$T_1$ 仅有 global offset  $\Delta t \sim \mathcal{U}(-200, 200)$\,ms。 $T_2$ 在 $T_3$去掉了jilter。
To simulate real-world conditions ranging from ideal to severe, we construct four test sets with varying degrees of temporal noise. Beyond the synchronized set $T_{clean}$ and the severe noise set $T_{3}$ (detailed in Sec.~\ref{sec:exp:protocol}), we introduce two intermediate sets. $T_{1}$ features only a mild global offset $\Delta t \sim \mathcal{U}(-200, 200)$\,ms. $T_{2}$ adopts the global offset and clock drift from $T_{3}$, but excludes transmission jitter to isolate the impact of irregular fluctuations.

% we introduce two intermediate levels:

% \begin{itemize}
%     \item \textbf{$\mathcal{T}_{clean}$ (Ideal):} The original time-synchronized dataset.
%     \item \textbf{$\mathcal{T}_{1}$ (Mild):} Contains only random transmission latency. We apply a global offset $\Delta t \sim \mathcal{U}(-200, 200)$\,ms.
%     \item \textbf{$\mathcal{T}_{2}$ (Moderate):} Simulates latency plus clock drift, but without packet jitter. This set includes the global offset from $T_3$ ($\pm 500$\,ms) and clock drift $\alpha \sim [0.95, 1.05]$, excluding Gaussian jitter.
%     \item \textbf{$\mathcal{T}_{3}$ (Severe):} The full noise protocol defined in Sec.~\ref{sec:exp:protocol}, combining aggressive global offset ($\pm 500$\,ms), clock drift, and transmission jitter.
% \end{itemize}

% \subsubsection{Result}
% 实验结果如 Tab.~\ref{tab:time_temporal_evaluation}所示，

% \subsubsection{Results and Analysis}
% The comparative results are reported in Tab.~\ref{tab:time_temporal_evaluation}. 
% In contrast, \methods~ maintains high accuracy across all noise levels, dropping only marginally from $\mathcal{T}_{clean}$ to $\mathcal{T}_{3}$. This empirically validates that spectral magnitude features provide inherent invariance to temporal noise, effectively decoupling recognition performance from precise synchronization.

\subsubsection{Results and Analysis.}
The comparative results in Tab.~\ref{tab:time_temporal_evaluation} reveal significant performance divergences under temporal noise. The ``w/o Spectral'' baseline deteriorates sharply on noisy sets, confirming the fragility of direct time-domain matching against misalignment. Although traditional pre-alignment methods (Linear, XCorr, DTW) mitigate simple global offsets ($T_1$), they fail to address the compounding drift and jitter in $T_3$. In contrast, \methods~ maintains high robustness across all noise levels with only marginal drops, empirically validating that spectral magnitude features provide inherent invariance to temporal noise, effectively decoupling recognition performance from precise synchronization.

% \textbf{Impact of Spectral Domain:} On $\mathcal{T}_{clean}$, the time-domain baseline achieves competitive performance. However, its accuracy degrades sharply on $\mathcal{T}_{1}$ and $\mathcal{T}_{2}$, confirming that direct time-domain matching is highly sensitive to even mild misalignments.
% \textbf{Failure of Pre-Alignment:} While XCorr and DTW improve performance on $\mathcal{T}_{1}$ compared to the raw baseline, they fail to recover accuracy on $\mathcal{T}_{3}$. This is likely because traditional alignment methods struggle with the multi-modal nature of the noise (e.g., drift and jitter distorting the signal shape).
% \textbf{Robustness of \methods:} In contrast, \methods~ maintains high accuracy across all noise levels, dropping only marginally from $\mathcal{T}_{clean}$ to $\mathcal{T}_{3}$. This empirically validates that spectral magnitude features provide inherent invariance to phase shifts and temporal scaling, effectively decoupling recognition performance from precise synchronization.

\begin{table*}
  \caption{\addl{\textbf{Evaluation Results on Temporal Misalignment.} We compare \methods~ against the time-domain ablation ("w/o Spectral"), pre-alignment methods (Linear, XCorr, DTW) applied to the time-domain backbone, and the VIPL baseline. Temporal misalignment increases progressively from $\mathcal{T}_{clean}$ to $\mathcal{T}_{3}$. Results demonstrate that \methods~ maintains high robustness even under severe synchronization noise ($\mathcal{T}_{3}$) where time-domain methods degrade sharply.}}
  \label{tab:time_temporal_evaluation}
  \begin{tabular}{c|cccc}
    \toprule
   \multirow{2}{*}{Methods}  & \multicolumn{4}{c}{Accuracy(\%) $\pm$ SD} \\
   \cline{2-5}
     & $T_{clean}$ & $T_1$ & $T_2$ & $T_3$ \\
    \midrule
    \methods & 97.01($\pm 0.26$)& 96.47($\pm 0.43$) & 95.70($\pm 0.95$) & 95.14($\pm 0.97$) \\
    \midrule
    w/o Spectral & 76.83($\pm 0.60$) & 65.19($\pm 0.64$) & 59.16($\pm 3.11$) & 50.38($\pm 6.82$) \\
    Linear \cite{furgale2013unified} & 78.18($\pm 1.17$) & 73.38($\pm 6.37$)  & 65.81($\pm 3.32$) & 51.82($\pm5.98$) \\
    XCorr \cite{knapp2003generalized} & 79.11($\pm 0.58$) & 74.95($\pm 6.79$)  & 63.82($\pm 3.36$) & 53.27($\pm5.91$) \\
    DTW \cite{berndt1994using} & 77.98($\pm 1.04$) & 74.91($\pm 1.32$)  & 72.09($\pm 1.62$) & 64.74($\pm4.32$)\\
    \midrule
    VIPL \cite{sun2020when} & 70.49($\pm 1.52$) &  65.89($\pm 2.23$)  & 65.06($\pm 1.66$) & 61.05($\pm2.70$)  \\
    \bottomrule
  \end{tabular}
\Description{Table comparing command source identification accuracy of HiSync and baselines under increasing temporal misalignment.
Across all conditions T_\text{clean} to T_3, HiSync achieves the highest accuracy, starting near 97\% and remaining above 95\% even at T_3. The ablation without spectral features and time-domain baselines (Linear, XCorr, DTW) begin around 77–79\% at T_\text{clean} and drop to roughly 50–65\% at T_3. The VIPL baseline performs worst overall, with accuracies consistently below the other methods.}
\end{table*}

\end{add}

% \section{User Study}

% \subsection{Settings}

% \subsection{Baselines}

% \subsection{Metrics}

% \subsection{Results and Analysis}

\begin{add}

\section{ONLINE EXPERIMENTS on REAL ROBOT}
\label{sec:real_robot}

To thoroughly evaluate the system's performance in real-world deployment and gather subjective user feedback, we deployed \methods~ on a quadruped robot and conducted both correctness tests and interactive usability experiments in dynamic indoor and outdoor environments.

\subsection{Participants and Apparatus}
\label{sec:real_robot:setup}
\begin{figure}[h]
  \centering
  \includegraphics[width=0.5\linewidth]{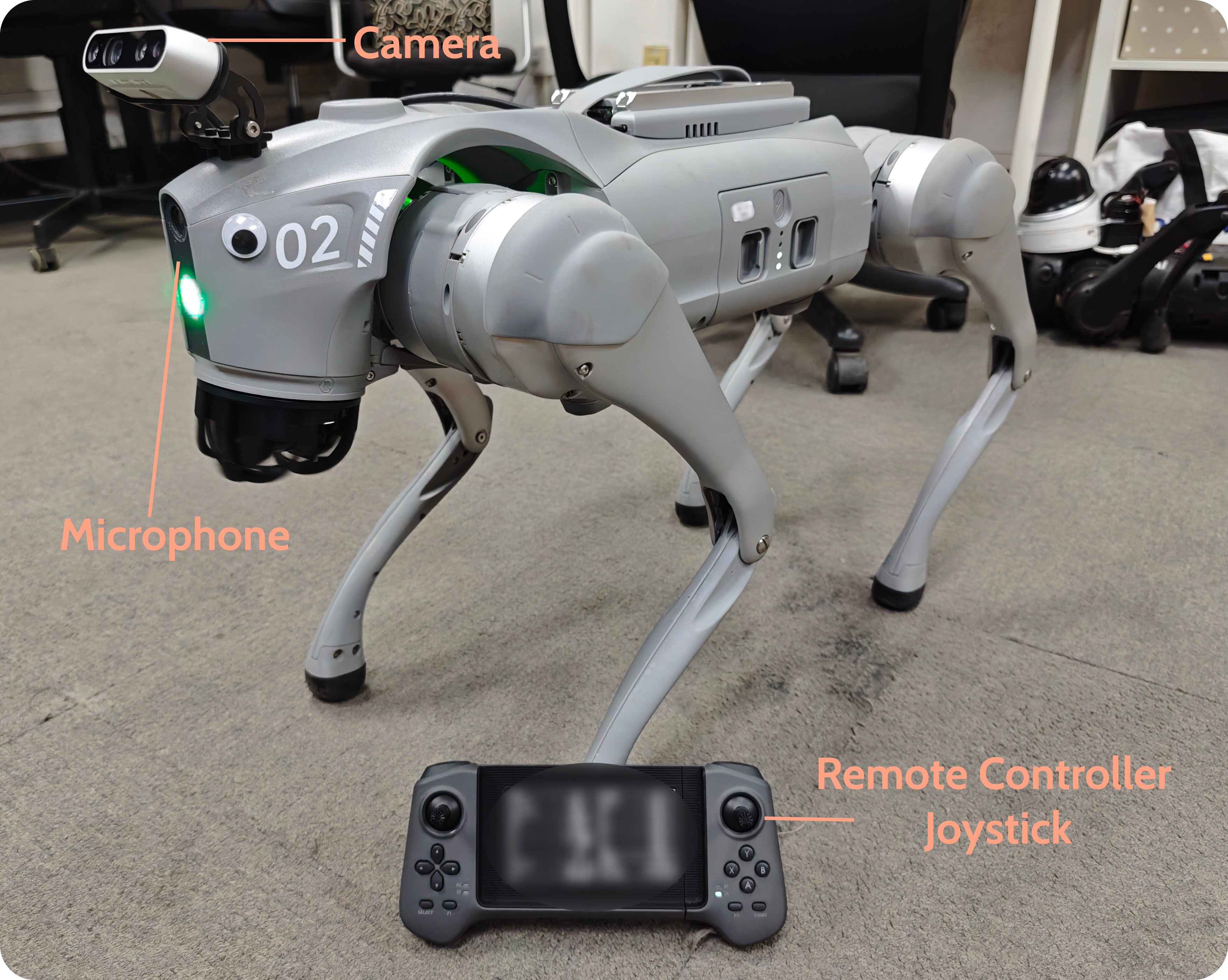}
  \caption{\addl{\textbf{Real Robot Apparatus.} The real-robot experiment is conducted on a \textit{Unitree Go2} quadruped robot, equipped with a \textit{Realsense D435} camera. The onboard microphone and official joystick are used in the \textit{Voice Control} and \textit{Remote Controller} baseline respectively.}}
  \label{fig:real_robot}
  \Description{Overview of the real-robot setup with Unitree Go2, camera, microphone, and handheld remote controller joystick. A Unitree Go2 quadruped robot stands indoors on a flat floor, facing slightly left. A depth camera is mounted at the front of the robot’s head, and a small microphone is positioned on the front side of the body. In front of the robot on the ground lies a handheld joystick-style remote controller. Text labels point to the camera, microphone, and remote controller joystick to highlight the hardware used for sensing and teleoperation baselines.}
\end{figure}

We recruited 12 participants ($N=12$, 6 male and 6 female; mean age = 23.9, $SD=3.7$) who had not participated in the formative study or dataset collection. Their expertise with robotics varied from novice to expert.

As shown in Fig.~\ref{fig:real_robot}, the system was deployed on a \textit{Unitree Go2} quadruped robot equipped with a \textit{Realsense D435} camera running in RGB mode at $1280\times720$ resolution (30\,fps). Based on the findings in Sec.~\ref{sec:discussion:position}, we utilized the wrist placement for the IMU sensor to balance signal quality and usability. IMU data was collected by a host computer and transmitted to the robot via the public network. This setup introduces realistic IOT network latency (measured at $\approx 387$,ms on average with $\approx 50$,ms jitter), strictly testing the system's robustness under non-ideal transmission conditions. Furthermore, unlike the batched inference used in offline evaluations, the real-robot system performs frame-by-frame inference, resulting in an effective processing rate of approximately 5,Hz due to computational overhead.

% \section{EXPERIMENTS on REAL ROBOT}

% To thoroughly evaluate 系统在实际使用中的效果以及获取用户主观反馈，we 部署我们的系统到了机器狗上，并进行了正确性测试和交互可用性实验。

% To thoroughly evaluate the system's performance in real-world usage and gather users' subjective feedback, we deployed our system on a quadruped robot and conducted correctness tests as well as interactive usability experiments.

% \subsection{Participants and Apparatus}

% 我们招募了没有参加过Formative study 和 dataset collection的 12 participants（6 male and 6 female, mean age=23.9 STD =3.7）用户对机器人的experitise various

% 实验采用unitree Go2 机器狗作为robot platform, the camera is Realsense D435 running at 1280x720@30FPS in RGB mode. IMU setting is the same as Sec.~\ref{sec:formative:apparatus}, 根据 Sec.~\ref{sec:discussion:position} 我们使用了 wrist IMU. The IMU sensor data is collected by a computer and  transmitted to the robot via the public network, 模拟真实的用wearable设备连接手机传输或者直接网络传输。

\subsection{Design and Procedure}

% The study consisted of two sessions: (1) an Interaction Usability Experiment and (2) a CSI Accuracy Experiment. Both sessions covered new indoor (3\,m, 5\,m) and outdoor (10\,m, 20\,m, 30\,m) scenarios not in the dataset.
The study consisted of two sessions: (1) an Interaction Usability Experiment and (2) a CSI Accuracy Experiment. Both sessions covered new indoor (3\,m, 5\,m) and outdoor (10\,m, 20\,m, 30\,m) scenarios that were not included in the dataset.

% 实验分为交互可用性实验和CSI正确率实验两个部分。实验涵盖了indoor 和outdoor的场景。对于用户和机器狗之间的距离，indoor包括 3m, 5m。outdoor包括10m, 20m, 30m

% \textbf{交互可用性实验}
\subsubsection{Interaction Usability Experiment.}
% In this experiment, the participant分别在不同的距离下，用遥控器和HiSync两种方式操纵机器狗到跟前。用户被告知使用右手做自行表示summon语义即可触发 \methods . 实验组和距离的顺序在用户间 counterbalance。在3m的距离下，还使用了 unitree Go2 自带的语音交互功能进行对比，用户可以说出唤醒词并让机器狗向前走到跟前，在更远的距离上用户都拒绝使用这种方式。

To assess interaction usability, we compared \methods~ against standard modalities in a robot summoning task across varying distances. (i) \textbf{Remote Controller (RC)}, utilizing the manufacturer's standard joystick; (ii) \textbf{Voice Control}, employing the robot's built-in wake-up word and command interface; and (iii) \textbf{\methods}, where participants triggered the approach via a natural right-hand wave mapped to the ``Summon'' semantic. Notably, Voice Control was exclusively evaluated at the short range (3\,m), as participants consistently declined to use voice at longer distances ($>$5\,m) due to the social embarrassment associated with shouting and the system's technical inability to capture distant audio. To minimize learning effects, modality and distance order were counterbalanced across participants.

% HiSync用summon的语义触发，用户自行决定采用什么手势

% \textbf{CSI正确率实验}
% 两个用户一组，其中一名用户作为机器狗的owner，自然的发出 Sec 中所示的六种语义动作。另一名用户作为strong negtive example，被告知尽可能的完全模仿owner的动作，越像越有可能将机器狗骗到自己这里来（详细指令见appendix）。同时作为实际真实场景中的非受控实验，实验中均有随机出现的路人，会产生部分遮挡和干扰，作为weak negtive ，

\subsubsection{CSI Accuracy Experiment.}
To evaluate robustness against active interference, participants worked in pairs at an \textbf{adversarial setting}. One participant acted as the \textbf{Owner}, issuing the five standard commands defined in Sec.~\ref{sec:formative:gesture}. The other acted as a \textbf{Strong Negative Sample (Mimic)}, instructed to mimic the owner's movements as closely as possible to ``trick'' the robot into responding to them. Additionally, the experiments were conducted in public spaces with uncontrolled passersby, who served as \textbf{Weak Negative Samples} (providing occlusion and background noise). 
% This setup creates a challenging adversarial scenario beyond standard datasets.

\subsection{Results and Analysis}

% 记得算P值

\subsubsection{Objective}

% As shown in Tab.~\ref{tab:csi_combined_performance}, we measured CSI accuracy and the response time from gesture onset to robot reaction. Under real robot settings and realistic strong and weak negative samples, \methods~ still remains high accuracy and high efficiency. Even at the distance of 30\,m, \methods~ still has the accuracy of 90.9\% and react in 3.88\,s.
As shown in Tab.~\ref{tab:csi_combined_performance}, we evaluated the CSI accuracy and the response time (from gesture onset to robot reaction). Despite the challenging real-world settings involving adversarial mimics (strong negatives) and uncontrolled pedestrians (weak negatives), \methods~ maintains high accuracy and operational efficiency. Notably, even at the extended distance of 30\,m, \methods~ achieves an accuracy of 90.9\% with an average response time of 3.88\,s, demonstrating its viability for long-range field deployment. These results demonstrate that \methods~ remains effective when deployed on a real quadruped robot and with strong bystander interference. Compared to the training dataset, the real-robot study involves new participants, a different camera configuration, and previously unseen indoor and outdoor environments, suggesting that \methods~ can generalize from curated offline data to in-the-wild deployments.

\subsubsection{Subjective}
We evaluated subjective feedback using the NASA-TLX \cite{hart1988development} and a subset of the System Usability Scale (SUS) \cite{brooke1996sus}. A Wilcoxon Signed-Rank Test was conducted to determine statistical significance between \methods~ and the baselines (Voice and Remote Controller), as shown in Figure~\ref{fig:sus}. Note that specific NASA-TLX metrics were inverted for consistency, so higher scores always indicate better performance.

Quantitatively, \methods~ achieved the highest mean scores across all metrics. Participants reported a significantly higher willingness to use \methods~ compared to both Voice ($p < 0.00049$) and the Remote Controller (RC) ($p = 0.034$). \methods~ was significantly easier to learn ($p < 0.0004 9$) and use ($p< 0.005$) than the RC, effectively combining the intuitiveness of natural voice interaction ($n.s.$ in ease of use) with the reliable performance of the RC ($n.s.$ in performance). Regarding task load (TLX), \methods~ significantly outperformed both baselines across all sub-scales (all $p < 0.03$). Deeper analysis revealed that interaction distance influenced these preferences (Fig.~\ref{fig:sus_dis}: at close range, participants felt the RC was ``overkill'' (P4, P7). Conversely, as the distance increased, distinct issues emerged: participants noted they ``could not see the robot dog clearly'' (P4) and that ``control errors with the remote controller increased'' (P9). Moreover, specifically for users unfamiliar with robots, the workload of operating at long distances increased significantly ($p < 0.05$). 

Participants praised the system's \textbf{naturalness} and \textbf{social presence}, noting it ``looks like a real dog'' (P5, P8, P11) and is ``suitable for public spaces'' (P7, P12), with emphasis that ``It was excellent, and I felt very confident'' (P4, P12).

\aptLtoX{\begin{table*}
\centering
\caption{\textbf{Quantitative Results of Real-World Experiments.} We report the mean CSI accuracy and response time (from gesture onset to robot reaction) under adversarial settings with strong mimics. Data are presented as Mean $\pm$ Standard Error (SE).} 
\label{tab:csi_combined_performance} 
\begin{tabular}{c|cc}
\multicolumn{3}{c}{\textbf{(a) Performance vs. Interaction Distance. This table illustrates the effectiveness of \methods~ under different real-world distance conditions.}} \\
\hline
Distance(m) & CSI Accuracy (\%) $\pm SE$ & Time(s) $\pm SE$ \\
 % 替换 \hline 更美观，与booktabs风格一致
3           & $96.4 \pm 2.55$            & $3.46 \pm 0.159$ \\
5           & $94.5 \pm 4.03$            & $3.23 \pm 0.130$ \\
10          & $92.7 \pm 3.53$            & $3.07 \pm 0.150$ \\
20          & $92.7 \pm 3.53$            & $3.09 \pm 0.130$ \\
30          & $90.9 \pm 3.91$            & $3.88 \pm 0.175$ \\
\hline
\end{tabular}

\begin{tabular}{c|cc}
\multicolumn{3}{c}{\textbf{(b) Performance vs. Gesture Category. This table demonstrates the generalization of the proposed gestures under \methods.}}\\
\hline
Gesture   & CSI Accuracy (\%) $\pm SE$ & Time(s) $\pm SE$ \\
\hline
Right     & $92.7 \pm 3.534$           & $3.12 \pm 0.132$ \\
Left      & $96.4 \pm 2.547$           & $3.47 \pm 0.175$ \\
Approach  & $90.9 \pm 3.912$           & $3.33 \pm 0.157$ \\
Retreat   & $90.9 \pm 3.912$           & $3.43 \pm 0.161$ \\
Summon    & $96.4 \pm 2.547$           & $3.38 \pm 0.176$ \\
\hline
\end{tabular}
\end{table*}}{\begin{table*}
\centering
\caption{\addl{\textbf{Quantitative Results of Real-World Experiments.} We report the mean CSI accuracy and response time (from gesture onset to robot reaction) under adversarial settings with strong mimics. Data are presented as Mean $\pm$ Standard Error (SE).}}  % 总标题（可根据需求修改或删除）
\label{tab:csi_combined_performance}  % 总表格引用标签
% 第一个子表格（距离相关）
\begin{subtable}[t]{0.48\linewidth}  % 宽度占页面45%，t表示顶部对齐
\centering
\caption{\addl{\textbf{Performance vs. Interaction Distance.} This table illustrates the effectiveness of \methods~ under different real-world distance conditions.}}  % 子表格标题
\label{tab:csi_distance}    % 子表格引用标签
\begin{tabular}{c|cc}
\toprule
Distance(m) & CSI Accuracy (\%) $\pm SE$ & Time(s) $\pm SE$ \\
\midrule  % 替换 \hline 更美观，与booktabs风格一致
3           & $96.4 \pm 2.55$            & $3.46 \pm 0.159$ \\
5           & $94.5 \pm 4.03$            & $3.23 \pm 0.130$ \\
10          & $92.7 \pm 3.53$            & $3.07 \pm 0.150$ \\
20          & $92.7 \pm 3.53$            & $3.09 \pm 0.130$ \\
30          & $90.9 \pm 3.91$            & $3.88 \pm 0.175$ \\
\bottomrule
\end{tabular}
\end{subtable}
\hfill  % 两个子表格之间的水平间距
% 第二个子表格（手势相关）
\begin{subtable}[t]{0.45\linewidth}  % 宽度与第一个一致，确保并排对齐
\centering
\caption{\addl{\textbf{Performance vs. Gesture Category.} This table demonstrates the generalization of the proposed gestures under \methods.}}  % 子表格标题
\label{tab:csi_gesture}     % 子表格引用标签
\begin{tabular}{c|cc}
\toprule
Gesture   & CSI Accuracy (\%) $\pm SE$ & Time(s) $\pm SE$ \\
\midrule
Right     & $92.7 \pm 3.534$           & $3.12 \pm 0.132$ \\
Left      & $96.4 \pm 2.547$           & $3.47 \pm 0.175$ \\
Approach  & $90.9 \pm 3.912$           & $3.33 \pm 0.157$ \\
Retreat   & $90.9 \pm 3.912$           & $3.43 \pm 0.161$ \\
Summon    & $96.4 \pm 2.547$           & $3.38 \pm 0.176$ \\
\bottomrule
\end{tabular}
\end{subtable}
\Description{Real-world command source identification accuracy and response time across distance and gesture. Across interaction distances from 3 to 30 meters, Command Source Identification accuracy stays high (about 91–96 percent) with only a slight drop at 30 meters, while response time remains around 3.1–3.9 seconds. Across gesture categories, accuracy is similarly stable (about 91–96 percent), with Left and Summon slightly higher than Right, Approach, and Retreat, and response times again clustered near 3.2–3.5 seconds, indicating robust and consistent performance over distance and gesture type.}
\end{table*}}

\begin{figure*}[h]
  \centering
  \includegraphics[width=0.9\linewidth]{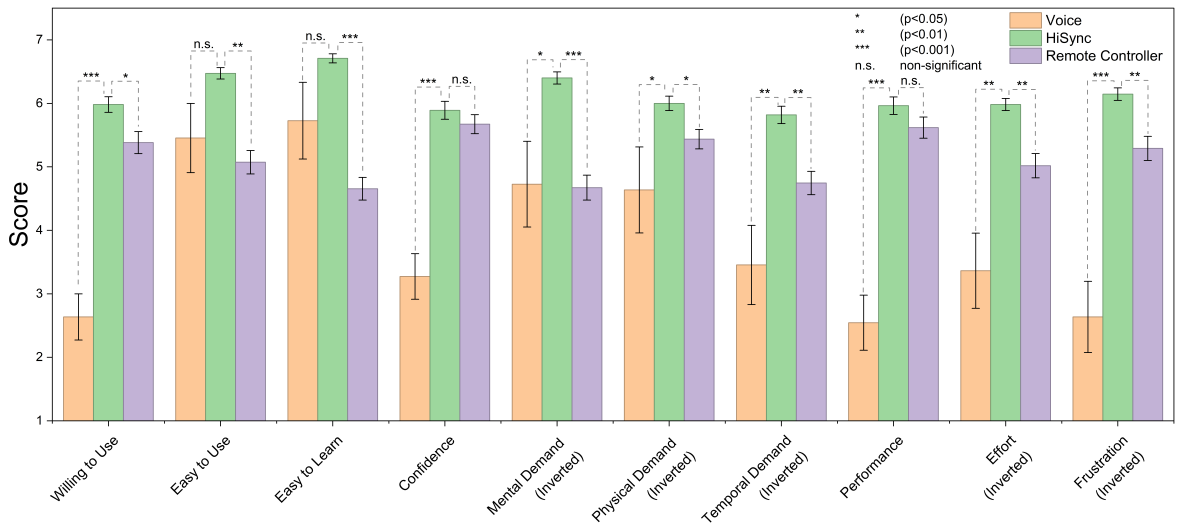}
  \caption{\addl{\textbf{Objective Results of Real Robot Experiment.} Mean subjective scores with standard error for the three interaction methods. Statistical significance is marked based on the Wilcoxon Signed-Rank Test. Note: Scores for NASA-TLX subscales are inverted, meaning higher values consistently represent a better user experience across all indicators. Overall, \methods~ consistently outperforms both baselines across all metrics, demonstrating superior usability and significantly reduced user workload.}}
  % \Description{\todo{画一条33\%的参考线}}
  \label{fig:sus}
  \Description{Aggregated subjective user scores for HiSync and a remote controller across distance. A line chart plots score versus distance from 3 to 30 meters, with HiSync as a higher curve above the remote controller at all distances. Both curves gently decline as distance increases. Each line is surrounded by a shaded band indicating standard error, showing that HiSync’s advantage is consistent over distance.}
\end{figure*}

\begin{figure}[h]
  \centering
  \includegraphics[width=0.8\linewidth]{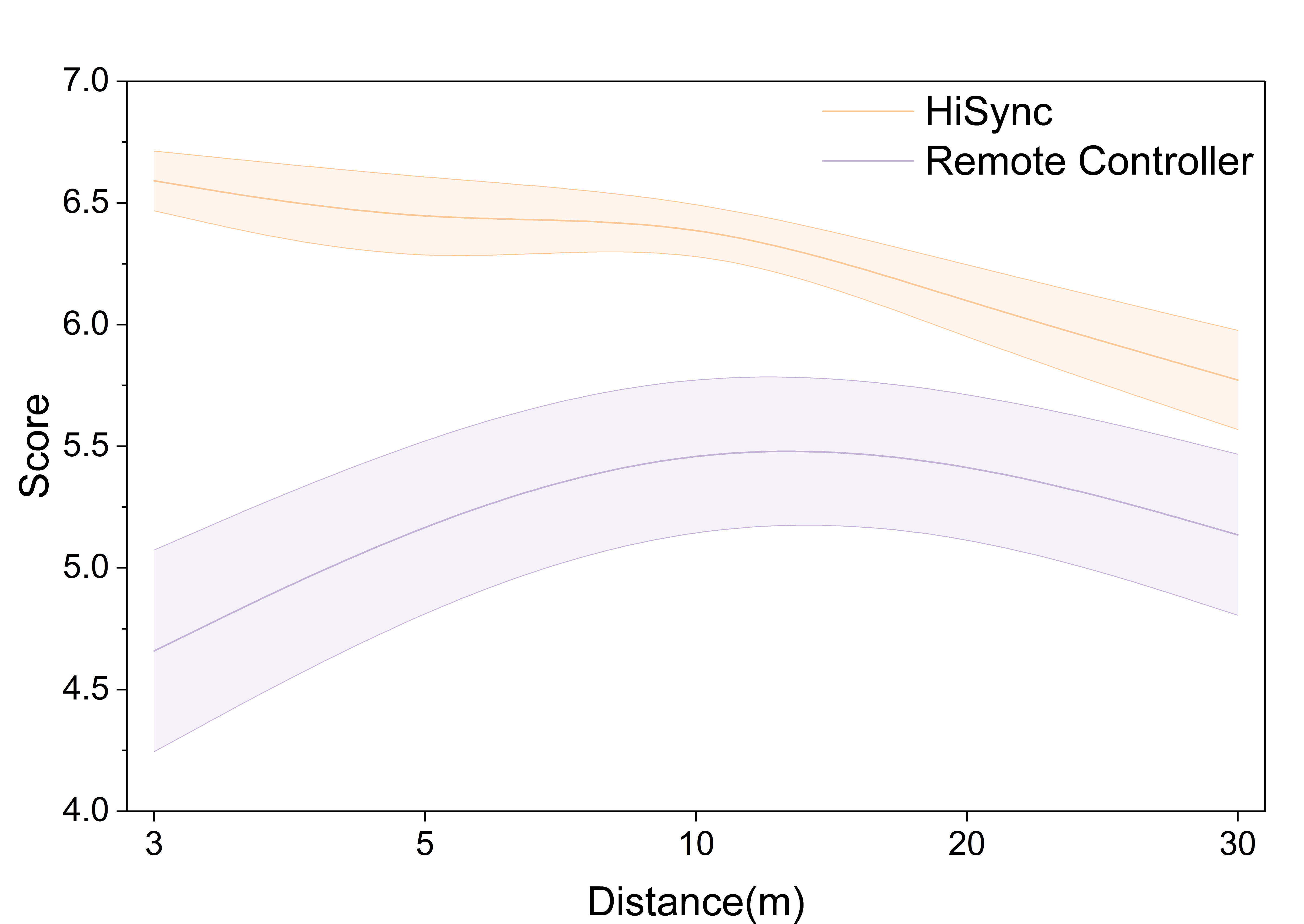}
  \caption{\addl{\textbf{Aggregated subjective scores as a function of distance.} This figure illustrates the average of SUS and NASA-TLX results. The higher the score, the better the system. \methods~ maintains a significant ($p < 0.0007$) advantage over the Remote Controller. Shaded areas denote standard error.}}
  % \Description{\todo{画一条33\%的参考线}}
  \label{fig:sus_dis}
  \Description{Subjective user ratings for three interaction methods in the real robot experiment. A grouped bar chart compares Voice, HiSync, and Remote Controller across nine metrics (Willing to Use, Easy to Use, Easy to Learn, Confidence, Mental Demand, Physical Demand, Temporal Demand, Performance, Effort, and Frustration). HiSync bars are consistently the tallest on all metrics, indicating better usability and lower workload than the two baselines. Voice scores lowest overall, while the Remote Controller lies between Voice and HiSync. Error bars show standard error, and significance markers above the groups indicate that many of HiSync’s improvements over the baselines are statistically significant.}
\end{figure}

\end{add}

% 无法解决遮挡的问题
\section{APPLICATIONS}

\begin{figure}[h]
    \centering
    \includegraphics[width=\linewidth]{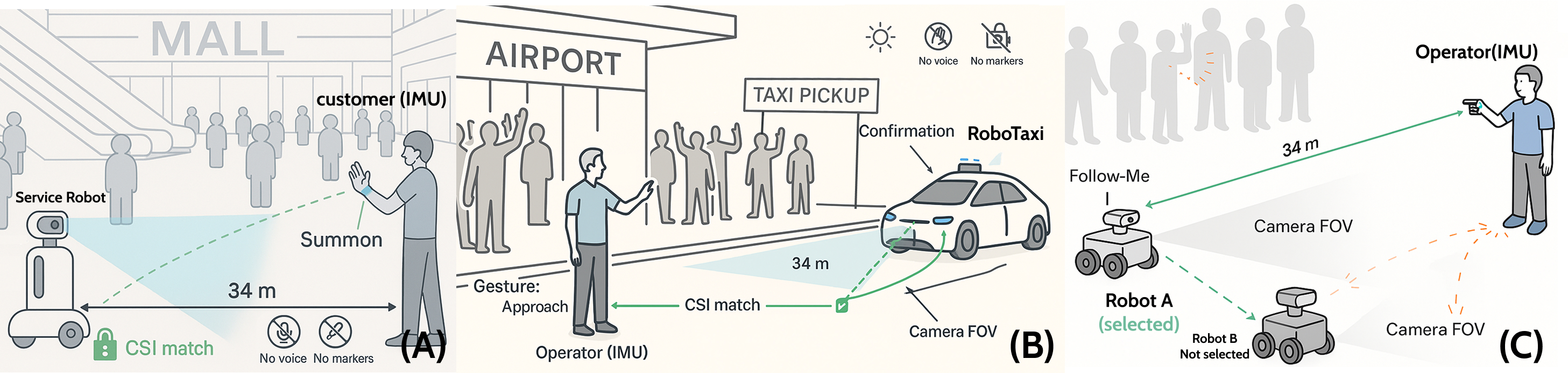}
    \caption{\textbf{Applications.} (A) Service-robot interactions at a mall. (B) Summoning a robotaxi. (C) ``Hi and Sync'' paradigm.}
    \label{fig:app}
    \Description{This figure illustrates various real-world applications of the system. (A) Service-robot interactions at a mall, where the customer uses a gesture to summon a service robot from 34 meters away. The CSI (Command Source Identification) system matches the customer’s gesture to the robot. (B) Summoning a RoboTaxi at a taxi pickup area using a gesture ("Approach"), with the system matching the customer’s gesture and confirming the selection. (C) The "Hi and Sync" paradigm, where an operator uses a gesture to select Robot A (from multiple robots), demonstrating the system's ability to identify and engage specific robots based on the operator’s movement.}
\end{figure}

% \subsection{Public-venue service-robot interactions}
% In public indoor settings such as airports, train stations, and shopping malls, our CSI pipeline enables users to summon and direct service robots from distances using brief hand gestures (e.g., \textit{Summon}, \textit{Approach}, \textit{Left}, \textit{Right}). The method jointly aligns far-field optical-flow spectra from robot-mounted cameras with IMU signals worn by the user, allowing the system to identify which person issued a command in crowded scenes without explicit per-user enrollment, fiducial markers, or spoken input. As a result, the approach preserves socially appropriate interaction in reverberant or quiet environments and reduces the need for time-consuming pairing procedures that can lead to incorrect device bindings in the presence of bystanders.

\subsection{Public-venue service-robot interactions}
In public indoor settings such as airports, train stations, and shopping malls, our CSI pipeline enables users to summon and direct service robots from a distance using brief hand gestures (for example, \textit{Summon}, \textit{Approach}, \textit{Left}, \textit{Right}). \emph{Imagine entering a shopping mall, making a short wrist wave with a smartwatch and within moments a nearby service robot recognizes the gesture, navigates to the user for further instruction}. Technically, the system aligns far-field optical-flow spectra captured by robot-mounted cameras with IMU traces from the user's wearable, which lets the pipeline identify who issued a command in crowded scenes without explicit per-user enrollment, fiducial markers, or spoken input. This maintains socially appropriate interaction in reverberant or quiet environments.

\subsection{Robotaxi summons}
In outdoor parking lots and similar open environments, a user can invoke a \textit{summon$\rightarrow$approach} gesture from tens of meters away to request a vehicle or low-speed carrier. Operating in the frequency domain, cross-modal spectral matching is inherently more tolerant to long-range visual degradation and variable illumination. The proposed setup that requires only a monocular camera on the vehicle and a wearable IMU minimizes infrastructure requirements while still supporting explicit confirmation cues and escalation procedures when the initial signal is ambiguous.

% \subsection{Robotaxi summons}
% In outdoor parking lots and other open environments, a user can issue a \textit{summon$\rightarrow$approach} gesture from tens of meters away to request a vehicle or low-speed carrier at a predefined pickup location. \emph{Imagine stepping out of an airport, giving a brief wave with your phone, and a robotaxi pulls up directly in front of you, eliminating any walk after a tiring trip}. Operating in the frequency domain, cross-modal spectral matching is inherently more tolerant to long-range visual degradation and variable illumination. The proposed setup that requiries only a monocular camera on the vehicle and a wearable IMU minimizes infrastructure requirements while still supporting explicit confirmation cues.

\subsection{``Hi and Sync'' Paradigm}
CSI can serve as a lightweight selection mechanism to bind an individual to a specific robot and subsequently enable behaviors such as \textit{Follow-Me} and \textit{Approach \& Align}. The same spectral alignment that disambiguates commands in multi-user scenes also reduces interference from nearby bystanders, supporting reliable selection, re-binding, and handover among co-located robots, making far-range ``Hi and Sync'' a feasible primitive for applications such as guided tours, patrols, and parcel delivery.

\section{DISCUSSION}

% \subsection{Potential Applications for Future HCI}

\subsection{Choice of IMU position}
\label{sec:discussion:position}

% 如图\todo{ref}所示，我们采用了三种IMU位置， 手腕、手掌、手指，分别模拟了智能手表，手持手机，智能戒指三种常见的IMU设备。而在适用性方面，智能手表作为现在最广泛的穿戴设备，智能戒指是一种新型的未来的佩戴最无感的设备。而如果手持手机就不需要额外的设备。针对不同设备的实验结果如Fig. \ef{acc_position}所示
% 手掌的效果最好，戒指可能噪声过大，而手腕运动信息相对过少。

As illustrated in Fig.~\ref{fig:dataset}, we compared three placements on the dominant hand. Beyond sensing fidelity, these choices reflect applicability in practice: smartwatches are the most widely adopted wearables, smart rings promise the least obtrusive ``always-on'' wear in the near future, and hand-held phones require no additional hardware.

\begin{figure}[h]
    \centering
    \includegraphics[width=0.7\linewidth]{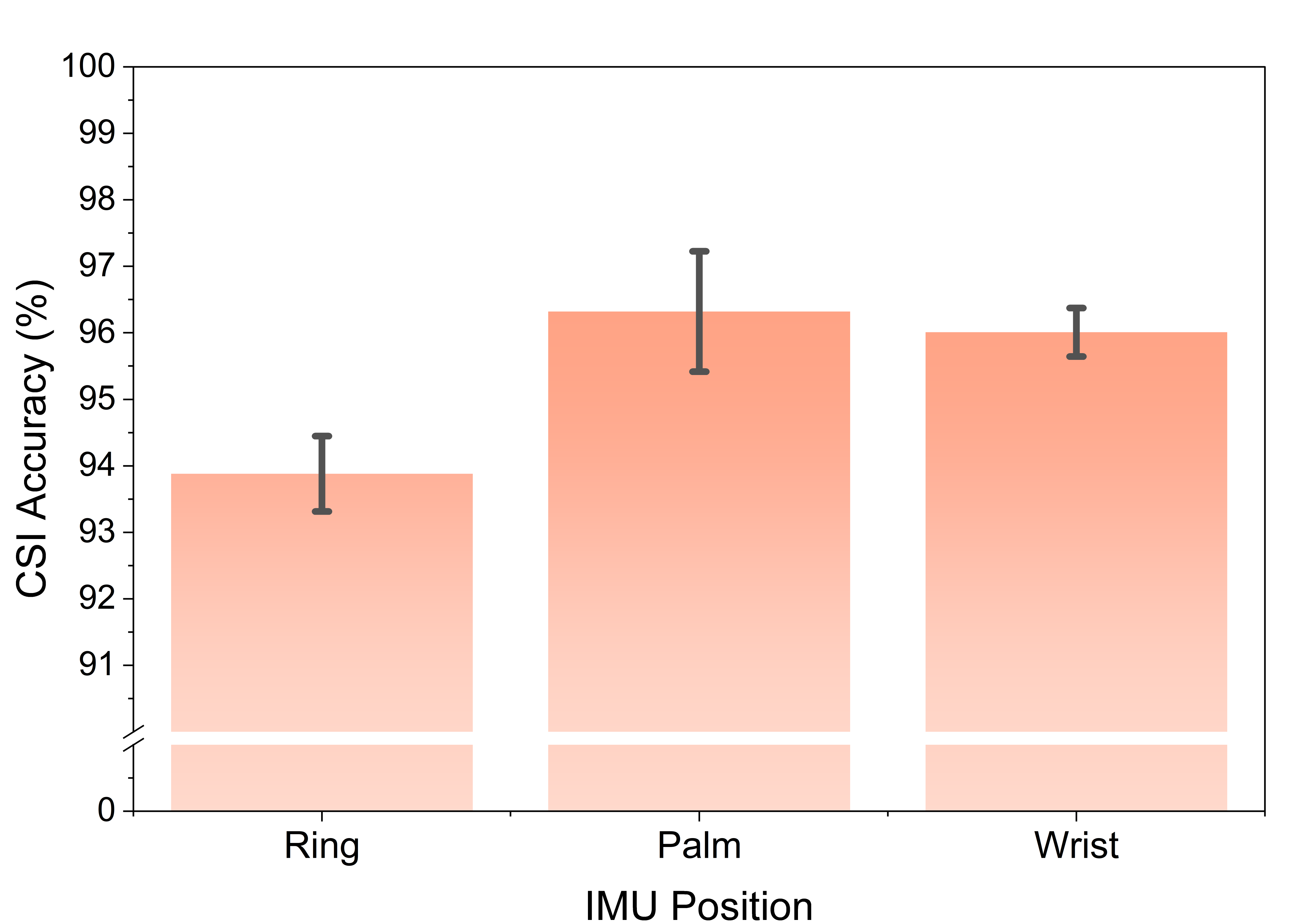}
    \caption{\textbf{CSI Accuracy vs. IMU Position.} \addl{While the Palm and Wrist provide slightly stronger motion cues than the Ring, the system achieves high accuracy ($>93\%$), validating the system's usability across diverse wearable form factors. Error bar denotes standard deviation.}}
    \label{fig:acc_position}
    \Description{This figure shows the Command Source Identification (CSI) accuracy for different IMU positions: Ring, Palm, and Wrist. The bar chart illustrates the accuracy percentages, with the Palm position yielding the highest CSI accuracy (~98\%) and the Ring and Wrist positions showing slightly lower but similar results. }
\end{figure}

Performance results (Fig.~\ref{fig:acc_position}) show that the palm IMU achieves the highest CSI accuracy, likely due to stronger distal motion and firmer sensor-gesture coupling that preserves high-frequency components exploited by our spectral encoder. The ring underperforms, plausibly due to higher micro-motion noise, while the relatively poor performance for wrist-mounted IMU may be attributed to the wrist's less discriminative motion patterns, as well as noise introduced by forearm movements.

 %proximal presenting less discriminative motion and mixed with forearm dynamics.

\textbf{Design Implications}. When robustness at long range is paramount, prefer palm-held. Smartwatches trade some accuracy for ubiquity and convenience, and rings favor wearability but require tighter stabilization and higher-quality sensing; phones offer strong accuracy without extra devices when hands are occupied.

% \subsection{统一的测试指标}
% 我们也找到了一些有相似模态的数据集例如，\cite{sun2020when}. 这让我们意识到，不同的摄像头参数会对CSI任务的正确性有很大的影响。我们认为需要一个摄像头无关的正确率来综合评估方法在不同距离下的感知能力。 所以定义
% 我们方法的正确率这这种setting下是

% 摄像头极限和统一测试指标的建议
% \subsection{极限距离}
% 我们将我们方法的极限距离设置在了34m，是因为在这个距离下人已经看不清，由于在画面中占比过小且像素过低VideoFlow不能正常工作，如Fig. \todo{} 所示。更好的光流估计模型可能可以提升这个距离，也可能可以先进行超分辨率处理然后在用。不过我们认为在更远的距离下用户可能也不能很好的捕捉到交互对象，经过简单的用户访谈(N=9)，大家一致认为在这个34m的距离下，日常状态下不会有交互意愿。还有相关的HCI理论\cite{ballendat2010proxemic,vogel2004interactive}也支持了这一点。
\subsection{Maximum Interaction Range}
\begin{figure}
    \centering
    \includegraphics[width=0.9\linewidth]{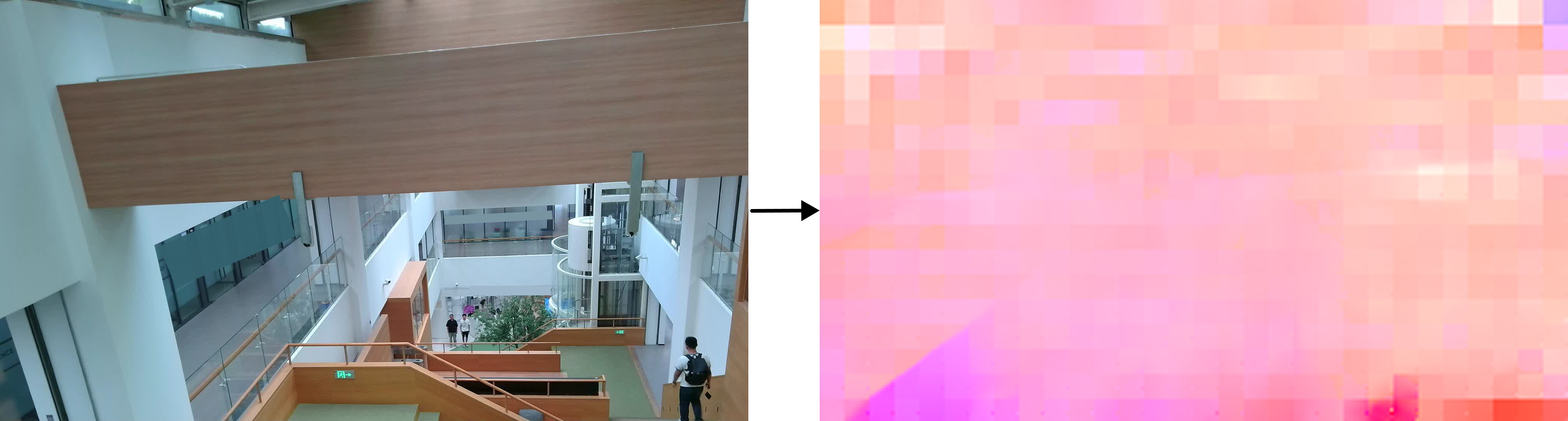}
    \caption{\textbf{Illustration of VideoFlow Failure Case.} \addl{At this extreme distance, human movements have become so minute that they are obscured by noise.}}
    \label{fig:flow}
    \Description{This figure demonstrates a failure scenario in the VideoFlow. The left image shows a normal scene captured from a camera, while the right image illustrates the result after processing, where the output is heavily pixelated.}
\end{figure}
We set the maximum interaction range to 34 m. At this distance, a person occupies too few pixels for reliable motion estimation; consequently, our optical-flow pipeline (VideoFlow) no longer produces stable hand-motion cues (Fig.~\ref{fig:flow}). Although a stronger flow estimator or a super-resolution pre-processing step could, in principle, extend this range slightly, we deliberately retain 34 m as a conservative upper bound for two reasons. First, beyond this range, users themselves find it difficult to visually acquire and keep the intended interlocutor in view. Second, in a brief interview study (N=9), participants consistently reported that, under everyday conditions, they would not attempt to initiate interaction at or beyond 34 m. This observation aligns with proxemics-based HCI research: engagement willingness drops sharply in the far public zone and typically requires additional signifiers or explicit invitation \cite{ballendat2010proxemic,vogel2004interactive}.

\subsection{Amplitude over Time}
\label{sec:formative:quant:time}
% 尽管有的用户会在非常远的距离下使用非常subtle的手势。但是我们发现，如果机器人无响应，all participants will consistently amplified their gestures, presumably to increase recognizability. 这给HRI系统启示了一种graceful escalation的方式，低置信度时反馈给到用户，提示他们做出更大的动作

\begin{figure}[h]
    \centering 
    \includegraphics[width=0.7\linewidth]{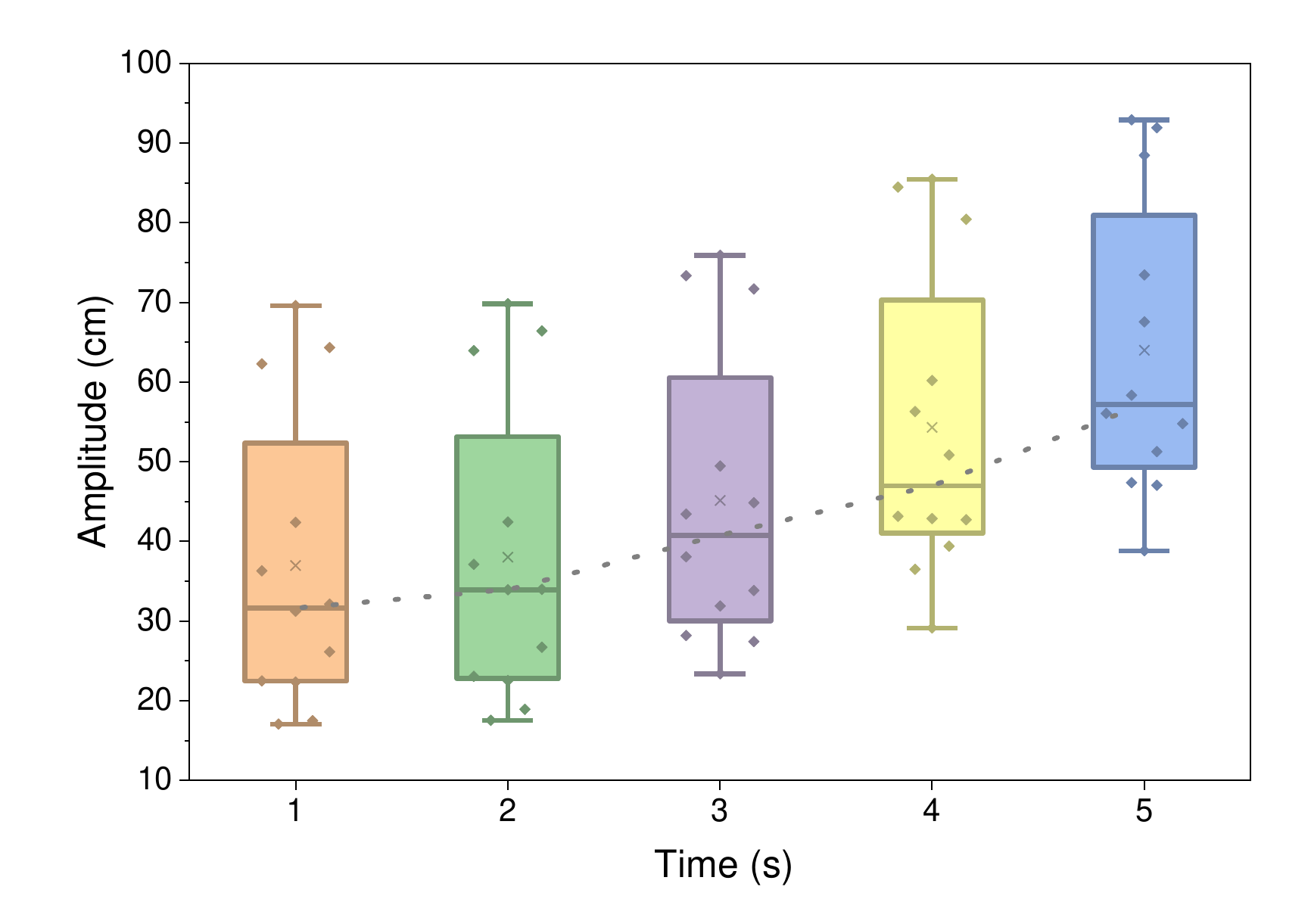}
    \caption{\textbf{Gesture Amplitude vs. Time.} \addl{Under non-response conditions, participants gradually amplify their gesture amplitude, suggesting a natural "graceful escalation" strategy for interaction recovery. Error bar shows standard deviation.}}
    \label{fig:amp_time}
    \Description{This figure shows a boxplot of gesture amplitude (in centimeters) over time (1-5 seconds). The x-axis represents time, while the y-axis shows the amplitude, with the boxes displaying the interquartile range, and the lines indicating the range of values. The dotted line illustrates the general trend, showing an increase in amplitude with time. Each time point (1 to 5 seconds) shows a distinct distribution of gesture amplitudes, suggesting that gesture strength varies with duration.}
\end{figure}

Despite some participants using low-amplitude gestures even at very long ranges, we observed a reliable self-adaptation: when the robot did not respond, all participants amplified their gestures as in Fig.~\ref{fig:amp_time}. This points to a confidence-contingent, graceful-escalation strategy for HRI systems: \emph{when recognition confidence drops below a calibrated threshold, the system should deliver cues (e.g., a brief LED blink or ring vibration) that explicitly invites the user to repeat the command with a larger or more distinctive motion}. Such graded feedback preserves natural interaction in the default case while reducing misrecognition at long range, without requiring users to learn a new gesture vocabulary or make burdensome movements unless needed.

\begin{add}
\subsection{Practical Deployment and Efficiency}
While our work focuses on the specific challenge of Command Source Identification (CSI), real-world integration also requires the ability of gesture classification, which is already a mature field for single IMU systems~\cite{shen2024mousering,vatavu2018q,dahiya2024efficient}. Although \methods~ processes continuous streaming by matching motion energy correlation independent of specific gesture classes, running the full visual-inertial fusion continuously is computationally redundant for idle periods. We therefore advocate for a cascaded processing pipeline where \methods~ operates as a downstream binding module, triggered only after a valid gesture candidate is detected. In this setup, the wearable IMU operates in a low-power mode using acceleration thresholds or lightweight on-device classifiers to filter background noise and establish communicative intent. This strategy leverages the maturity of IMU-based recognition for intent detection while reserving HiSync’s spectral capabilities for the distinct task of resolving the command source in multi-user environments, thereby balancing responsiveness with operational duration.
% To address the distinction between detecting the issuance and of a command and identifying its source, we clarify that HiSync is designed as a modular component specifically for the latter—Command Source Identification (CSI)—operating independently of the upstream gesture detection task. While gesture classification using IMUs is a mature field, HiSync's spectral alignment mechanism inherently supports continuous streaming by matching motion energy correlation rather than specific gesture classes, although robustness to unconstrained non-gesture motion remains a direction for future work. In practical deployments, to balance energy efficiency and intent understanding, we advocate for a cascaded wake-up strategy: a low-power IMU thread first monitors acceleration thresholds to detect activity onset and validates the command intent via a lightweight classifier; only upon confirming a purposeful gesture does the system trigger the computationally more intensive HiSync module to process the buffered window, thereby ensuring the robot responds only to directed commands while minimizing power consumption during idle periods.
\end{add}
% While our work focuses on the specific challenge of Command Source Identification (CSI), effective real-world integration necessitates coupling this with gesture classification—a mature field for single-IMU systems~\cite{shen2024mousering,vatavu2018q,dahiya2024efficient}. Although \methods~ inherently supports continuous streaming by matching motion energy correlation independent of specific gesture classes, running the full visual-inertial fusion continuously is computationally inefficient for idle periods. We therefore advocate for a cascaded processing pipeline where \methods~ operates as a downstream binding module, triggered only after a valid gesture candidate is detected. In this setup, the wearable IMU operates in a low-power mode using acceleration thresholds or lightweight on-device classifiers to filter background noise and establish communicative intent. This strategy leverages the maturity of IMU-based recognition for intent detection while reserving \methods~'s spectral capabilities for the distinct task of resolving the command source in multi-user environments, thereby balancing responsiveness with operational duration.

\begin{fadd}
\subsection{The Necessity of Visual Binding}
% Human-robot interaction in long-range scenarios relies heavily on localization.
% A command such as ``Come here'' is semantically incomplete without the user's location. Although the bound IMU link provides the intent, it lacks the spatial coordinates required for navigation or gaze orientation. 
% Therefore, the visual-inertial binding capability afforded by HiSync enables the robot to correctly respond to distant inertial commands.
Human-robot interaction in long-range scenarios relies heavily on localizing the user who issues a command. Commands such as ``Come here'' are semantically incomplete without knowing the user’s location. In our setting, users convey intent through IMU-based gestures, which do not encode positional information, while user location is captured separately by the robot’s onboard cameras. This separation introduces ambiguity in identifying the command source when multiple visual distractors are present. HiSync addresses this challenge by reframing user localization as an intent grounding problem across heterogeneous sensing modalities. By synchronizing motion-based intent signals with visual observations, HiSync enables the robot to disambiguate command sources and attribute commands to a specific user in the environment.

% For embodied agents, identification implies localization. 

% HiSync serves as the bridge that grounds the floating IMU signal to a physical entity in the visual scene. Therefore, even in our evaluated setup of a single active IMU, the system provides indispensable value by converting a digital signal into a spatial target.

% However, HiSync still tackles the critical challenge of visual-inertial binding. Without binding visual input to inertial data, the robot would receive a command issued as an inertial signal without knowing where the command came from. Therefore, the visual-inertial binding capability afforded by HiSync enables the robot to correctly respond to distant inertial commands.

\end{fadd}

\section{LIMITATIONS and FUTURE WORK}

\textbf{Limitations.} Our study was conducted in semi-controlled corridors and atriums with two concurrent participants and only occasional passersby; the resulting low crowd density (about 3 people per frame) falls short of the congestion and occlusion found in malls and airports, limiting ecological validity. The participant pool under-represents older adults and children, constraining generalizability across age groups. Moreover, the current pipeline is sensitive to robot ego-motion and often benefits from the platform pausing before identification, which may restrict responsiveness and practical applicability. \faddl{Finally, we assumed a setup where only the target user transmits inertial data to the robot. While this setup allows us to validate the system's ability to spatially localize a \emph{known command source}, it does not cover multi-user scenarios where multiple IMU-equipped users simultaneously signal to a shared robot (e.g., in public spaces where a communal robot is controlled by nearby users). Although one could, in principle, run the HiSync procedure independently for each IMU stream, we have not yet validated this extension experimentally.
% but it does not cover multi-user scenarios where multiple IMU-equipped users simultaneously signal to a shared robot (e.g., in public spaces where a communal robot is controlled by nearby users). Although one could, in principle, run the visual-inertial binding procedure independently for each IMU stream, we have not yet validated this extension experimentally, and issues such as identity disambiguation and interference between concurrent signals remain open.
}
% it limits our assessment of signal separation. Specifically, it does not account for potential spectral collisions or interference that might arise in scenarios where multiple active IMU streams compete for the robot's attention.
% }

% Consequently, our current evaluation does not account for potential spectral collisions or interference that might arise in scenarios where multiple active IMU streams compete for the robot's attention.}

% While this setup allows us to validate the system's ability to spatially localize a known signal source, it limits our assessment of signal separation. Specifically, it does not capture the challenges of 'spectral collisions' that could occur if multiple bystanders were simultaneously transmitting inertial commands.

\textbf{Future Work.} Next steps include leveraging on-board multi-sensor fusion (e.g., robot IMU/odometry/visual-inertial SLAM) to compensate ego-motion and sustain CSI while the platform is moving; extending the approach to capture command-source identification from natural, unconstrained, and small-amplitude gestures rather than a fixed gesture set \addl{and even infer from user intentions other than gesture}; exploring human-prior-free, pixel-level sync to better handle occlusion and relax assumptions about fixed IMU placement; 
\faddl{conducting systematic evaluations in multi-IMU settings and refining the pipeline to mitigate unforeseen failures that emerge during validation;} 
and conducting in-situ deployments in high-density venues (e.g., large shopping malls) to assess performance, latency, and failure modes under real-world crowding.

% \section{Limitations and Future Work}

% Limitation
% 数据集，主要在走廊/大厅，且每次两名参与者同场，偶有路人路过；总体人群密度较低（平均≈3人/帧）。这些设置与机场/商场高密度人群有差距。缺少年长者/儿童。缺乏真机实验。现在的系统对机器人自身的运动敏感，可能限制了应用场景以及效率（需要先停下来识别）。

% Future Work

% 可以考虑利用机器人上的传感器fusion进一步去掉运动

% 在自然的非固定动作下捕捉小幅度运动的CSI

% 没有human-prior的情况下直接像素级的进行匹配，这样可以解决遮挡问题以及不要求在固定的位置佩戴IMU

% 在真实场域（大型商场/航站楼）做原位评估

% \todo{证明时间鲁棒性的图}

\section{CONCLUSION}

We introduced \methods, an optical-inertial fusion framework for Command Source Identification (CSI) that binds ``who is commanding whom'' at long range without voice, markers, or user appearance prior. By matching a wearable IMU's motion spectrum to per-candidate flow spectra from a robot-mounted RGB camera, \methods~ enables robust hand gesture interactions across embodiments (service robots, drones, quadrupeds) and crowded scenes, expanding interaction range to 34\,m. Limitations include low-density test sites, constrained demographics, and sensitivity to robot ego-motion. Future work will (i) fuse on-robot sensors to operate reliably while moving, (ii) report camera-agnostic, distance-aware metrics tied to human pixel footprint, (iii) broaden to natural micro-gestures, and (iv) evaluate in situ in malls and transit hubs. Reframing long-range summoning as cross-modal matching, \methods~ advances far-range, many-to-many human-robot interaction.

\begin{acks}
This work was supported by the National Key R\&D Program of China under Grant No. 2024YFB4505500 \& 2024YFB4505501, Beijing Key Lab of Networked Multimedia, Institute for Artificial Intelligence, Tsinghua University (THUAI).
\end{acks}
\bibliographystyle{ACM-Reference-Format}
\bibliography{sample-base}

% %%
% %% If your work has an appendix, this is the place to put it.
\appendix

\begin{add}

\section{Quantitative Results of Formative Study of Example User}
\label{appendix:amp}

\begin{figure}[htbp]
  \centering 
    \includegraphics[width=0.5\linewidth]{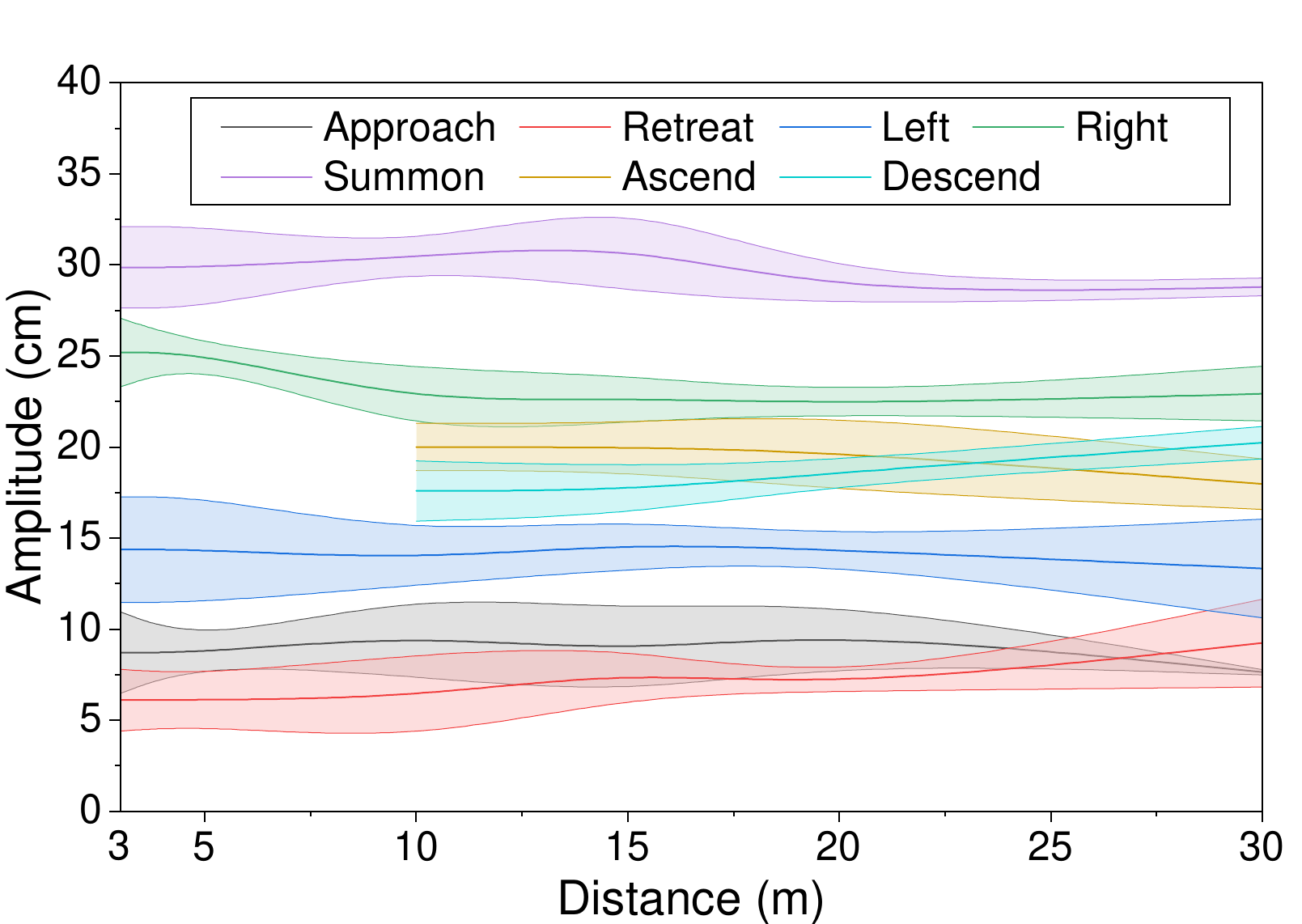}
    \caption{\textbf{Gesture Amplitude Consistency across Distances.} Data from User 1 illustrates a distance-invariant strategy, where the amplitude of gestures does not significantly correlate with the interaction distance. Error bar denotes STD over different robots.}
    \label{subfig:stable}
    \Description{This line chart, titled "Figure 24: Gesture Amplitude Consistency across Distances," plots gesture amplitude in centimeters (y-axis, 0–40 cm) against interaction distance in meters (x-axis, 3–30 m) for seven distinct robot control gestures: Approach (black), Retreat (red), Left (blue), Right (green), Summon (purple), Ascend (yellow), and Descend (cyan). Shaded error bands represent the standard deviation (STD) across different robots. The data from User 1 demonstrates a distance-invariant strategy, where gesture amplitude shows no significant correlation with interaction distance. Summon maintains the highest amplitude (around 30 cm) across all distances, while Retreat has the lowest (around 5–10 cm), with all gestures exhibiting consistent amplitude levels as distance changes.}
\end{figure}

\begin{figure}[htbp]
    \centering
    \includegraphics[width=0.5\linewidth]{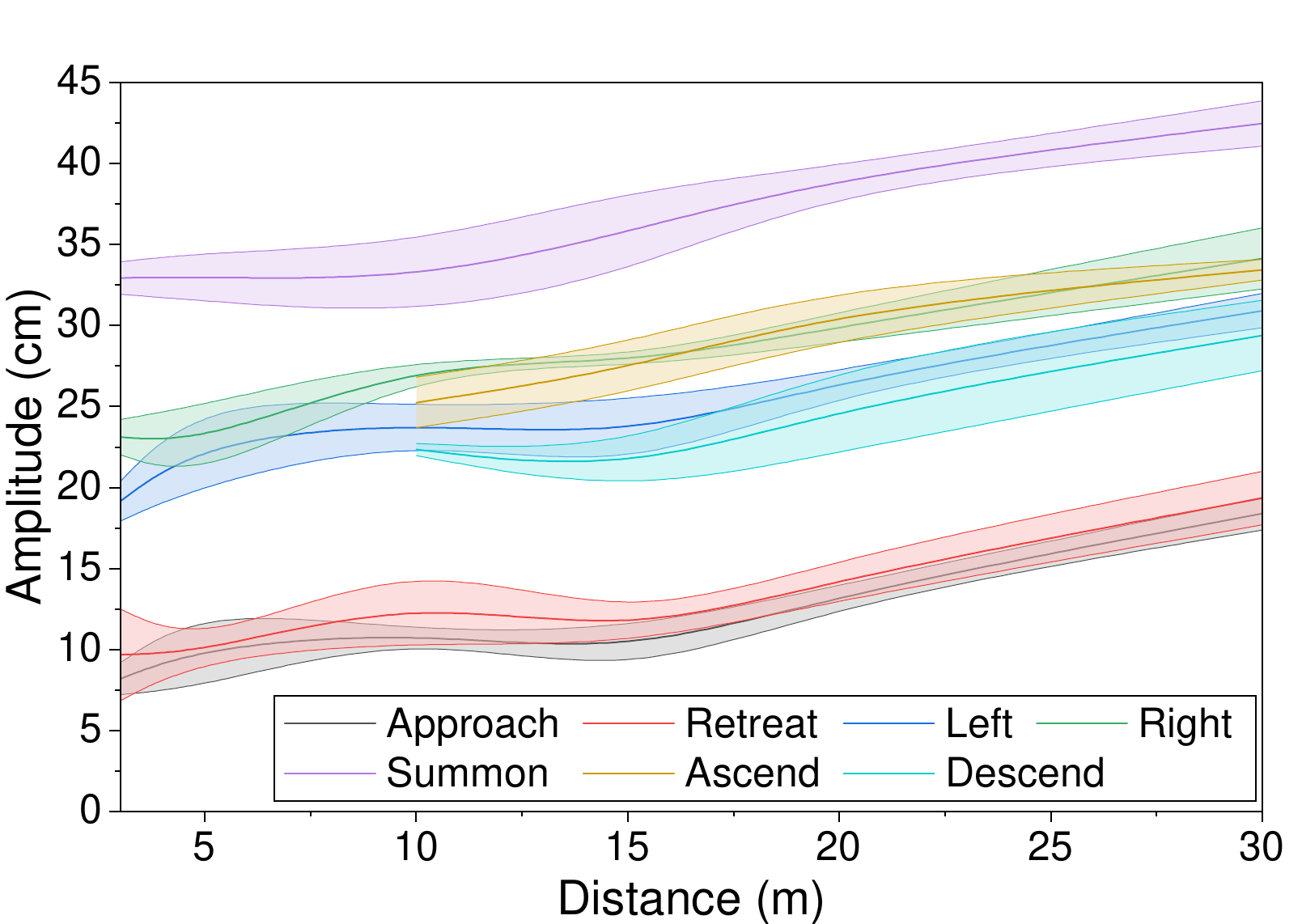}
    \caption{\textbf{Gesture Amplitude Scaling across Distances.} Data from User 2 illustrates a distance-adaptive strategy, where the amplitude of gestures positively correlates with the interaction distance. Error bar denotes STD over different robots.}
    \label{subfig:increase}
    \Description{This line chart, titled "Figure 25: Gesture Amplitude Scaling across Distances," plots gesture amplitude in centimeters (y-axis, 0–45 cm) against interaction distance in meters (x-axis, 0–30 m) for seven distinct robot control gestures: Approach (black), Retreat (red), Left (blue), Right (green), Summon (purple), Ascend (yellow), and Descend (cyan). Shaded error bands represent the standard deviation (STD) across different robots. The data from User 2 illustrates a distance-adaptive strategy, where gesture amplitude positively correlates with interaction distance: Summon maintains the highest amplitude (increasing from ~32 cm to ~42 cm), while Retreat has the lowest (increasing from ~10 cm to ~20 cm), with all gestures showing a clear upward trend as distance increases.}
\end{figure}

\section{Spectral Feature.}
\label{appendix:PSD_feature}

\subsection{Detailed Spectral Feature Definition}
% Given a length-$N$ sequence $x[n]$ sampled at $f_s$, we apply a window $w[n]$ and compute the discrete Fourier transform (DFT) $X[k]$. We then form the single-sided power spectral density (PSD) $S[k]=|X[k]|^2$ over a passband $\mathcal{B}$ that excludes a small neighborhood around DC. The \textbf{PSD} vector itself serves as our primary feature denoted $\mathbf{PSD}$. From this PSD we extract a compact set of interpretable descriptors: $p$, the \textbf{peak height} (the maximum spectral magnitude) indicating dominant rhythmic strength; $f$, the \textbf{peak frequency} (the frequency at which the peak occurs) representing the dominant tempo; $\kappa$, \textbf{clarity}, defined as the ratio of the peak to the band-mean PSD and thus quantifying periodicity relative to background noise; $H$, the \textbf{spectral entropy}, which measures the concentration of spectral energy (lower $H$ denotes a sharper, more focused peak); $\Delta f$, the \textbf{frequency spacing} between the two strongest peaks, which captures harmonic structure; $\boldsymbol{\mathrm{SNR}}$, an in-band signal-to-noise ratio comparing peak energy to a robust residual floor; and $P_{\mathrm{avg}}$, the \textbf{band-average power} reflecting overall energy. These descriptors, together with the full PSD vector, form the final \textbf{feature vector}:
\begin{equation}
\boldsymbol{\phi}(x)=[\mathbf{PSD}; \text{Feat]}=\left[\mathbf{PSD},\, p,\, f,\, \kappa,\, H,\, \Delta f,\, \mathrm{SNR},\, P_{\mathrm{avg}}\right].
\end{equation}

\begin{itemize}
    \item \textbf{Peak Height ($p$):} The maximum magnitude in the power spectral density. This metric captures the intensity of the dominant rhythmic component, facilitating the matching of motion strength between the visual and inertial modalities.
    
    \item \textbf{Peak Frequency ($f$):} The frequency at which the spectral peak occurs. It indicates the fundamental tempo of the gesture, serving as the primary cue for synchronizing the speed of the user's hand movement across different sensors.
    
    \item \textbf{Clarity ($\kappa$):} Defined as the ratio of the peak magnitude to the average spectral power. This quantifies the periodicity of the signal, effectively distinguishing deliberate, rhythmic command gestures from irregular or non-periodic background noise.
    
    \item \textbf{Spectral Entropy ($H$):} A measure of the concentration of spectral energy distribution. Lower entropy corresponds to a sharper peak, indicating a "purer" or more focused single-frequency movement, which helps in filtering out complex or chaotic motion artifacts.
    
    \item \textbf{Frequency Spacing ($\Delta f$):} The frequency difference between the two most significant peaks. This characterizes the harmonic structure of the motion, acting as a kinematic signature to differentiate between users who may be gesturing at similar fundamental speeds but with different motion styles.
    
    \item \textbf{Signal-to-Noise Ratio ($\mathrm{SNR}$):} The ratio of the in-band peak energy to the estimated residual noise floor. This evaluates the signal quality, ensuring that the system prioritizes clean, high-confidence motion signals over sensor noise or environmental interference.
    
    \item \textbf{Band-Average Power ($P_{\mathrm{avg}}$):} The mean power calculated across the passband. This represents the overall energy expenditure of the movement, aiding in the discrimination between high-energy (e.g., large arm waves) and low-energy (e.g., subtle wrist flicks) gestures.
\end{itemize}

\subsection{Spectral Feature Ablation Results}

\begin{figure}[H]
    \centering
    \includegraphics[width=0.7\linewidth]{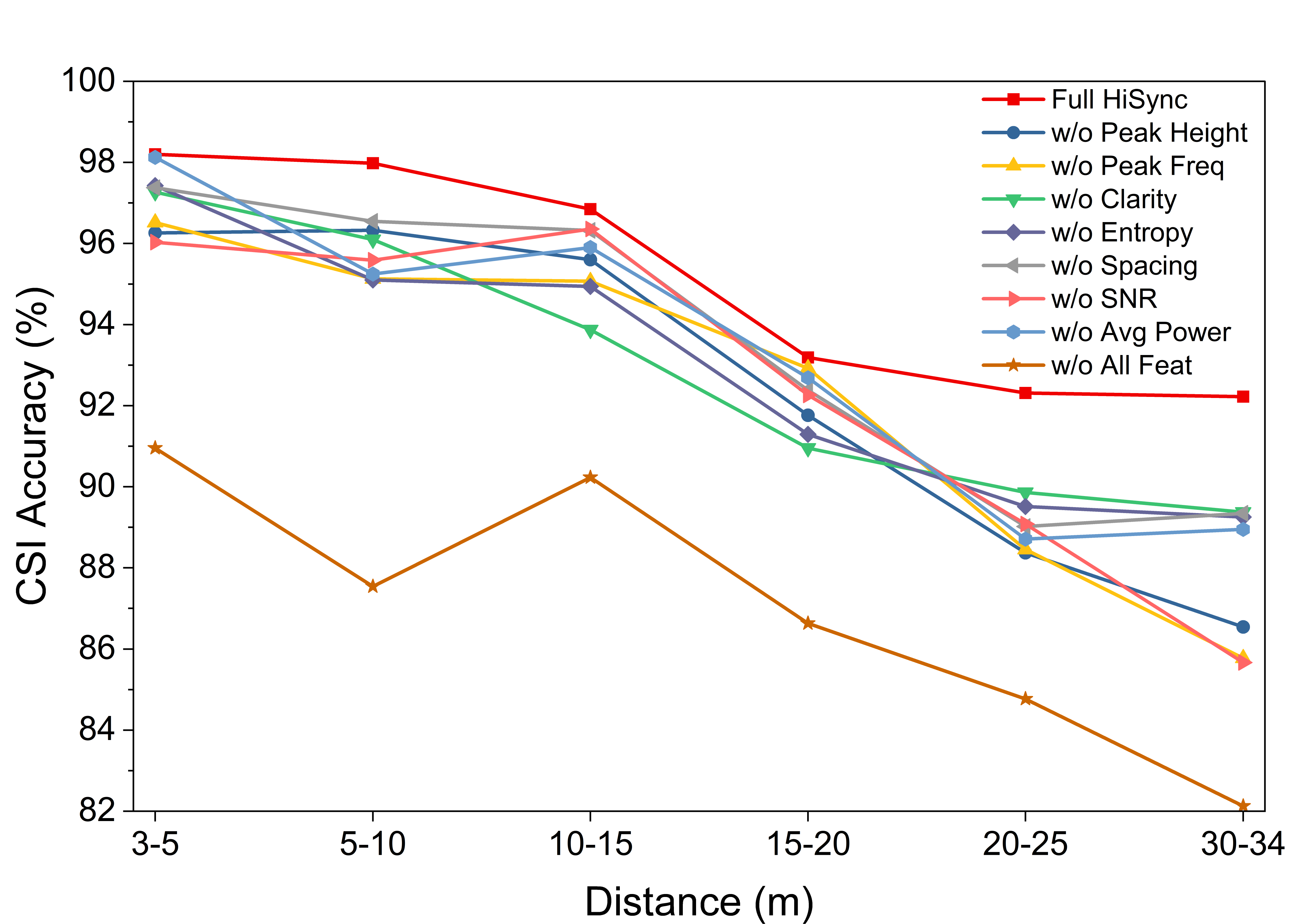}
    \caption{Spectral Feature Ablation Results}
    \label{fig:ablation_feat}
    \Description{This line chart, titled "Figure 26: Spectral Feature Ablation Results," plots CSI accuracy (\%) on the y-axis against interaction distance (m) on the x-axis, with distance bins at 3–5, 5–10, 10–15, 15–20, 20–25, and 30–34 meters. It compares 10 different spectral feature ablation conditions: Full HiSync (red), w/o Peak Height (blue), w/o Peak Freq (yellow), w/o Clarity (green), w/o Entropy (purple), w/o Spacing (gray), w/o SNR (pink), w/o Avg Power (light blue), w/o All Feat (orange), and the baseline. The results show that CSI accuracy generally decreases as distance increases, with the Full HiSync condition consistently achieving the highest accuracy across all distances, while removing all spectral features (w/o All Feat) leads to the lowest performance, particularly at longer distances, demonstrating the effectiveness of the spectral feature design.}
\end{figure}

As shown in the Fig.~\ref{fig:ablation_feat}, the experimental results demonstrate the effectiveness of the Spectral Feature design, with a particularly notable performance improvement observed at longer distances.

% \section{User Instructions.}

\end{add}









\end{document}